\documentclass[letterpaper,11pt]{article}

\AtBeginDocument{
	\addtocontents{toc}{\normalsize}}
\pdfoutput=1

\usepackage{amsmath,amssymb,amsfonts,amsthm,amscd,mathrsfs}
\usepackage{geometry}
\usepackage[noadjust]{cite}
\usepackage{array,mathrsfs,amsfonts,yfonts,dsfont,bbm,colonequals}
\usepackage{xcolor}
\usepackage[all]{xy}
\usepackage[colorlinks=true, 
linkcolor=teal!75!black,
urlcolor=teal!75!black,
citecolor=newyellow,
]{hyperref}

\usepackage{comment}
\usepackage{tikz}
\usetikzlibrary{shapes}
\usetikzlibrary{positioning}
\usepackage{tkz-euclide}

\usepackage{subcaption}
\usepackage{tikz}
\usepackage{pgfplots}
\usepackage{adjustbox}

\definecolor{newyellow}{RGB}{128, 10, 7}

\newcommand{\bD}{\mathbb{ D}}
\newcommand{\Ccal}{\mathcal{C}}
\newcommand{\s}{\sigma}
\newcommand{\be}{\begin{equation}}
\newcommand{\ee}{\end{equation}}
\newcommand{\Ocal}{\mathcal{O}}
\newcommand{\hOcal}{\hat{\mathcal{O}}}
\newcommand{\thetab}{{\bar \theta}}
\newcommand{\psib}{{\bar \psi}}

\newcommand{\hD}{{\hat \Delta}}

\newcommand{\bea}{\begin{eqnarray}}
\newcommand{\eea}{\end{eqnarray}}
\newcommand{\ba}{\begin{equation}\begin{aligned}}
\newcommand{\ea}{\end{aligned}\end{equation}}
\newcommand{\fchiu }{\,}

\newcommand{\hl}{{\hat {\ell}}}

\definecolor{red}{rgb}{0.7,0.3,0.3}
\definecolor{green}{rgb}{0.3,0.7,0.3}
\definecolor{blue}{rgb}{0.3,0.3,0.7}

\renewcommand{\u}{u}
\renewcommand{\v}{v}

\newcommand{\De}{\Delta}

\newcommand{\xperp}{x^{\perp}}
\newcommand{\xpar}{x^{\paral}}

\newcommand{\yperp}{y^{\perp}}
\newcommand{\ypar}{y^{\paral}}
\newcommand{\Xperp}{X^{\perp}}
\newcommand{\Xpar}{X^{\paral}}
\newcommand{\so}{SO}

\newcommand\myatop[2]{\genfrac{}{}{0pt}{}{#1}{#2}}

\def\XXint#1#2#3{{\setbox0=\hbox{$#1{#2#3}{\int}$}
     \vcenter{\hbox{$#2#3$}}\kern-.5\wd0}}

\def\s{\sigma}

\def\e{\epsilon}

\def\D{\Delta}

\def\l{\lambda}

\def\w{w}

\newcommand{\up}{u}
\newcommand{\ut}{v}
\newcommand{\z}{w}
\newcommand{\vt}{v}

\makeatletter
\newcommand\xleftrightarrow[2][]{%
  \ext@arrow 9999{\longleftrightarrowfill@}{#1}{#2}}
\newcommand\longleftrightarrowfill@{%
  \arrowfill@\leftarrow\relbar\rightarrow}
\makeatother

\newcommand\boxedB[1]{{\setlength\fboxsep{10pt}\boxed{#1}}}

\newcommand{\paral}{{\scalebox{0.7}[0.6]{$\parallel$}}}

\makeatletter

\@addtoreset{equation}{section}

\makeatother

\begin{document}
	
	\vspace*{-.6in} \thispagestyle{empty}
	\begin{flushright}
			\end{flushright}
	\vspace{1cm} {\Large
		\begin{center}
			{\bf 
            A Tale of Two Uplifts: Parisi–Sourlas with Defects
            }\\
	\end{center}}
	\vspace{1cm}
	\begin{center}
		{\bf Kausik Ghosh$^{\color{newyellow} \paral}$}
        {\bf and Emilio Trevisani$^{\color{teal} \perp}$}
        \\[1cm] 
		{
			{\em ${}^{\color{newyellow} \paral}$Department of Mathematics, King's College London, Strand, London, WC2R 2LS, United Kingdom
            }
}\\
 \vspace{.3cm}
{${}^{\color{teal}\perp}\!\!$
	{\em  Laboratoire de Physique Th\'eorique et Hautes \'Energies, CNRS \& Sorbonne Universit\'e, \\ 4 Place Jussieu, 75252 Paris, France \\
    \vspace{.3cm}
    ${}^{\color{teal}\perp}\!\!$ Department of Theoretical Physics, CERN, 1211 Meyrin, Switzerland
    }}
   \normalsize	
	\end{center}
	
	\begin{center}
		{  \texttt{\  kau.rock91@gmail.com, emilio.trevisani.et@gmail.com} 
		}
		\\
	\end{center}
	
	\vspace{8mm}
	
	\begin{abstract}
 \vspace{2mm}
Defects in conformal field theories (CFTs) play a key role in critical phenomena by modifying scaling behaviors and generating new universality classes. We introduce Parisi–Sourlas (PS) supersymmetry in the presence of extended operators and demonstrate that any $p$-dimensional defect in a CFT$_d$ can be uplifted to a defect in a PS-supersymmetric CFT$_{d+2}$. Surprisingly, there are actually two distinct uplifted defects—of dimensions $p$ and $p+2$—which reduce to the original one. 
We show how this reduction works for correlators with insertions both in the bulk and on the defect. 
As a byproduct, we find new relations between defect conformal blocks in dimensions $d$ and $d+2$. 
We further show that the reduction of the $p$-dimensional defect implies and extend a “global symmetry reduction” previously considered in the literature.
Finally, we provide various examples of these uplifts, including perturbative computations in epsilon expansion of the uplift of the Ising magnetic line defect, as well as   exact computations of observables in the four-dimensional uplift of minimal models with boundaries.

\end{abstract}
	\vspace{2in}
	\newpage
	{
		\setlength{\parskip}{0.05in}
		\tableofcontents
		\renewcommand{\baselinestretch}{1.0}\normalsize
	}
	\setlength{\parskip}{0.1in}
 \setlength{\abovedisplayskip}{15pt}
 \setlength{\belowdisplayskip}{15pt}
 \setlength{\belowdisplayshortskip}{15pt}
 \setlength{\abovedisplayshortskip}{15pt}
	\bigskip \bigskip
	\section{Introduction}

In \cite{Parisi:1979ka}, a new class of models invariant under Parisi-Sourlas (PS) supersymmetry was proposed. These supersymmetric theories exhibit several intriguing properties. First, they obey dimensional reduction: specifically, a PS theory in $d+2$ dimensions can be described by a $d$-dimensional non-supersymmetric model \cite{Parisi:1979ka, CARDY1983470, CARDY1985383, KLEIN1983473, Klein:1984ff, paper1}.\footnote{Historically, dimensional reduction was first proposed for random field models \cite{PhysRevLett.37.1364}, and was later rephrased in terms of Parisi–Sourlas supersymmetry \cite{Parisi:1979ka}. }
Second, this form of supersymmetry sometimes emerges at the fixed point of the renormalization group (RG) flow in models with random-field (RF) disorder, as demonstrated by numerical simulations and supported by various theoretical arguments. 

PS supersymmetry does not always arise in all RF models, for instance, it appears in the RF Ising model in five dimensions but not in $D=3,4$, while in the RF $\phi^3$ model, it emerges across all dimensions $2 \leq D < 8$ (for further details, see \cite{PhysRevLett.35.1399,
Parisi:1979ka,
PhysRevLett.37.1364, CARDY1985123, PhysRevLett.88.177202,
Fytas_2019,
PhysRevE.95.042117,
PhysRevLett.122.240603,
PhysRevLett.116.227201,
PhysRevLett.46.871,
PhysRevB.78.024203, Tarjus:2024mop, 
Kaviraj:2020pwv,
Kaviraj:2021qii, Kaviraj:2022bvd, Rychkov:2023rgq, Piazza:2024wll}
and references therein; see also \cite{PARISI1982321, 
Brezin:1984ie, Kaviraj:2024cwf, 
05b5fc40-d3a0-33cf-b199-5d8140f70420,
Cardy:2023zna,
Nakayama:2024jwq} for more applications of PS supersymmetry).
On the contrary, dimensional reduction seems to universally work. This motivated the study of the so called ``dimensional uplift'', namely the idea that any $d$-dimensional model could be embedded into a PS theory living in $d+2$ dimensions. This was investigated in \cite{paper1, Trevisani:2024djr} in the context of CFTs. These works suggest that any CFT$_d$ can be embedded in a  CFT with PS supersymmetry (PS CFT) living in $d+2$ dimensions. 
This was checked at the level of four-point functions of local scalar operators, which were shown to satisfy all the required CFT axioms as crossing symmetry and conformal block decomposition.

\begin{figure}[ht]
\centering
\begin{tikzpicture}[scale=2]
  \node (Em) at (1.8,0.46) {  $ 
  \myatop{\scriptstyle \textrm{Emergence}}{\scriptstyle \textrm{of SUSY}}
    $};
\node(Uplift) at (2.74,-0.27) {\scriptsize $ \textrm{Uplift} $};
\node(Red) at (3.2, -0.78) {\scriptsize$  \textrm{Reduction} $};
  \node (RF) at (0,0) {} ;
\tikzstyle{every node}=[draw, ellipse, fill=teal!25!white, ultra thick, minimum width=20pt, align=center]
 \node[right=-2cm of RF]  {
RF$_{d+2}$ f.p.
 \\
 with defect
 };
 \node[right=6 cm of RF] (PS) {PS CFT$_{d+2}$ \\with defect};
   \node[below right=1cm and 2.1cm of RF] (dhat) {CFT$_{d}$\\with defect};
     \draw[-latex,ultra thick,black] plot [smooth,tension=1] coordinates { (2.36,-0.66) (2.7,-0.46)(3.3,-0.25)};
    \draw[latex-,ultra thick,black] plot [smooth,tension=1] coordinates { (2.43,-0.83) (3.2,-0.54)(3.47,-0.32)};
    \draw[-latex,ultra thick,black] plot [smooth,tension=1] coordinates { (0.6,0.1) (1.75,0.2)(3.1,0.1)};
\end{tikzpicture}%
\caption{ \label{fig:relations}
Diagram of relations between the fixed point of a random field theories in $d+2$ dimensions (RF$_{d+2}$ f.p.) in the presence of a defect, a PS  CFT$_{d+2}$ with defects and a pure CFT$_{d}$ with a defect.}
\end{figure}
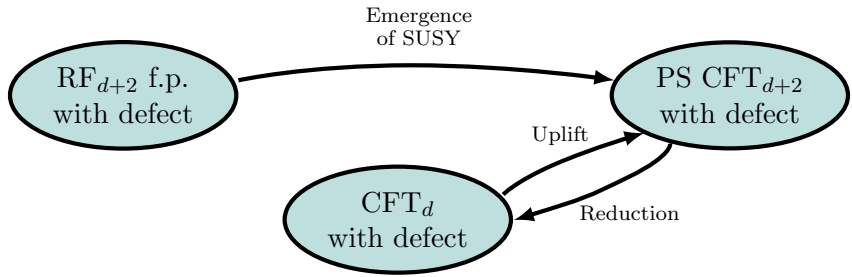

Nevertheless, a CFT also contains extended operators like $p$-dimensional conformal defects. Defect and boundary conformal field theories (DCFTs and BCFTs)  have long played a vital role in understanding systems with spatial inhomogeneities. Their conceptual roots go back to studies of critical phenomena near special surfaces in statistical mechanics, for example, how fluids or magnets behave at interfaces, which were effectively captured using BCFT techniques (see \cite{Diehl:1996kd} for a review). In condensed matter physics, BCFTs provide the theoretical underpinning for phenomena like the Kondo effect—where a magnetic impurity alters the resistivity of a metal at low temperatures—as elegantly explained by 
Affleck and Ludwig \cite{PhysRevB.48.7297}. More broadly, DCFTs describe the physics of localized impurities, junctions in quantum wires, and edge states in quantum Hall systems, many of which have been realized experimentally \cite{Kane:1992xse}. Recently, there has been renewed interest in studying the magnetic line defect in the Ising model \cite{allais2014magneticdefectlinecritical, Parisen_Toldin_2017} and its
Wilson-Fisher fixed point realization. Also see the recent work on clarifying the boundary universality landscape in 3D \cite{Metlitski:2020cqy}. 

Given the importance of DCFTs it is  natural  to ask whether defects in RF models also have emergent PS supersymmetry and whether these supersymmetric defects can be dimensionally reduced to a lower dimensional DCFT, as  shown in figure \ref{fig:relations}.
 This is not an academical question, indeed RF models do possess physical defects. For example, RF models are typically defined in finite volume and their boundary can be described by a BCFT. Also one can generalize the idea of the magnetic line defect in the Ising model when switching on the RF interaction. 
Whether these setups flow in the IR to a defect in a PS CFT will be the topic of a companion paper \cite{WIP}.

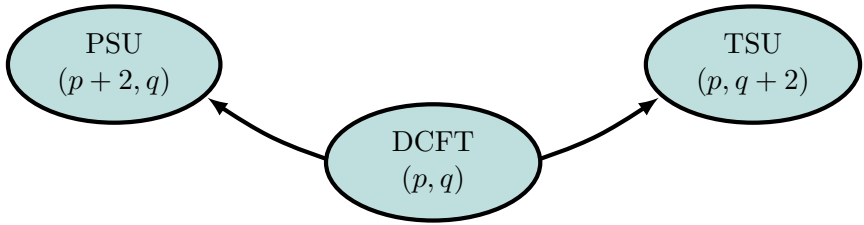
\begin{figure}[ht]
\centering
\begin{tikzpicture}[scale=2]
\tikzstyle{every node}=[
  draw, ellipse,   fill=teal!25!white,
  ultra thick, minimum width=20pt, align=center
]
\coordinate (Uplift1) at (0,0);

\node[left=2.8cm of Uplift1
] (RF) {
\hspace{0.36cm} PSU \hspace{0.36cm} \\ 
$(p+2,q)$
};

\node[right=2.8cm of Uplift1
] (PS) {
\hspace{0.36cm} TSU \hspace{0.36cm} \\ 
$(p,q+2)$
};

\node[below=0.5cm of Uplift1] (dhat) {
\hspace{0.20cm} DCFT \hspace{0.20cm} \\
$(p,q)$
};

\draw[-latex,ultra thick,black] plot [smooth,tension=0.8] coordinates { (0.7,-0.63) (1.1,-0.47)(1.5,-0.22)};

\draw[-latex,ultra thick,black] plot [smooth,tension=0.8] coordinates { (-0.7,-0.63) (-1.1,-0.47)(-1.5,-0.22)};

\end{tikzpicture}
\caption{\label{fig:twouplift}Two different PS uplifts of the same $(p,q)$ defect theory.}
\end{figure}

In this work we focus on the other part of the question, namely whether defects in PS CFT undergo dimensional reduction and uplift. 
There are various reasons why this is important. 
First of all, since a full-fledged CFT$_p$ lives on top of a $p$-dimensional defect, proving the existence of uplifted defects provides a very non-trivial check of the existence of the uplifted theory.
Moreover, the uplift of extended operators seems puzzling. Indeed, given a $p$-dimensional defect in a CFT$_d$, how are we supposed to embed it in a PS CFT$_{d+2}$? Should we still consider a $p$-dimensional defect or should we instead uplift it to a $p+2$-dimensional one? 
Interestingly enough, we will show that both uplifts are possible. This is by itself a very surprising result, since the uplifted theory hosts two extended objects of different dimensionalities which 
dimensionally reduce to the same object.  
More concretely we will show the following results:
\begin{itemize}
    \item 
    Let us label a DCFT by the $(\text{dimension},\text{codimension})$ of the defect. Any given DCFT labelled by $(p,q)$  can be uplifted to two distinct PS DCFTs labelled by $(p+2,q)$ and $(p,q+2)$ which we  dub respectively as the parallel space uplift (PSU) and the transverse space uplift (TSU), as shown in figure \ref{fig:twouplift}.
    We provide perturbative and non perturbative arguments supporting this claim.
    \item We derive novel results for defect conformal blocks of $(p,q+2)$ and $(p+2,q)$ theories in terms of those in the $(p,q)$ defect theory.
    \item The TSU construction implies a reduction/uplift at the level of global symmetry which relates $OSp(m|n)$ to $O(m-n)$. This has been observed in many examples case by case, whereas our construction implies that such relation should hold more generally.
    \item We apply the uplift to compute 
    exact observables of 
    an infinite class  of solvable non unitary DCFTs in $d=4$, which corresponds to the uplift of boundaries in minimal models.
\end{itemize}

The plan of the paper is as follows. In section \ref{sec:conventions} we will review the main characters of this work, namely PS supersymmetry and DCFTs.
In section \ref{sec:uplifts} we will define the two possible uplifted defects in a $d+2$-dimensional PS theory and we will show why they both dimensionally reduce to the same defect in a $d$-dimensional model. 
In section \ref{sec:correlators} we study correlators of local operators in the presence of defects in a PS CFT$_{d+2}$. We will show that if the they arise by uplifting correlators of a CFT$_{d}$, then they automatically satisfy the bootstrap axioms of crossing symmetry and conformal block decomposition. As a byproduct this gives rise to numerous identities for the associated conformal blocks. In section \ref{sec:examples} we provide various examples of the uplifted defects: we start from the simple checks of trivial defects and free theory defects, we then proceed to the perturbative example of a magnetic line in the Wilson-Fisher fixed point and finally give the non-perturbative example of BCFTs in minimal models.
We conclude the paper with some final remarks in section \ref{sec:discussion} and 
we provide extra details in the appendices and in an ancillary Mathematica file. 

\section{Conventions and review}
\label{sec:conventions}
In this section we review the topics of 
PS supersymmetry and DCFT, while   introducing the conventions used in the rest of the paper.  
\subsection{Review of PS supersymmetry, dimensional reduction and uplift}
\label{sec:review_PS}
In the following we briefly review some features of theories with PS supersymmetry and we explain how they are related by dimensional reduction to theories living in two less dimensions. For more details see e.g.  \cite{paper1, Trevisani:2024djr}. 

PS QFTs can be thought as usual QFTs where the spacetime is promoted to a superspace whose fermionic directions transform as spacetime scalars.
In particular superspace coordinates are defined as $y=(X^1,\dots,X^{d+2},\theta,\thetab)\in \mathbb{R}^{d+2|2}$, where  $\theta,\thetab$ are the Grassmann variables and we chose to have  $d+2$ bosonic directions $X\in \mathbb{R}^{d+2}$ for future convenience. 
A very explicit example of these theories is defined by the following scalar action in superspace,
\be
\label{PS_scalar_action}
\int  d^{d+2}X [d \theta] [d \thetab] \left[-\frac{1}{2}\Phi(y) \square_y  \Phi(y) + V(\Phi(y))\right] \, ,
\ee
where $V$ is a potential, $\square_y \equiv \partial^2_x+2\partial_{\thetab}\partial_{\theta}$ and we defined $2\pi [d \theta][d \thetab]\equiv d \theta d \thetab$.
This action is invariant under superspace translations and rotations which define a super-Poincaré algebra.

At the fixed points of RG flow  these theories are expected to describe PS CFTs, which enjoy PS superconformal symmetry  $OSp(d+2,1|2)$. In practice we can can realize these transformation as translations, rotations, dilatations and special conformal transformations in superspace. 
One can then classify superprimary operators $\Ocal(y)$ (and states) in terms of their dimension $\Delta$ under superdilations and spin $\ell$ under $OSp(d+2|2)$. One  finds that correlation functions of the PS CFTs look very much as standard CFT correlators, with the only distinction that the coordinate in spacetime are replaced by coordinates $y$ in superspace. E.g. scalar two- and three-point functions take the form
\be
\! \! \! \!  
\langle \Ocal(y_1)\Ocal(y_2) \rangle= \tfrac{1}{(y^2_{12})^{\Delta} }
\, ,
\;
\langle \Ocal_1(y_1)\Ocal_2(y_2)\Ocal_3(y_3) \rangle= \tfrac{\displaystyle\lambda_{123}}{(y^2_{12})^{\frac{\Delta_1+\Delta_2-\Delta_3}{2}}
(y^2_{23})^{\frac{\Delta_2+\Delta_3-\Delta_1}{2}}
(y^2_{13})^{\frac{\Delta_1+\Delta_3-\Delta_2}{2}}
} 
\, ,
\ee
where $\lambda_{123}$ is an OPE coefficient and 
 $y_{ij}^2=X_{ij}^2-2\theta_{ij} \thetab_{ij}$ where subscripts $ij$ define differences of coordinates at point $i$ and $j$. Similarly also higher point functions and correlators of spinning operators would look like canonical correlators as functions of $y$. 

\begin{figure}[htbp]
\centering
\begin{tikzpicture}[scale=1.5]
\tkzInit[xmin=0,xmax=5.5,ymin=0,ymax=1.5] 
    \tkzClip[space=.5] 
    \tkzDefPoint(0.3,0){A} 
    \tkzDefPoint(4.5,0){B} 
    \tkzDefPoint(5.5,1){C} 
    \tkzDefPointWith[colinear= at C](B,A) \tkzGetPoint{D}
       \tkzDrawPolygon[fill=black!10](A,B,C,D)
    \tkzDefPoint(2. ,0.4){X1} 
      \tkzDefPoint(2. ,0.44){X1p} 
       \tkzDefPoint(2. ,1.4){Y1} 
     \tkzDefPoint(4. ,0.7){Xn} 
        \tkzDefPoint(4. ,0.74){Xnp} 
          \tkzDefPoint(4. ,1.3){Yn} 
          \draw[fill=black, draw=black] (X1) circle (0.9pt);
           \draw[fill=black, draw=black] (Xn) circle (0.9pt);
            \draw[fill=black, draw=black] (Y1) circle (0.9pt);
           \draw[fill=black, draw=black] (Yn) circle (0.9pt);
        \draw[dashed, -stealth](Y1)->(X1p);
       \draw[dashed, -stealth](Yn)->(Xnp);
    \node[below] at (X1) { 
    };
  \node[below] at (Xn) { 
  };
  \node[above] at (Y1) { $ y_1$};
  \node[above] at (Yn) { $ y_n$};
  \node[above,rotate=-2] at (3.,1.35) { $ \dots$};
  \node[above,rotate=7] at (3.,0.4) { $ \dots$};
   \node[above] at (1.2,1.35) {$\mathbb{R}^{d+2\mid 2}$};
    \begin{scope}[every node/.append style={xslant=0, yslant=0}]
     \node[above] at (1.2,0.1) {$\mathbb{R}^{d}$}; 
               \end{scope}
\end{tikzpicture}%
\caption{ \label{fig:dim_red}
Dimensional reduction. }
\end{figure}
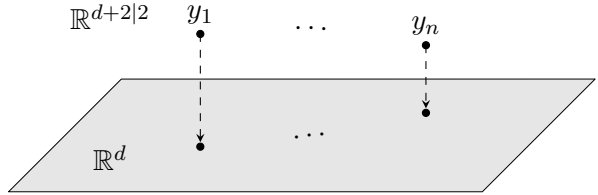

A very interesting feature of PS SUSY theories (this is valid also for PS QFTs) is that they satisfy the so called dimensional reduction. The idea is simple. Let us consider a correlator of $n$ superfields, which we shall consider as scalars for simplicity, placed at positions $y_1,\dots,y_n$. Dimensional reduction states that if we place all these insertions in $\mathbb{R}^{d}$, as shown in figure \ref{fig:dim_red}, the resulting correlator describes a conventional $d$-dimensional model. 
Let us be more specific. If we apply dimensional reduction to the theory \eqref{PS_scalar_action}, we find that the resulting model is described by reduced $d$-dimensional action 
\be
\label{scalar_action}
 \int  d^{d}x \left[-\frac{1}{2}\phi \square_x  \phi + V(\phi(x))\right] \,  ,
\ee
which is the same scalar action as before but now defined on $\mathbb{R}^{d}$.
The same idea can be applied to correlators of PS CFTs.
While by reducing a usual CFT we expect each operator in the higher dimensional theory to give rise to a tower of infinitely many operators in lower dimensions, for PS CFTs it can be proved that in the reduced CFT infinite towers of operators decouple leaving a spectrum of primaries which matches the spectrum of the superprimaries of the original PS CFTs. To be precise some PS CFT operators are mapped to zero under dimensional reduction, but it never happens that they give rise to infinitely many operators.
A surprising feature worth mentioning is that if the PS theory is local, also the reduced one is local, meaning that it will contain a conserved stress tensor.
The fact that the spectrum of operators matches in the PS CFT and in the reduced CFT can be also seen at the level of conformal blocks. Indeed it was found that superconformal blocks in superspace are equal to the usual conformal blocks in the reduced theory. 

The fact that the PS theories would give rise to lower dimensional models, motivated the investigation of the opposite question: is  a generic QFT always associated to a higher dimensional PS theory which reduces to it? This is called ``dimensional uplift''. From the scalar action \eqref{scalar_action} point of view it seems that this is indeed the case, since we can define an uplifted model \eqref{PS_scalar_action} which reduces to  \eqref{scalar_action}. There are also perturbative indications that more generic models with gauge fields and spinors would also have an uplifted description \cite{MCCLAIN1983430}. 
One can also study this question non-perturbatively in the context of CFTs. In particular in \cite{Trevisani:2024djr} it is shown that from any scalar four-point function of a CFT$_d$ it is possible to generate a set of 43 correlators of a PS CFT in $d+2$ dimensions, related by PS supersymmetry which satisfy the required bootstrap axioms of crossing and conformal block decomposition. This is conjectured to work for any four-point function, also of operators with spin, which would thus imply that the CFT$_d$ can be fully mapped to a PS CFT in $d+2$.

In the present paper we want to extend the dimensional reduction and uplift to observables which include non-local operators.

\subsection{Review of DCFT}
\begin{figure}[ht]
\centering
\begin{tikzpicture}[scale=1.5,>={Stealth[scale=1.1]}]
\tkzInit[xmin=0,xmax=5.5,ymin=0,ymax=1]  \tkzClip[space=.5] 
    \tkzDefPoint(0.3,0){A} 
    \tkzDefPoint(4.5,0){B} 
    \tkzDefPoint(5.5,1){C} 
    \tkzDefPointWith[colinear= at C](B,A) \tkzGetPoint{D}
\tkzDrawPolygon[fill=black!10,opacity=0.9](A,B,C,D)
\tkzDefPoint(3.3,0){A1} 
    \tkzDefPoint(4.3,1){B1}     
     \draw[newyellow,thick](A1)->(B1);   
    \tkzDefPoint(2.5 ,0.7){X1} 
          \tkzDefPoint(3.8 ,0.5){XPP}    
          \draw[fill=black, draw=black] (X1) circle (0.9pt);
               \draw[fill=black, draw=black] (XPP) circle (0.9pt);
    \node[below] at (X1) { $ x$
    };
   \node[below] at (XPP) { $\xpar$
   };
    \begin{scope}[every node/.append style={xslant=0, yslant=0}]\node[above] at (3.3,-0.4) {$\mathbb{R}^{p}$};
    \end{scope}
    \begin{scope}[every node/.append style={xslant=0, yslant=0}]
     \node[above] at (1.2,0.1) {$\mathbb{R}^{d}$}; 
               \end{scope}
\end{tikzpicture}%
\caption{ \label{fig:defect}
Extended $p$-dimensional defects in CFT$_d$.  }
\end{figure}
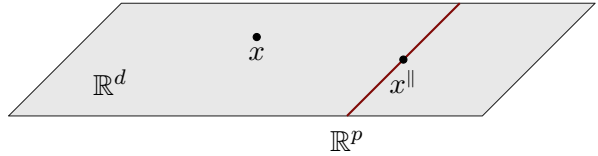

In this section, we give a brief overview of DCFTs and introduce the relevant correlation functions studied in this paper. See \cite{Billo:2016cpy,Lauria:2017wav,Lauria:2018klo} for more details.
We consider  a $p$-dimensional flat defect in $\mathbb{R}^d$ as shown in figure \ref{fig:defect}.
The presence of such defect selects special directions in spacetime $\xpar \in \mathbb{R}^{p}$, $\xperp\in \mathbb{R}^{q}$ respectively parallel and orthogonal to the defect, 
where $q\equiv d-p$ denotes the co-dimension of the defect.
Accordingly, the full conformal group $\so(d+1,1)$ breaks to the following subgroup $\so(d+1,1)\rightarrow \so(p+1,1)\times \so(q)$.
We can classify operators into two categories: bulk operators and defect operators, which live in the bulk and on the defect, respectively. 
Bulk operators $O$ are not affected by the presence of the defect and are thus labelled by a dimension $\Delta$ and an  $\so(d)$-spin $\ell$.
The usual OPE relations between bulk operators away from the defect remain unchanged, e.g.
\begin{equation}
    O_1(x_1)O_2(x_2)\sim \sum_{\mathcal{O}
    } \lambda_{O_1O_2O
    } |x_{12}|^{\Delta
    -\Delta_1-\Delta_2}O
    (x_2)+\dots \, .
\end{equation}
The defect operators carry ``parallel'' spin under the preserved subgroup \( \so(p) \) and transform in representations of the transverse rotation group \( \so(q) \). Each defect operator is therefore labeled by its conformal dimension \( \hat{\Delta} \), parallel spin \( \hl\), and transverse spin \( s \).
In the presence of a defect, we can also quantize the theory around a point on the defect. The local primary and descendant operators living on the defect form a complete basis for the Hilbert space. As a result, bulk operators can be expanded in terms of defect operators,
\be \label{eq:opebulktodefect}
O(x_1)\sim \sum_{\hat{O}}b_{O\hat{{O}}} |\xperp_1|^{\hat{\Delta}-\Delta_{O}}\hat{O}(\xpar_{1}) \, .
\ee
This introduces new OPE data: the bulk to defect OPE coefficient $b_{{O}\hat{{{O}}}}$, which defines the normalization of the two-point function of a bulk and defect operator $ \langle O \hat{O}\rangle $. In particular, when $\hat{O}$ is the defect identity, $b_{{O}\hat{1}}\equiv a_{O}$ is the normalization of the one-point function of $O$. Namely,
\be
\langle O(x_1) \hat{O}(\xpar_{2}) \rangle = \frac{b_{O \hat{O}} }{|\xperp_1|^{\Delta-\hD} |\xpar_{12}|^{2\hD}} \, , \qquad \qquad 
\langle O(x) \rangle = \frac{a_O}{|\xperp|^{\Delta}} \, .
\ee
The defect operators are constrained by the $p$-dimensional conformal group, and therefore their OPE is fixed in the usual way in terms of the data $(\hat{\Delta},\hat \ell,s), \lambda_{\hat{O}_i \hat{O}_j \hat{O}_k}$. The set of CFT data
\[
\left\{ (\hat{\Delta},\hat \ell,s), \lambda_{\hat{O}_i \hat{O}_j \hat{O}_k}, b_{{O} \hat{O}}, a_O \right\}
\]
is new, in addition to the usual bulk CFT data (i.e., data in the absence of a defect).

The two-point function of bulk operators is the first non trivial correlation function in presence of defect. It is fixed up to an arbitrary function of two conformal cross ratios, $\up$ and $\ut$,
\begin{equation}
\label{2ptD}
  \langle O_1(x_1) O_2(x_2) \rangle  =\frac{f(\up,\ut)}{|\xperp_1|^{\Delta_1}|\xperp_2|^{\Delta_2}} \, ,
\end{equation}
where 
\be
\label{def_CR}
\up \equiv \frac{1}{2}\frac{|\xpar_{12}|^2 + (\xperp_1)^2 + (\xperp_2)^2 }{|\xperp_1| |\xperp_2| },
 \, 
\qquad
\qquad
\ut \equiv \frac{\xperp_1 \cdot \xperp_2 }{|\xperp_1| |\xperp_2| } \, .
\ee
The function $f(\up,\ut)$ can be expanded both in the defect channel and bulk channel conformal blocks which we respectively picture as,
\be
\label{blocks_schematic}
\hat{g}^{(p,q)}_{\hat{\Delta},s}(\up,\ut)\equiv
 \raisebox{3.3em}{
$
\xymatrix@=3pt{
&&& _{(p,q)} 
\ar@{-}@<-0.1mm>@[newyellow][dd] 
\ar@{-}@<0.1mm>@[newyellow][dd] 
\ar@{-}@[newyellow][dd]
& \\
&&&& \\
\ar@{-}[rrr]& &&*+[o][F]{} 
\ar@{-}@<-0.1mm>@[newyellow][dd] 
\ar@{-}@<0.1mm>@[newyellow][dd] 
\ar@{-}@[newyellow][dd]  \\  
&  && &\! \! \! \! \! {\mbox{\scriptsize$\hD,s$}} 
\\
\ar@{-}[rrr]&&&*+[o][F]{}
\ar@{-}@<-0.1mm>@[newyellow][dd] 
\ar@{-}@<0.1mm>@[newyellow][dd] 
\ar@{-}@[newyellow][dd]& \\
&&&& \\
&&&& }
$
}
\; ,
\qquad
\qquad
g^{(p,q)}_{\Delta,\ell}(\up,\ut)\equiv 
 \raisebox{2.3em}{
$
\xymatrix@=5.3pt{
{
}\ar@{-}[rdd]& &&& _{(p,q)} 
\ar@{-}@<-0.1mm>@[newyellow][dd] 
\ar@{-}@<0.1mm>@[newyellow][dd] 
\ar@{-}@[newyellow][dd]
&  \\  
&&&&\\
& *+[o][F]{}  \ar@{.}[rrr]^{\Delta,\ell } & &&
*+[o][F]{}  
\ar@{-}@<-0.1mm>@[newyellow][dd] 
\ar@{-}@<0.1mm>@[newyellow][dd] 
\ar@{-}@[newyellow][dd]
\\
&&&\\
\ar@{-}[ruu]&&&& }
$
} \, .
\ee
The defect channel conformal blocks are known in closed form \cite{Billo:2016cpy},
\be
\label{defect_blocks}
 \hat g_{\hD,s}(\up,\ut)=\up^{-\hat{\Delta}}  {}_2F_1\left(\frac{\hat{\Delta }}{2},\frac{1+\hat{\Delta }}{2} ;2-\frac{p+2}{2}+\hat{\Delta };\frac{1}{\up ^2}\right) \, \ut ^s \, _2F_1\left(-\frac{s}{2},\frac{1-s}{2};2-\frac{q}{2}-s;\frac{1}{\ut ^2}\right) \, .
\ee
Conversely, the bulk channel blocks are not known in closed form, but there are many ways to compute them (see e.g.~\cite{Billo:2016cpy, Isachenkov:2018pef}); 
in our work we will use the Mathematica file provided in~\cite{Lauria:2017wav}.
By equating the expansions in the two channels, we obtain the crossing relation for the two-point function of bulk operators,
\be
f(\up,\ut)=\sum_{\Delta,\ell}\lambda_{O_1O_2O_{\Delta,\ell}} a_{O_{\Delta,\ell}} g^{(p,q)}_{\Delta,\ell}(\up,\ut) = \sum_{\hat{\Delta},s}b_{O_1\hat{O}_{\hat{\Delta},s}} b_{O_2\hat{O}_{\hat{\Delta},s}} \hat{g}^{(p,q)}_{\hat{\Delta},s}(\up,\ut)
\, .
\ee
We will study this correlation function in the uplifted theory. Another quantity that we consider, which is relatively simpler yet nontrivial, is the \emph{defect-defect-bulk} three-point function of scalar operators \cite{Lauria:2020emq}. This is fixed up to an arbitrary function of a single  cross ratio,
\bea
\label{BDD}
\langle O_{\Delta_1}(x_1)\hat{O}_{\hat{\Delta}_2}(x^{||}_2)\hat{O}_{\hat{\Delta}_3}(x^{||}_3)\rangle=
\frac{f(\z)}{|x_1^\perp|^{\Delta_1}||x_1-x^{||}_2|^{\hat{\Delta}_2-\hat{\Delta}_3}|x_1-x^{||}_3|^{\hat{\Delta}_3-\hat{\Delta}_2}|x^{||}_2-x^{||}_3|^{\hat{\Delta}_2+\hat{\Delta}_3}} \, ,
\eea
where the cross ratio is given by
\be
\z \equiv \frac{|x_1^\perp|^2 |x^{||}_{23}|^2}{|x_1-x^{||}_2|^2|x_1-x^{||}_3|^2} \, .
\ee
We can expand $f(\z)$ in the defect channel as,
\begin{equation}
    f(\z)=\sum_{\hat{O}} \lambda_{\hat{O}_2\hat{O}_3\hat{O}} b_{O\hat{O}} \hat{f}_{\hat{O}}^{(p)}(\z),
\end{equation}
where $\hat{f}_{\hat{O}}^{(p)}(\z)$ is the bulk-defect-defect conformal block defined as 
\be
\hat{f}_{\hat{\Delta}}^{(p)}(\z)\equiv 
\raisebox{
1.5em
}{
$\xymatrix@=10pt{
&&& 
\ar@{-}@<-0.1mm>@[newyellow][d] 
\ar@{-}@<0.1mm>@[newyellow][d] 
\ar@{-}@[newyellow][d] \\
\ar@{-}[rrr]^{\hat{\Delta}} \ar@{}[rrr]
& & & *+[o][F]{} 
\ar@{-}@<-0.1mm>@[newyellow][d] 
\ar@{-}@<0.1mm>@[newyellow][d] 
\ar@{-}@[newyellow][d] \\
&&&
*{\bullet}
\ar@{-}@<-0.1mm>@[newyellow][d] 
\ar@{-}@<0.1mm>@[newyellow][d] 
\ar@{-}@[newyellow][d] \\
&&& 
*{\bullet}
\ar@{-}@<-0.1mm>@[newyellow][d] 
\ar@{-}@<0.1mm>@[newyellow][d] 
\ar@{-}@[newyellow][d]\\
&&& \quad \quad \\
&&& *{\vphantom{\bullet}} \save "4,4"*+!U(-8){\scriptstyle (p)} \restore
}$
}
\equiv \z^{\frac{\hat{\Delta}}{2}} \,_2F_1(\frac{\hat{\Delta}+\hat{\Delta}_{23}}{2},\frac{\hat{\Delta}-\hat{\Delta}_{23}}{2};\hat{\Delta}+1-\frac{p}{2},\z)\, ,
\vspace{-2.5em}
\ee
where we dropped the label $q$ since the block only depends on $p$ and we labelled the defect operator $\hat{O}$ uniquely in terms of its dimension $\hD$ since the only allowed transverse spin is $s=0$.
We will explore the consequences of PS uplift for this observable as well.
\section{Dimensional reduction and uplift in presence of defects}
\label{sec:uplifts}

Using the PS uplift we describe a CFT$_d$ in terms of CFT$_{d+2}$ invariant under PS superconformal symmetry $OSp(d+1,1|2)$. In practice this uplift allows one to extend the spacetime coordinate by two bosonic and two fermionic directions, namely by
\be
(x_{d+1},x_{d+2},\theta,\bar\theta) \in \mathbb{R}^{2|2}  \, .
\ee
Given a coordinate $x\in \mathbb{R}^{d}$,
its uplifted location is obtained by the map 
\be
x \to y 
\equiv (X,\theta,\bar \theta) \, ,
\ee
where $X=(x,x_{d+1},x_{d+2}) \in \mathbb{R}^{d+2}$.
As explained in \cite{paper1, Trevisani:2024djr} 
a bulk superprimary $\Ocal(y)$ is labelled by dimension $\Delta$ and spin  $\ell$ under $OSp(d+2|2)$. From a single scalar superprimary  $\Ocal(y)$ inserted in superspace we  obtain four different primaries components\footnote{Notice that the primary operator $\Ocal_{\theta \bar \theta}$ was called $\widetilde{\Ocal}_{\theta \bar \theta}$ in \cite{Trevisani:2024djr}, to distinguish it with the highest component which is obtained by just taking derivatives in $\theta \thetab$. Here we drop the tilde to make the notation less heavy. 
} inserted at $X$,
\begin{align}
\Ocal_0(X)\equiv \Ocal(y)|_{0} \, , 
\quad 
\Ocal_\thetab (X)\equiv \partial_{\theta} \Ocal(y)|_{0} \, , 
\quad 
\Ocal_\theta (X)\equiv \partial_{\thetab_i} \Ocal(y)|_{0} \, ,
\quad
\Ocal_{\theta \thetab} (X)\equiv {\bf D} \Ocal(y)|_{0}
 \, , \label{def:O_comp}
\end{align}
where $|_{0}$ sets to zero all Grassmann variables $\theta,\thetab$ and the differential operator ${\bf D}$  is defined as
\be
\label{boldD}
{\bf D} \equiv \partial_{x}^2- (2 \Delta-d+2) \partial_{\bar\theta} \partial_{\theta} \, ,
\ee
where $\Delta$ is the dimension of the operator $\Ocal(y)$.
The operator $\Ocal_0(X)$ has scaling dimension $\Delta$, $\Ocal_\theta ,\Ocal_\thetab $ have $\Delta+1$, while $\Ocal_{\theta\thetab}$ has $\Delta+2$. The first and fourth operator have bosonic statistic, while the middle two are fermionic. We will keep the convention that subscript on the operators will refer to the components \eqref{def:O_comp} while superscripts will refer to the spin. For example the super stress tensor $\mathcal{T}^{MN}(y)$ has indices $M,N=1,\dots,d+2,\theta,\thetab$ (it has  $OSp(d+2|2)$-spin $\ell=2$), so e.g. its lowest component will be named $\mathcal{T}^{MN}_0$.

Extended operators are more involved. 
There are  two separate ways to perform a PS uplift  of flat $p$-dimensional defects: 
either by uplifting the space parallel or orthogonal to the defect. 
Accordingly, the uplifted PS CFT contains two different defects, of dimensionality either $p+2$ or $p$. In the following we shall first explain how to define the two uplifts. Then we will explain why these two seemingly very different extended operators both dimensionally reduce to the same original  $p$-dimensional defect. 

\subsection{Two different uplifts of defect operators}
\label{sec:PSU_TSU}
\begin{figure}[ht]
\centering
\begin{tikzpicture}[scale=1.5,>={Stealth[scale=1.1]}]
\tkzInit[xmin=0,xmax=5.5,ymin=-1,ymax=2] 
    \tkzClip[space=.5] 
\tkzDefPoint(3.3,-1.5){A2} 
    \tkzDefPoint(3.3,0){B2} 
    \tkzDefPoint(4.3,1){C2} 
    \tkzDefPointWith[colinear= at C2](B2,A2) \tkzGetPoint{D2}       \tkzDrawPolygon[fill=newyellow!10](A2,B2,C2,D2)
    \tkzDefPoint(0.3,0){A} 
    \tkzDefPoint(4.5,0){B} 
    \tkzDefPoint(5.5,1){C} 
    \tkzDefPointWith[colinear= at C](B,A) \tkzGetPoint{D}
\tkzDrawPolygon[fill=black!10,opacity=0.9](A,B,C,D)
\tkzDefPoint(3.3,0){A1} 
    \tkzDefPoint(3.3,1.5){B1} 
    \tkzDefPoint(4.3,2.5){C1} 
    \tkzDefPointWith[colinear= at C1](B1,A1) \tkzGetPoint{D1}       \tkzDrawPolygon[fill=newyellow!10](A1,B1,C1,D1)
    \tkzDefPoint(2.5 ,0.7){X1} 
      \tkzDefPoint(2.5 ,0.74){X1p} 
       \tkzDefPoint(2.5 ,1.4){Y1} 
        \tkzDefPoint(2.5 ,1.36){Y1p} 
        
          \tkzDefPoint(3.8 ,0.5){XPP} 
        \tkzDefPoint(3.8 ,0.54){XPPp}     
          \tkzDefPoint(3.8 ,1.3){YPP}
           \tkzDefPoint(3.8 ,1.26){YPPp}
               \draw[newyellow,thick](A1)->(D1);
          \draw[fill=black, draw=black] (X1) circle (0.9pt);
             \draw[fill=black, draw=black] (Y1) circle (0.9pt);
               \draw[fill=black, draw=black] (XPP) circle (0.9pt);
            \draw[fill=black, draw=black] (YPP) circle (0.9pt);
        \draw[dashed, <-](Y1p) ->(X1p);
        \draw[dashed, <-](YPPp) ->(XPPp);
    \node[below] at (X1) { $ x$
    };
   \node[below] at (XPP) { $ \xpar$
   };
  \node[above] at (Y1) { $ y$};
   \node[above] at (YPP) {$\ypar$};
  \node[above] at (1.2,1.35) {$\mathbb{R}^{d+2\mid 2}$};
    \begin{scope}[every node/.append style={xslant=0, yslant=0}]\node[above] at (3.8,-0.7) {$\mathbb{R}^{p+2\mid 2}$};
    \end{scope}
    \begin{scope}[every node/.append style={xslant=0, yslant=0}]
     \node[above] at (1.2,0.1) {$\mathbb{R}^{d}$}; 
               \end{scope}
\end{tikzpicture}%
\caption{ \label{fig:PSU}
Parallel space uplift (PSU).  }
\end{figure}
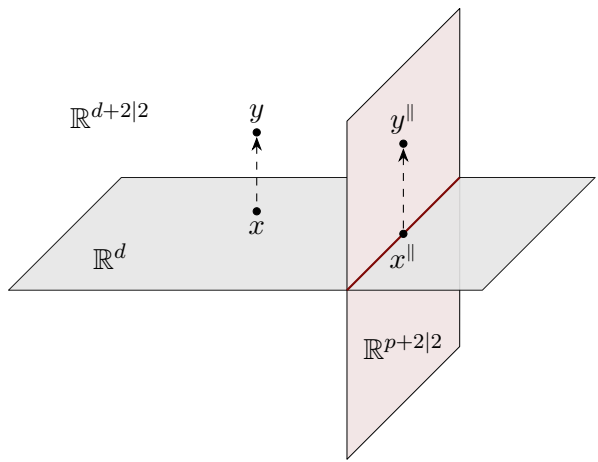

\paragraph{Parallel Space Uplift (PSU)} 
A PSU is defined by uplifting the space parallel to the defect
\be
\label{PSU_coords}
\xpar \to \ypar
\equiv (\Xpar,\theta, \bar \theta)
\, ,
\qquad\qquad
\xperp \to \yperp=\xperp \, ,
\ee
where $\Xpar=(\xpar,x_{d+1},x_{d+2}) \in \mathbb{R}^{p+2}$. The defect itself is uplifted to $\mathbb{R}^{p+2|2}$ and with it all defect insertions, as shown in figure \ref{fig:PSU}. This implies that the transverse space is kept fixed. Of course bulk insertions will now live in $\mathbb{R}^{d+2|2}$.  

The symmetry preserved by the PSU is $OSp(p+3,1|2)\times \so(q)$. We understand $OSp(p+3,1|2)$ as the symmetry of a CFT$_{p+2}$ living on the $p+2$-dimensional defect with PS supersymmetry. The transverse $\so(q)$ symmetry is understood as a global symmetry under which the defect operators transform. 
By the PSU, the original CFT$_{p}$ living on the defect is uplifted to a CFT$_{p+2}$ with PS SUSY. This thus corresponds to a standard PS uplift applied to the theory living on the defect. So if we believe that the uplift of \cite{Trevisani:2024djr} is correct, we automatically obtain that the PSU should be consistent. Nevertheless the story here has more structure since the PSU also contains a bulk Hilbert space which was not considered in \cite{Trevisani:2024djr} and should therefore be studied.

Let us detail the properties of the local 
operators in the PSU. The characterization of bulk operators does not depend on the presence of defects, thus it is the same as discussed before, e.g. around equation \eqref{def:O_comp}.
A defect superprimary $ \hOcal$ is labelled by the scaling dimension $\hD$, the parallel spin $\hat{\ell}$ and the transverse (global symmetry) spin $s$.
In the PSU case the spin $\hat{\ell}$ labels a representation of $OSp(p+2|2)$, while  the transverse spin $s$ labels a representation of $\so(q)$.
Defect superprimaries are inserted  at positions $\ypar=(\Xpar,\theta,\thetab)$,  so  
$ \hOcal(\Xpar,\theta,\thetab)$ can be expanded in components. 
For example for a scalar $ \hOcal$ we obtain the following primary components
\begin{align}
\hOcal_{0  }(\Xpar) \equiv \hOcal(\ypar) |_0 
\, , 
\ \;
\hOcal_{\theta }(\Xpar) \equiv \partial_{\bar \theta} \hOcal(\ypar) |_0 \, , 
\ \;
\hOcal_{\bar \theta  }(\Xpar) \equiv \partial_{ \theta} \hOcal(\ypar) |_0
\, , 
\ \;
\hOcal_{\theta \bar \theta  }(\Xpar) \equiv \hat{\bf D} \hOcal(\ypar) |_0,
\label{def:hO_comp}
\end{align}
where  we defined 
\be
\label{boldhD}
\hat{\bf D} = \partial_{\Xpar}^2 - (2\hD-p) \partial_{\bar \theta}\partial_{\theta} \, .
\ee  
As before $\hOcal_{0  }$ is a boson of dimension $\hD$, $\hOcal_{\theta},\hOcal_{\thetab}$ are fermions of dimensions $\hD+1$ and $\hOcal_{\theta \thetab}$ is a boson of dimension $\hD+2$.

As an example, let us consider the super displacement operator $\mathcal{D}^i$ which arises from the Ward identity of the super stress tensor $\mathcal{T}^{MN}(y)$ that expresses the violation of translations in transverse directions,
\begin{equation}
    \frac{\partial}{\partial y^M}\mathcal{T}^{Mi}(y)=\mathcal{D}^{i}(\ypar)\delta^{(q)}(x^\perp) \, ,
    \label{PSU_displ}
\end{equation}
where $M$ is summed over all $d+2|2$ directions, while index $i$ runs over the transverse directions  $i=p+1,\dots,d$. The super stress tensor  $\mathcal{T}^{MN}(y)$ has super dimension $d$, therefore from \eqref{PSU_displ} the quantum numbers of the super displacement are
$\hD=p+1, \hat{\ell}=0, s=1$.
Let us discuss the primary components in its multiplet. 
We have
\begin{equation}
   \mathcal{D}^i_0(\Xpar) \, , 
   \quad 
      \mathcal{D}^i_{\theta}(\Xpar) \, , 
   \quad 
      \mathcal{D}^i_{\thetab}(\Xpar) \, ,
       \quad 
      \mathcal{D}^i_{\theta \thetab}(\Xpar) \, , 
\end{equation}
where all the operators are vectors of $\so(q)$. $\mathcal{D}^i_0$ is bosonic and has dimension $p+1$, $\mathcal{D}^i_{\theta},\mathcal{D}^i_{\thetab}$ are fermionic and have dimension $p+2$, while $\mathcal{D}^i_{\theta \thetab}$ is bosonic with dimension $p+3$. 
Notice that the
highest component $\mathcal{D}^i_{\theta \thetab}$ has the correct quantum numbers for a displacement of the uplifted $p+2$-dimensional defect, while the lowest component $\mathcal{D}^i_0$ has the  quantum number  of the displacement in a dimensionally reduced theory. The latter is not a coincidence, as we will explain in the following section.

Finally it is useful to keep in mind a way to realize the PSU in the case when the original defect is defined by integrating a scalar bulk operator $O$ on $\mathbb{R}^p$. 
In the uplifted theory there exists a bulk scalar superprimary $\Ocal$ which reduces to $O$. The PSU is then obtained by integrating $\Ocal$ on $\mathbb{R}^{p+2|2}$. The fermionic part of the integral selects the highest component of $\Ocal$ which then  gets integrated over the remaining  $p+2$ bosonic dimensions. 

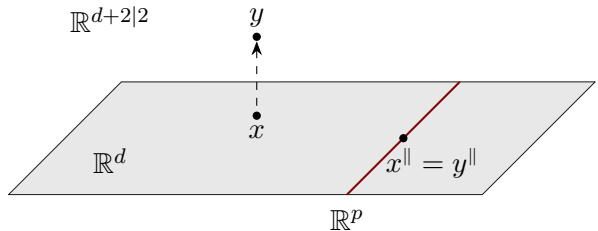
\begin{figure}[ht]
\centering
\begin{tikzpicture}[scale=1.5,>={Stealth[scale=1.1]}]
\tkzInit[xmin=0,xmax=5.5,ymin=0,ymax=1.5] 
    \tkzClip[space=.5] 
    \tkzDefPoint(0.3,0){A} 
    \tkzDefPoint(4.5,0){B} 
    \tkzDefPoint(5.5,1){C} 
    \tkzDefPointWith[colinear= at C](B,A) \tkzGetPoint{D}
\tkzDrawPolygon[fill=black!10,opacity=0.9](A,B,C,D)
\tkzDefPoint(3.3,0){A1} 
    \tkzDefPoint(4.3,1){B1}     
     \draw[newyellow,thick](A1)->(B1);    
    \tkzDefPoint(2.5 ,0.7){X1} 
      \tkzDefPoint(2.5 ,0.74){X1p} 
       \tkzDefPoint(2.5 ,1.4){Y1} 
        \tkzDefPoint(2.5 ,1.36){Y1p} 
          \tkzDefPoint(3.8 ,0.5){XPP} 
        \tkzDefPoint(3.8 ,0.54){XPPp}     
          \tkzDefPoint(3.8 ,1.3){YPP} 
          \draw[fill=black, draw=black] (X1) circle (0.9pt);
             \draw[fill=black, draw=black] (Y1) circle (0.9pt);
               \draw[fill=black, draw=black] (XPP) circle (0.9pt);
        \draw[dashed, <-](Y1p) ->(X1p);
    \node[below] at (X1) { $ x$
    };
   \node[below] at (XPP) { $\ \ \ \ \ \ \xpar=\ypar$
   };
  \node[above] at (Y1) { $ y$};
  \node[above] at (1.2,1.35) {$\mathbb{R}^{d+2\mid 2}$};
    \begin{scope}[every node/.append style={xslant=0, yslant=0}]\node[above] at (3.3,-0.4) {$\mathbb{R}^{p}$};
    \end{scope}
    \begin{scope}[every node/.append style={xslant=0, yslant=0}]
     \node[above] at (1.2,0.1) {$\mathbb{R}^{d}$}; 
               \end{scope}
\end{tikzpicture}%
\caption{ \label{fig:TSU}
Transverse space uplift (TSU).  }
\end{figure}

\paragraph{Transverse space uplift (TSU)}
A TSU is defined by uplifting the space orthogonal to the defect
\be
\label{TSU_coords}
\xpar \to \ypar=\xpar
\, ,
\qquad\qquad
\xperp \to \yperp
\equiv (\Xperp,\theta, \bar \theta) \, ,
\ee
where $\Xperp=(\xperp,x_{d+1},x_{d+2}) \in \mathbb{R}^{q+2}$. 
In this case the uplifted defect looks like the original $p$-dimensional defect, however the full theory lives in $\mathbb{R}^{d+2|2}$ as shown in figure \ref{fig:TSU}.

The symmetry preserved by this configuration is $\so(p+1,1)\times OSp(q+2|2)$. Here  $\so(p+1,1)$ is the symmetry of the CFT$_p$ leaving on the $p$-dimensional defect. The defect operators transform according to the global symmetry $OSp(q+2|2)$.
 It is important  to emphasize that this  $OSp(q+2|2)$ symmetry is not the usual PS supersymmetry,  but rather a global type of  ``supersymmetry'',   which commutes with the spacetime symmetries. It still maps bosons to fermions, but crucially, it does not alter scaling dimensions. This contrasts with e.g. PS supertranslations, which shift the scaling dimensions by one unit.
It also follows that unlike standard PS supersymmetry, which is typically formulated in superspace, 
a TSU defect operator $ \hOcal(\xpar)$ is inserted at a point $\xpar$ which lives in the conventional bosonic space $\mathbb{R}^p$. The spin $\hl$ is accordingly just a representation of $\so(p)$. However the transverse spin $s$ now labels a representation for $OSp(q+2|2)$. E.g. an operator  with transverse $OSp(q+2|2)$-spin $s$ can be decomposed into $\so(q+2)$ representations in terms of two bosonic operators $\hat \Ocal^{i_1 \dots i_{s}},\hat \Ocal^{i_1 \dots i_{s-2} \theta \thetab}$ with transverse $\so(q+2)$-spin respectively $s$ and $s-2$,  and two fermionic ones  $\hat \Ocal^{i_1 \dots i_{s-1}\theta},\hat \Ocal^{i_1 \dots i_{s-1} \thetab}$ with $\so(q+2)$-spin $s-1$. All these operators have the same scaling dimensions and $OSp(q+2|2)$ transformations can rotate bosonic fields into fermionic ones and vice versa. 

As an example let us consider the displacement operator $\mathcal{D}^I$, 
which arises from the Ward identity of the super stress tensor as follows
\begin{equation}
    \frac{\partial}{\partial y^M}\mathcal{T}^{MI}(y)=\mathcal{D}_0^{I}(\xpar)\delta^{(q+2|2)}(y^\perp) \, ,
    \label{TSU_Disp}
\end{equation}
where $I=\{i,\theta,\bar{\theta}\}$ with 
 $i=p+1,\dots,  d+2$, while $M$ is summed over all $\mathbb{R}^{d+2|2}$ directions.
This operator can be decomposed into $\so(q+2)\times Sp(2)$ representations as 
\begin{equation}
   \mathcal{D}^i_0(\xpar) \, , 
   \quad 
      \mathcal{D}^\theta_0(\xpar) \, , 
   \quad 
      \mathcal{D}^\thetab_0(\xpar) \, , \qquad  (\text{TSU})
\end{equation}
where all components have conformal dimension $p+1$. The first operator is a vector of $\so(q+2)$, while the second two operators are anticommuting scalars.

Let us discuss how to realize the TSU when the original defect is defined by integrating a bulk operator $O$ over  $\mathbb{R}^p$. 
As for the PSU, we shall use the bulk scalar superprimary $\Ocal$ which reduces to $O$. 
While the PSU was obtained by integrating the highest component of $\Ocal$ over $\mathbb{R}^{p+2}$, the TSU is realized by integrating the lowest component of $\Ocal$ over $\mathbb{R}^{p}$. Notice that the TSU from this point of view should trivially provide dimensional reduction. Indeed by dimensional reduction the lowest component of $\Ocal$ maps to  $O$, so the defect in the TSU is exactly equal to the original integrated operator.

Therefore we expect that the theory living on the TSU defect and the one living on the reduced defect should be equivalent. However we notice right away that this equivalence might look a bit puzzling since the reduced defect theory is  a CFT$_p$ with global symmetry  $\so(q)$, while the TSU defect theory is a CFT$_p$ with global symmetry  $OSp(q+2|2)$.
This raises the question: how can two 
CFT$_{p}$ be equivalent, if they have different global symmetries?
To answer this question we should point out that it was often observed 
that a given model with global symmetry
$OSp(m|n)$ yields the same results as one with global symmetry 
$O(m-n)$ (for any $m$ and any even value of
$n$).
The relation between global symmetries $OSp(m|n) \leftrightarrow O(m-n)$ is  well established. It is often studied
in the context of Wilson Fisher fixed point in epsilon expansion 
for global symmetry $Sp(n)$  and $O(-n)$ (which amounts to setting $m=0$ above). The latter can also be used to give a prescription to compute observables in models with $O(n)$ symmetry where $n$ is analytically continued to zero or to negative values. For some discussions and applications  see e.g.  \cite{
parisi:jpa-00231808,
LeClair:2006kb,
LeClair:2007iy,
Fei:2015kta,
Caracciolo:2017xyc,
Klebanov:2021sos, Narovlansky:2022ijq, 
 Anninos:2011ui,
Piazza:2024wll, Kaviraj:2024cwf}. 
In appendix \ref{app:global_sym_uplift}  we review some one-loop computation in epsilon expansion for $OSp(m|n)$ global symmetry and show how these indeed relate to equivalent computations for $O(m-n)$.
The main idea is very general and simple to understand. In practice  $O(m-n)$ vector model has the same diagrammatic expansion as the $OSp(m|n)$ model. In particular, any time a closed loop of the indices appears in the  $O(m-n)$ model it gives a factor of $m-n$. The same factor appears in the $OSp(m|n)$ model, since the fermionic components contribute with a minus sign.\footnote{\label{footnote:Z2}
The relation between $OSp(m|n) \leftrightarrow O(m-n)$ is known for 
the full orthogonal group $O$, while the transverse symmetry of the TSU is only the proper subgroup of rotations $\so$. 
In this paper we are mostly concerned on continuous symmetries, so we are not careful on  $\mathbb{Z}_2$ factors, but these can be opportunely fixed as follows. 
Instead of 
considering CFT invariant under $\so(d+1,1)$---the familiar conformal group---
 we can focus on CFTs that are invariant under inversion/parity/reflections, so that their symmetry group is enhanced to $O(d+1,1)$ (for more discussions see e.g. \cite{Nakayama_2012}). 
Similarly,  we can focus on
defect setups that break this symmetry to $O(p+1,1)\times O(q)$, in such a way to have transverse symmetry $O(q)$, as desired.

Let us also mention that the notation $OSp(d+3,1|2)$ typically refers to the full group which contains both proper and improper super rotations, this thus should be related to the uplift of  $O(d+1,1)$ and not $\so(d+1,1)$. 
One could consider a symmetry subgroup $OSp_+(d+3,1|2)$ with only elements connected with the identity, but whether this defines an uplift of the $\so(d+1,1)$-invariant theories is an open problem. 
In fact the uplift/reduction of parity breaking models is more subtle since the epsilon tensor in different dimensions has different number of indices. We will defer this subtlety to future investigations. 
 }

The global symmetry reduction works both for singlet operators and more complicated representations. In the latter case the prescription is the same as usual dimensional reduction when dealing with operators with spin, see e.g. \cite{paper1}. Namely, given a representation of $OSp(m|n)$, we take its branching to $O(m-n)\times OSp(n|n)$ and we select only the singlet piece under $OSp(n|n)$. In practice this means that we are taking the indices associated to  $OSp(m|n)$ and restricting them to $O(m-n)$. E.g. a rank $k$  graded-symmetric traceless tensor of  $OSp(m|n)$  projects to a rank $k$  symmetric traceless tensor of $O(m-n)$, see also  appendix \ref{app:global_sym_uplift}.

The TSU seems to suggest that the duality between models with $OSp(m|n)$ and $O(m-n)$ global symmetry should be much more general than the instances considered in the literature. According to the TSU, this relation should hold non-perturbatively when applied to the transverse symmetry of all existent defect theories. This could be an indication that the validity of this relation is even larger. 
It would be interesting 
to prove to which extent generic theories with $OSp(m|n)$ global symmetry can be described by theories with $O(m-n)$ global symmetry.

\subsection{Dimensional reduction}
\label{sec:reduction}
In this section we shall give two arguments for dimensional reduction for defect setups. The first argument is non-perturbative and is based on \cite{paper1} while the second one is perturbative and based on the original idea of Parisi-Sourlas \cite{Parisi:1979ka} (rephrased  in the language of appendix A of \cite{paper1}).

First of all let us define what is the prescription for dimensional reduction. 
The idea is exactly the same as we showed in figure \eqref{fig:dim_red} for theories without defects, namely we only need to plug all operator insertion in $\mathbb{R}^d$. E.g. for a correlator of bulk and defect scalar operators we have 
\be
\label{corr_red}
\langle \Ocal_1(y_1) \dots  \Ocal_n(y_n)  \hOcal_1(\ypar_1) \dots  \hOcal_m(\ypar_m) \rangle \bigg |_{\mathbb{R}^d}\, ,
\ee
where $y_i\to x_i \in \mathbb{R}^d$ and $\ypar_i\to \xpar_i \in \mathbb{R}^p \subset \mathbb{R}^d$. In practice this means that we should consider figures \ref{fig:PSU} and \ref{fig:TSU} and reverse the direction of the arrows.
Notice that in the TSU case all defect operators already live in $\mathbb{R}^p$ so we do not have to project those insertions.

In the case that the local operators have spin we should also project the spin indices of bulk operators to $\mathbb{R}^d$, while the indices of defect operators in such a way that the parallel spin becomes $\so(p)$ and the transverse spin $\so(q)$. Namely for PSU we need to set the parallel indices to $\mathbb{R}^p$, while for the TSU we need to set the transverse indices to $\mathbb{R}^q$. While projecting the spin indices we preserve the shape of the Young tableau  of the representation, e.g. a graded symmetric and traceless representation of $OSp(p+2|2)$ is mapped to a symmetric and traceless representation of $\so(p)$. Sometimes it happens that the reduced representation is zero-dimensional. In this case we say that the operator is projected to zero by dimensional reduction, namely it does not exist in the lower dimensional theory. To see this phenomenon, one should fix the value of $p,q$ and consider representations which involve enough antisymmetrizations. In this paper we will mostly work with $p,q$ generic, so we will not see this problem.

\paragraph{Non perturbative decoupling}
Following \cite{paper1}, we want to classify the possible kinematic invariants for observables of the type \eqref{corr_red}. We want to prove that when performing dimensional reduction, nothing can depend on the directions orthogonal to $\mathbb{R}^d$. 
This type of argument is extremely powerful and proves that infinitely many operators must decouple in the reduced theory.

Let us start by quickly reviewing the case of \cite{paper1}, when the defect is not present. In this case we can imagine that  $\mathbb{R}^d$ behaves like a trivial defect which breaks the PS superconformal symmetry as $OSp(d+1,1|2)\to \so(d-1,1)\times OSp(2|2)$. Accordingly it is convenient to split the full metric $g_{d+2|2}$ into a parallel and a transverse metric, respectively $g_{d}$ and  $g_{2|2}$. 
The prescription of dimensional reduction of figure \ref{fig:dim_red} generates automatically correlators of local operators in the CFT$_d$ which are singlet under the global symmetry $OSp(2|2)$. 
Now the idea is that if one writes correlators of $OSp(2|2)$-singlets inserted in $\mathbb{R}^d$, then anything involving the metric $g_{2|2}$ decouples. 
 The reason is that the correlator must be invariant under $OSp(2|2)$, but there is no possible invariant that can be written using the metric $g_{2|2}$, since it is supertraceless and it is transverse to all vectors,
namely  $\text{str } g_{2|2} =0$, $x \cdot g_{2|2} =0$ for any vector $x\in\mathbb{R}^p$. This argument can be used to prove that given a superprimary  $\Ocal(y)$, all the $OSp(2|2)$-singlet primaries in the dimensionally reduced theory decouple but one,  $\Ocal(x)$.
E.g. operators of the form $(\partial \cdot g_{2|2}  \cdot \partial)^k \Ocal(x)$ decouple for $k\neq 0$ because they depend on $ g_{2|2}$. 
This explains the decoupling statements reviewed in section \ref{sec:review_PS}. 
Notice that this argument is very generic and can be used to argue that any time we find $g_{2|2}$ inside any $OSp(2|2)$-invariant expression arising from dimensional reduction, then we are allowed to set $g_{2|2}\to 0$ inside this expression. This can be applied to  equations which describe the kinematics or the dynamics of PS theories to show that they are equal to equations appearing the dimensionally reduced models.
E.g. in \cite{paper1} this type of argument was essentially employed to show that the Casimir equation that defines PS superconformal blocks in superspace is the same as the dimensionally reduced Casimir equation which defines conventional blocks.

This argument can be easily adapted to both PSU and TSU uplifts. Let us start by the TSU. 
In this case we want to consider a $d$-dimensional trivial defect that contains the  $p$-dimensional plane selected by the TSU, so the pattern of preserved symmetries is 
$ \so(p-1,1)\times \so(q) \times OSp(2|2)$.
Accordingly we shall split the bulk $g_{d+2|2}$ metric into three parts: $g_{p}$, $g_{q}$ and $g_{2|2}$. 
As before, dimensional reduction \eqref{corr_red} imposes on us to consider correlators of singlets of $OSp(2|2)$ and again no possible invariant can be built out of $g_{2|2}$. 
Now let us consider the PSU.
We define a $d$-dimensional trivial defect which intersects the  $p+2|2$-dimensional plane on $\mathbb{R}^p$. This setup also preserves the symmetry $ \so(p-1,1)\times \so(q) \times OSp(2|2)$, where again we can split the bulk metric  into the three pieces $g_{p}$,  $g_{q}$ and $g_{2|2}$. As before, for all $OSp(2|2)$-singlet observables there is no possible way to build invariants out of $g_{2|2}$. 

For both PSU and TSU, this argument proves the infinite decoupling of  operators built using derivatives in the $2|2$-direction and similarly the fact that we can drop $g_{2|2}$ in all equations which arise from dimensional reduction. 
E.g. one can use this to show that the Ward identities \eqref{PSU_displ} and \eqref{TSU_Disp} for the super displacements in the PSU and TSU imply the Ward identity for the dimensionally reduced displacement.
Let us show this in more detail for the PSU. Setting   \eqref{PSU_displ} to $\mathbb{R}^{d}$, we find the usual Ward identity of the original DCFT in $d$ dimensions up to operators that decouple since are contacted with $g_{2|2}$, namely 
\begin{equation}
    \frac{\partial}{\partial x^\mu}T^{\mu i}(x)=\mathcal{D}_0^{i}(\xpar)\delta(x^\perp)+\text{decoupled} \, ,
\end{equation}
where $\mu=1,\dots,d$ and $T$ is the lowest component of the super stress tensor $\mathcal{T}_{0}$ opportunely rewritten in terms of  $\so(d)$ irreducible representations,  namely $T^{\mu \nu}=\mathcal{T}_0^{\mu \nu}+\frac{g_d^{\mu \nu}}{d} \mathcal{T}_0^{MN} (g_{2|2})_{MN}$.
The decoupled terms in the equation come from the traceless condition above together with the superspace divergence $\partial^{M}_y  (g_{2|2})_{MN} \mathcal{T}_{0}^{Mi} $. Both terms are explicitly built out of $g_{2|2}$. By similar manipulations on $\eqref{TSU_Disp} $ one finds the respective dimensionally reduced Ward identity  for  the TSU.

Similarly one can reuse the same argument of \cite{paper1} to show that also in presence of defects the Casimir equation that defines all possible PS superconformal blocks in superspace is equal to the dimensionally reduced one. This holds because the  Casimir operator is explicitly the same up to terms that are proportional to $g_{2|2}$. 
Because of this reasoning  we  find that all the superconformal blocks in superspace  for both PSU and TSU are the same functions as the usual conformal conformal blocks in the dimensionally reduced theories.

\paragraph{Reduction of Feynman diagrams }
While the argument above is very powerful, it is also abstract. For this reason, in the following we focus on a perturbative setup and show how dimensional reduction works at the level of Feynman diagrams for a scalar model \eqref{PS_scalar_action}.
The main idea is  the original one of \cite{Parisi:1979ka}, namely that integrals in the fermionic directions have the effect of eating bosonic directions, e.g. $\int d^{d+2}X [d \theta][d \thetab] F(X^2-2 \theta \thetab)= \int d^{d}x  F(x^2)$. In the next we show how to apply this logic to all Feynman diagrams in the PSU and TSU case following appendix A of \cite{paper1}.

Let us first consider the PSU case where the defect coordinates $\ypar$ are defined in \eqref{PSU_coords}.
A generic Feynman diagram is given by the following expression,
\begin{equation} \label{eqn:Feynman}
   G(y_1,\dots y_N,\ypar_1,\dots \ypar_n)= \int [dy] \prod_{I<J}(G^{BB}_{\Phi\Phi}(y_{IJ}))^{q_{IJ}}\prod_{K,i}(G^{BD}_{\Phi\Phi}(y_{Ki}))^{q_{Ki}}\prod_{j<k}(G^{DD}_{\Phi\Phi}(y_{jk}))^{q_{jk}},
\end{equation}
with $[dy]=\prod_{r,s} dy_r d\ypar_s$, where $r$ refers to internal bulk points, while $s$ to all the internal defect points. Points $y$ with labels $I,J,K$ are in the bulk, while $i,j,k$ are on the defect (so $y_i$ should be in principle $y_i^{\parallel}$ but we drop $\parallel$ to shorten the notation). In both cases they can be either external or integrated point. 
The numbers $q_{IJ}$, $q_{Ki}$, and $q_{jk}$ denote the power with which bulk to bulk, bulk to defect, and defect to defect propagators appear. $G^{BB}_{\Phi\Phi}(y_{IJ})$ is a bulk to bulk free propagator, similarly, $G^{BD}_{\Phi\Phi}(y_{Ij})$ denotes the bulk to defect free propagator and $G^{DD}_{\Phi\Phi}(y_{ij})$ denotes the defect to defect free propagators. The free propagator is given by,
\begin{equation}
\label{def:GPhiPhi}
    G_{\Phi\Phi}(y_{ab})=\frac{\mathcal{N}_d}{(X_{ab}^2-2\theta_{ab}\bar{\theta}_{ab})^{\frac{d-2}{2}}} \, ,
\end{equation}
where $\mathcal{N}_d=\frac{1}{(d-2)\Omega_{d-1}} $ with $\Omega_{d-1}= 2 \frac{\pi^{d/2}}{\Gamma(d/2)}$.
 All propagators ---bulk to bulk, bulk to defect and defect to defect--- are obtained by restricting the points $a$, and $b$ to either bulk or defect accordingly. Now we use the Schwinger parametrization to manipulate the propagators,
\begin{equation}
    \frac{1}{A^k}=\frac{1}{\Gamma[k]}\int du u^{k-1} e^{-u A}.
\end{equation}
With this parametrization the dependence on all the points is gaussian.
 In appendix \ref{app:RedFeyn} we show the details of the computation, but the main idea is that now the Gaussian integrals can  be performed exactly and that the contribution of two bosonic directions cancel the one of two fermionic 
directions proving dimensional reduction. 
To give a bit more of details, let us select for all points two bosonic  directions  $\alpha=d+1,d+2$ (transverse to the $d$ dimensions of the reduced theory) and compute the Gaussian integral over all these directions together  with all ferionic directions. The result takes the form
\be
\label{eq:2|2integral}
\int \prod_{c\in\mathcal{I}} dX_{c} [d\bar{\theta}_c] [d\theta_c] \exp[-\sum_{a,b\in \mathcal{I}} (X_a^{\alpha} \tilde{M}_{ab} X_{b\alpha}+2 \bar{\theta}_a \tilde{M}_{ab} \theta_b)]=\frac{\det M}{\pi^{n_{t}}}\frac{\pi^{n_{t}}}{\det M}=1\, ,
\ee
where $\mathcal{I}$ is the set of all integrated internal points, whose total number is denoted by $n_t$. The matrix $\tilde{M}_{ab}$ is defined in the appendix \ref{app:RedFeyn}.
 The appearance of the same matrix $\tilde{M}_{ab}$ in both the bosonic and fermionic integrals ultimately stems from the fact that the coordinate dependence arises solely through the distances $|y_{ab}|$.
We thus find that the full dependence on $X^\alpha_a,\theta_a,\thetab_a$ drops out from the integral. Moreover the leftover integral can be shown to exactly match the correspondent lower dimensional Feynman integral, 
\begin{equation} \label{eqn:Feynman_dimred}
 G(x_1,\dots x_N,\xpar_1,\dots \xpar_n)= \int [dx] \prod_{I<J}(G^{BB}_{\phi\phi}(x_{IJ}))^{q_{IJ}}\prod_{K,i}(G^{BD}_{\phi\phi}(x_{Ki}))^{q_{Ki}}\prod_{j<k}(G^{DD}_{\phi\phi}(x_{jk}))^{q_{jk}} \, ,
\end{equation}
defined using the same conventions as in \eqref{eqn:Feynman_dimred}, e.g. capitalized labels for  bulk points and lowercase for defect ones. Here the propagators are given by the usual two-point function
$G_{\phi\phi}(x_{ab})=\mathcal{N}_d (x_{ab}^2)^{-\frac{d-2}{2}}$. This thus proves the dimensional reduction of Feynman integrals in the PSU case.

Let us now turn to the TSU case where the inserions are defined by \eqref{TSU_coords}. This amounts to setting $X^{\alpha}_i=\theta_i=\bar{\theta_i}=0$ for defect operators in the expressions above. Then, following the same steps, we see the dimensional reduction in this case as well.

 \section{Uplifted defects and   correlators}
 \label{sec:correlators}
 
In the following we  consider correlators of local operators in the presence of  PSU and TSU defects. We study different primary components of such correlators and check that they take the correct form implied by conformal symmetry. 
Moreover we show that the uplifted correlators automatically satisfies CFT axioms if the dimensionally reduced counterparts do.

To extract a given primary component from a superprimary we need to act with the superspace differential operators
$\partial_{\theta},\partial_{\thetab},{\bf D}$ (or $\hat{\bf D}$ if the point is on the defect) according to \eqref{def:O_comp} and \eqref{def:hO_comp}.
For a correlator with many operator insertions a generic component is extracted by acting with a product of such differential operators at different points. 
Following the logic of \cite{Trevisani:2024djr}, we shall refer to this product as $\bD_{[\s]}$.  
The notation $[\s]$ stands for a string of different numbers which can be barred, unbarred or bold, e.g. $[\bar{5} 1 {\bf 6} 3 \bar{2}]$. Every unbarred number $k$ stands for the action of $\partial_{\theta_k}$ while barred numbers $\bar{k}$ stand for $\partial_{\bar{\theta_k}}$. Bold numbers ${\bf k}$ symbolize the action of either ${\bf D}$ or $\hat{\bf D}$ depending on whether the operator at position $k$ is a bulk or defect operator.  The order of the string is the order of the differential operators.

Altogether the extraction of a component for a generic correlator is implemented by 
\be
\label{bDS_def}
\bD_{[\s]} \langle \Ocal_{1}(y_1) \dots \Ocal_{n}(y_n)  \hOcal_{n+1}(\ypar_{n+1}) \dots \hOcal_{n+m}(\ypar_{n+m}) \rangle |_{0} \, .
\ee
For example if $n=1$ and $m=2$, we can have $\bD_{[1\bar{2} {\bf 3}]}$, which corresponds to $\partial_{{\theta_1}} \partial_{\bar{\theta_2}} \hat{\bf D}_3$, where the third operator is hatted because it lives on the defect.
Let us point out that since the differential operators at different points graded-commute, all permutations of a given string $[\s]$ are equal (up to overall minus signs) and will not be considered. Moreover, because of the $Sp(2)$ global symmetry (see \cite{Trevisani:2024djr} for more details), the total number of barred and unbarred indices must be equal (while the number of bold indices is not restricted). 
With this in mind, we shall now study the primary components of various correlation functions.

\subsection{One-point function and bulk-defect two-point function}
As a warm up we start by considering the uplift of correlators that are fully fixed by symmetry. This is the case for the bulk one-point function and for the bulk-defect two-point function. Also the correlators of up to three defect insertions are fixed by symmetry, but these match the ones of ordinary CFTs which were already studied considered in \cite{paper1, Trevisani:2024djr} and will not be reviewed here.
\paragraph{One-point function} 
Let us consider the one-point function in superspace $\langle \mathcal{O}(y) \rangle = a_\Delta {|\yperp|^{-\Delta}}$.
In the two different uplifts this takes the form
\begin{align}
&\langle \mathcal{O}(y) \rangle  = \frac{a_\Delta }{(|\xperp|^2)^{\frac{\Delta}{2}}} \, , &\text{(PSU),} 
\label{eq:1pt_PSU}
\\
&\langle \mathcal{O}(y) \rangle = \frac{a_\Delta }{(|\Xperp|^2-2 \theta \thetab)^{\frac{\Delta}{2}}} \, ,  &\text{(TSU).}
\label{eq:1pt_TSU}
\end{align}
We can now expand $\mathcal{O}$ in components. 
In the PSU case we find that only the lowest component of $\mathcal{O}(y)$ acquires a non-zero one-point function, namely $\langle \mathcal{O}(y) \rangle =\langle \mathcal{O}_0(X) \rangle$.
Conversely in the TSU case, besides the lowest component $\langle \mathcal{O}_0(X) \rangle=a_\Delta |X^\perp|^{-\Delta}$, also the highest component acquires a one-point function,  which can be computed by acting on \eqref{eq:1pt_TSU} with \eqref{boldD}, namely
\be
\langle 
{\mathcal{O}}_{\theta \bar \theta}(X)
\rangle 
= {\bf D} \langle \mathcal{O}(y) \rangle |_{\theta, \bar \theta=0} =
\Delta (p-\Delta) \frac{a_\Delta }{|X^\perp|^{\Delta+2}} \, , \qquad \text{(TSU)}. \label{1pt_STU} 
\ee

\paragraph{Bulk-defect two-point function}
We now study the uplift of a bulk-boundary two-point function 
$
\langle \Ocal(y_1) \hOcal(\ypar_2) \rangle = b_{\Ocal \hOcal} |\yperp_{1}|^{\hD-\Delta} |\ypar_{12}|^{-2\hD}
$
where $\D,\hD$ are respectively the dimensions of the bulk and the boundary operators. 
Let us consider the different components of this observable in both TSU and PSU case.

In the TSU case $\hOcal$ has single component. So  the observable in total has two possible components which arise by considering the lowest and highest components of $\Ocal$, namely 
\be
\langle 
{\Ocal}_{0}(X_1) \hOcal(\xpar_{2})
\rangle=\frac{b_{\Ocal \hOcal} }{|\Xperp_1|^{\Delta-\hD} |\xpar_{12}|^{2\hD}} \, ,
\quad
\langle 
{\Ocal}_{\theta \bar \theta}(X_1) \hOcal(\xpar_{2})
\rangle 
= \frac{b_{\Ocal \hOcal} (\Delta -\hat{\Delta }) (-\Delta -\hat{\Delta }+p) } {|\Xperp_1|^{\Delta+2-\hD} |\xpar_{12}|^{2\hD}} \, , \quad (\text{TSU})\, .
\nonumber
\ee
The PSU case is richer because $\hOcal$ has  various components. We can compactly write them as
\be
\label{def:Bd_comp}
\bD_{[\sigma]} \langle \Ocal(y_1) \hOcal(\ypar_2) \rangle|_0 = c_{[\sigma]}  b_{\Ocal \hOcal}  \Sigma_{[\sigma]} \frac{1} {(\xperp_{1}{}^2)^{\frac{\Delta-\hD}{2}} (\Xpar_{12}{}^2)^{\hD}} \, , \quad (\text{PSU})
\ee
where $c_{[\sigma]}$ is a coefficient fixed by supersymmetry through the action of the differential operator $\bD_{[\sigma]}$ as defined in \eqref{bDS_def}. 
Here we further define $\Sigma_{[\sigma]}$ following \cite{Trevisani:2024djr} as an operator that shifts the conformal dimensions according to the component selected by the operator $\bD_{[\sigma]}$, e.g. $\Sigma_{[{\bf 1}]}$ shifts $\Delta_1 \to \Delta_1+2$, while $\Sigma_{[{1\bar 2}]}$ shifts $(\Delta_1,\hD_2) \to (\Delta_1+1,\hD_2+1)$ and so on. 
There are in total five non-zero inequivalent components and their coefficients $c_{[\sigma]}$ read
\begin{align}
&c_{[\cdot]}=1 \, ,\quad
c_{[{\bf 1}]}=(\hD-\D)(\hD-\D+q-2) \, ,\quad
c_{[{\bf 2}]}=-4 \hat{\Delta } (\hat{\Delta }+1) \, ,
\\
&c_{[{1 \bar{2}}]}=-2 \hat{\Delta }\, ,
\quad
c_{[{\bf 12}]}=4 \hat{\Delta } (\hat{\Delta }+1) (\Delta +\hat{\Delta }-p) (p+q-\Delta -\hat{\Delta }-2)\, .
\end{align}

\subsection{Bulk-bulk  2pt function }
\label{sec:BB_2pt}
A two-point function of bulk scalar super primaries $\mathcal{O}_i$ with dimensions $\Delta_i$ is fixed as 
\begin{equation}
\label{2ptD_uplift}
  \langle \mathcal{O}_1(y_1) \mathcal{O}_2(y_2) \rangle  =\frac{f(U,V)}{|\yperp_1|^{\Delta_1}|\yperp_2|^{\Delta_2} } \, ,
\end{equation}
where $f$ is a function of the two super-space cross ratios
\be
\label{def_CR_Uplift}
U \equiv \frac{1}{2}\frac{|\ypar_{ 12}|^2 + |\yperp_1|^2 + |\yperp_2|^2 }{|\yperp_1| |\yperp_2| }
 \, ,
\qquad
V \equiv \frac{\yperp_1 \cdot \yperp_2 }{|\yperp_1| |\yperp_2| } \, .
\ee
This superspace observable contains the information of various two-point functions of primaries, which can be extracted from the action of the differential operators $\bD_{[\s]}$ as in \eqref{bDS_def}. 
For both PSU and TSU, after the action of each $\bD_{[\s]}$, the resulting primary two-point function is canonically written as in \eqref{2ptD} in terms of a function of cross ratios  \eqref{def_CR}
\be
\label{def:fks}
f^{k}_{[\s]}(\u,\v) \equiv 
D^{k}_{[\s]} f(\u,\v) \, , \qquad k=P,T \, ,
\ee
where $P,T$ labels whether we are considering PSU or TSU and $D^{k}_{[\s]}$ are differential operators in the variables $\u,\v$  which we define below.
Strictly speaking, the dependence on the position in the prefactor \eqref{2ptD} and in the cross ratios  \eqref{def_CR} also depends on $P,T$, since in PSU we should replace $\xpar \to \Xpar$, while in TSU  $\xperp \to \Xperp$ according to the definitions in section \ref{sec:PSU_TSU}. We leave this implicit to simplify the notation.
Notice that when stripping out the dependence on positions following  \eqref{2ptD}, we should remember that the conformal dimension of the primary components might be shifted by some units as explained after formula \eqref{def:O_comp}.

The differential operators take the following form. For the lowest components we have $D^P_{[\cdot]}=D^T_{[\cdot]}=1$. Higher components are defined by
\begin{align}
D^P_{[1\bar{2}]}&=\partial _{\up }  \fchiu \, ,\\
D^T_{[1\bar{2}]}&=-\partial _{\ut }  \fchiu ,\\
D^P_{[{\bf i}]}&=\Delta _i \left(\Delta _i-q+2\right) \fchiu +\up  \left(2 \Delta _i-q+3\right) \partial _{\up }  \fchiu +(1 -q )\ut \partial _{\ut }  \fchiu -\left(\ut ^2-1\right) \partial _{\ut }^2  \fchiu +\left(\up ^2-1\right) \partial _{\up }^2  \fchiu  \, ,  \\
D^T_{[{\bf i}]}&=\Delta _i \left(p-\Delta _i\right) \fchiu +\ut  \left(-2 \Delta _i+p-1\right) \partial _{\ut }  \fchiu +(p+1) \up  \partial _{\up }  \fchiu -\left(\ut ^2-1\right) \partial _{\ut }^2  \fchiu +\left(\up ^2-1\right) \partial _{\up }^2  \fchiu \, ,
\end{align}
with $i=1,2$. Finally the differential operators associated to the highest components take a fairly involved form which however can be compactly written by composing the differential operators that we just defined,
\begin{align}
D^P_{[{\bf 1 2}]}&=D^P_{[{\bf 2}]}D^P_{[{\bf 1}]}-
\left(-2 \Delta _1+p+q-2\right) \left(-2 \Delta _2+p+q\right) \partial _{\up }^2  \fchiu 
\, ,
\\
D^T_{[{\bf 1 2}]}&=D^T_{[{\bf 2}]}D^T_{[{\bf 1}]}-\left(-2 \Delta _1+p+q-2\right) \left(-2 \Delta _2+p+q\right) \partial _{\ut }^2  \fchiu \, .
\end{align}
The order of the differential operators in the right-hand side of the expressions above can be easily reversed by using the following commutators,  
\be
[D^P_{[{\bf 2}]},D^P_{[{\bf 1}]}]=
4 \left(\Delta _2-\Delta _1\right) \partial _{\up }^2  \fchiu 
\, ,
\qquad
[D^T_{[{\bf 2}]},D^T_{[{\bf 1}]}]=4 \left(\Delta _2-\Delta _1\right) \partial _{\ut }^2  \fchiu 
 \, .
\ee
It is also interesting to mention that the TSU and PSU differential operators happen to be  related by the following map
\be
\label{mapD_T_P}
\begin{array}{l}
D^P_{[{\bf i}]}|_{(\ut,q)\leftrightarrow(\up,p+2)
}=-D^T_{[{\bf i}]}
\, , \quad  D^P_{[{1 \bar{2}}]}|_{(\ut,q)\leftrightarrow(\up,p+2)
}=- D^T_{[{1 \bar{2}}]}
\, , 
\quad
D^P_{[{\bf 12}]}|_{(\ut,q)\leftrightarrow(\up,p+2)
}= D^T_{[{\bf 12}]} \, .
\end{array}
\ee
The reason why equation \eqref{mapD_T_P} holds is explained in detail in Appendix \ref{App:T-Pmap}. 
The key idea is that this relation originates from a kinematic map, which generally applies in DCFT setups, based on an interchange of what we define as parallel and orthogonal directions.

\subsection{Uplifted CFT axioms and relations for bulk-bulk  conformal blocks } \label{subsection:blockrelation}
In this section we want to prove that the uplifted two-point functions of bulk operators respect the required CFT axioms. In particular we need to verify that all the uplifted two-point functions are decomposable in the higher dimensional conformal blocks. There are two different OPE channels ---bulk and defect--- and for each one of them we expect a correct conformal block decomposition. 
We find that, if $f(\up,\ut)$ is decomposable in bulk and defect conformal blocks in lower dimensions, then  the functions $f^P_{[\s]}(\up,\ut)$ and $f^T_{[\s]}(\up,\ut)$ defined in  \eqref{def:fks} will be automatically decomposable in the respective uplifted conformal blocks. Since the uplifted functions $f^k_{[\s]}(\up,\ut)$ are obtained by the action of $D^{k}_{[\sigma]}$ on $f(\up,\ut)$, the above statement can be proved if one finds that $D^{k}_{[\sigma]}$ on lower dimensional blocks is a linear combination of blocks in higher dimensions.

We indeed find a set of remarkable properties that the conformal blocks \eqref{blocks_schematic} satisfy.
For the defect channel they can be written as 
\begin{align}
\label{def_rel}
D^{k}_{[\sigma]}
  \raisebox{3.3em}{
$
\xymatrix@=3pt{
&&& _{(p,q)} 
\ar@{-}@<-0.1mm>@[newyellow][dd] 
\ar@{-}@<0.1mm>@[newyellow][dd] 
\ar@{-}@[newyellow][dd]
& \\
&&&& \\
\ar@{-}[rrr]& &&*+[o][F]{} 
\ar@{-}@<-0.1mm>@[newyellow][dd] 
\ar@{-}@<0.1mm>@[newyellow][dd] 
\ar@{-}@[newyellow][dd]  \\  
&  && &\! \! \! \! \! {\mbox{\scriptsize$\hD,s$}} 
\\
\ar@{-}[rrr]&&&*+[o][F]{}
\ar@{-}@<-0.1mm>@[newyellow][dd] 
\ar@{-}@<0.1mm>@[newyellow][dd] 
\ar@{-}@[newyellow][dd]& \\
&&&& \\
&&&& }
$
} = \sum_{(i,j) \in \hat S^k_{[\sigma]}} \hat c^{k}_{[\s],i,j} 
\raisebox{3.3em}{
$
\xymatrix@=3pt{
&&& _{(p,q)^k} 
\ar@{-}@<-0.1mm>@[newyellow][dd] 
\ar@{-}@<0.1mm>@[newyellow][dd] 
\ar@{-}@[newyellow][dd]
& \\
&&&& \\
\ar@{-}[rrr]& &&*+[o][F]{} 
\ar@{-}@<-0.1mm>@[newyellow][dd] 
\ar@{-}@<0.1mm>@[newyellow][dd] 
\ar@{-}@[newyellow][dd]
\\  
&  && &\! \! \! \! \! {\mbox{\scriptsize$\hD+i,s+j$}} 
\\
\ar@{-}[rrr]&&&*+[o][F]{} 
\ar@{-}@<-0.1mm>@[newyellow][dd] 
\ar@{-}@<0.1mm>@[newyellow][dd] 
\ar@{-}@[newyellow][dd]
& \\
&&&& \\
&&&& }
$
}
\end{align}
where $k=T,P$. We defined $(p,q)^T\equiv (p,q+2), (p,q)^P \equiv(p+2,q)$. The sets in the sum contain at most two elements, in particular $\hat S^T_{[\sigma]}=\{(0,0),(0,-2)\}$ for all $[\sigma]$ besides $\hat S^P_{[1\bar 2]}=\{(0,-1)\}$,  similarly 
$\hat S^P_{[\sigma]}=\{(0,0),(2,0)\}$ for all $[\sigma]$ besides $\hat S^P_{[1\bar 2]}=\{(1,0)\}$. The coefficients $c^{k}_{[\s],i,j} $ are defined below. 

The labels in $\hat S^T_{[\sigma]}$ are easily explained by noticing that in the TSU all defect operators do not live in superspace so their dimension equals the dimensionally reduced operator, on the other hand the transverse spin in now defined in terms of the supergroup $OSp(q+2|2)$ and thus when the external operators are bosonic we expect defect exchanges  with indices $\hat \Ocal^{i_1 \dots i_s }$ and $\hat \Ocal^{i_1 \dots i_{s-2} \theta \thetab}$. 
In the case $[\sigma]=[1\bar{2}]$ the external operators are fermionic, thus the exchanged operators must also be fermionic and will thus take the form $\hat \Ocal^{i_1 \dots i_{s-1} \theta }$ or $\hat \Ocal^{i_1 \dots i_{s-1} \thetab}$.
Conversely for the PSU the defect operators have operators with global symmetry  $\so(q)$, so they always have the same spin $s$, but they are inserted in superspace so by expanding them in components we find shifts in their conformal dimensions. Specifically $\hat \Ocal^{i_1 \dots i_{s}}(\Xpar,\theta,\thetab)$ will contain two bosonic components with dimensions $\hD$ and $\hD+2$ which are exchanged when the external operator are bosonic. In the case $[\sigma]=[1\bar{2}]$, the two operator insertions are fermionic, thus also the operators in the defect expansions are all fermionic, namely they correspond to the components $\hat \Ocal^{i_1 \dots i_{s}}_{\theta}(\Xpar)$, $\hat \Ocal^{i_1 \dots i_{s}}_{\thetab}(\Xpar)$ which have dimensions $\hD+1$.

A similar formula holds for the bulk channel blocks
\be
\label{bulk_rel}
D^{k}_{[\sigma]}
 \raisebox{2.3em}{
$
\xymatrix@=5.3pt{
{
}\ar@{-}[rdd]& &&& _{(p,q)} 
\ar@{-}@<-0.1mm>@[newyellow][dd] 
\ar@{-}@<0.1mm>@[newyellow][dd] 
\ar@{-}@[newyellow][dd]
&  \\  
&&&&\\
& *+[o][F]{}  \ar@{.}[rrr]^{\Delta,\ell } & &&
*+[o][F]{}  
\ar@{-}@<-0.1mm>@[newyellow][dd] 
\ar@{-}@<0.1mm>@[newyellow][dd] 
\ar@{-}@[newyellow][dd]
\\
&&&\\
\ar@{-}[ruu]&&&& }
$
}
 = \sum_{(i,j) \in  S}  c^{k}_{[\s],i,j} \Sigma_{[\s]}
\raisebox{2.3em}{
$
\xymatrix@=5.3pt{
\ar@{-}[rdd]& &&& _{\; (p,q)^k} { 
\ar@{-}@<-0.1mm>@[newyellow][dd] 
\ar@{-}@<0.1mm>@[newyellow][dd] 
\ar@{-}@[newyellow][dd]
}\\  
&&&&\\
& *+[o][F]{}  \ar@{.}[rrr]^{ \Delta+i,\ell+j  } & && *+[o][F]{}  
\ar@{-}@<-0.1mm>@[newyellow][dd] 
\ar@{-}@<0.1mm>@[newyellow][dd] 
\ar@{-}@[newyellow][dd]
\\
&&&\\
\ar@{-}[ruu]&&&& }
$
}
\ ,
\ee
where $\Sigma_{[\sigma]}$ implements shifts in the conformal dimension as  explained in \eqref{def:Bd_comp}, e.g. 
 $\Sigma_{[{\bf i}]}$ acts as $\Delta_i \to \Delta_i+2$. 
The set $S$ always contains four elements $S=\{(0,0),(0,-2),(2,0),(2,-2)\}$ independently of the component $[\sigma]$. 
This set defines the possible bulk exchanges that have a non-zero one-point function. In particular, they must always  be bosonic operators with even spin. This selects the cases  $\Ocal^{\mu_1 \dots \mu_\ell}_{0},\Ocal^{\mu_1 \dots \mu_{\ell-2}\theta \thetab}_{0}, \Ocal^{\mu_1 \dots \mu_\ell}_{\theta \thetab},\Ocal^{\mu_1 \dots \mu_{\ell-2}\theta \thetab}_{\theta \thetab}$ which exactly match the shifts in the set $S$.

Let us finally stress that bulk  and defect blocks are very different functions that solve very different Casimir differential equations. Nevertheless we find that these different functions satisfy similar properties under the action of a common set of differential operators. This is quite surprising. 
Indeed it is actually crucial that in the left hand side of equations \eqref{bulk_rel} and \eqref{def_rel}  the same differential operator $D^{k}_{[\s]}$ appears. This is what ensures that, given any two-point function $f(\u,\v)$ which can be decomposed in bulk and defect blocks in lower dimensions, then also all the components $D^{k}_{[\s]} f(\u,\v)$ 
  of the uplifted theory will be decomposed in both bulk and defect blocks. Altogether this is a very strong check that both PSU and TSU always define consistent uplifts.

\paragraph{Explicit examples: defect blocks }
Let us show in more detail the relations for the defect blocks.
In the PSU case for the defect blocks we have\footnote{In principle we should shift the dimensions of the operators $\Delta_1,\Delta_2$ of the conformal blocks on the right, but the defect blocks do not depend on those labels so we can omit the shifts.}
\begin{equation}
\begin{array}{lll}
&D^P_{[\s]} \hat g^{(p,q)}_{\hat \Delta, s}=\hat c^{P}_{[\s],0} \ \hat g^{(p+2,q)}_{\hat \Delta, s} +\hat c^{P}_{[\s],2} \ \hat g^{(p+2,q)}_{\hat \Delta+2, s} 
\, ,
\qquad \qquad 
&[s]=[\cdot],[{\bf 1}],[{\bf 2}],[{\bf 1 2}]\, ,
\\
&D^P_{[1\bar 2]} \hat g^{(p,q)}_{\hat \Delta, s}=- \hat \Delta \ \hat g^{(p+2,q)}_{\hat \Delta+1, s} 
\, , 
\qquad 
&
\end{array}
\end{equation}
where   we dropped the 
 the dependence on the cross ratios in the blocks and the label $j$ in $c^{P}_{[\s],i,j}$ since it is always zero. The coefficients are defined as follows 
\be
\!\!\!
{
\medmuskip=0mu
\thinmuskip=0mu
\thickmuskip=0mu
\begin{array}{ll}
c^{P}_{[\cdot],0}=1
\, ,
& c^{P}_{[\cdot],2}=-\frac{ \hat{\Delta } (\hat{\Delta }+1)}{(-2 \hat{\Delta }+p-2) (p-2 \hat{\Delta })}  
\, ,
\\
c^{P}_{[{\bf i}],0}=-(-\hat{\Delta }+\Delta _i+s) (\hat{\Delta }-\Delta _i+q+s-2) 
\, ,
& c^{P}_{[{\bf i}],2}=(-\hat{\Delta }-\Delta _i+p-s) (-\hat{\Delta }-\Delta _i+p+q+s-2)
c^{P}_{[\cdot],2}
\, ,
\\
c^{P}_{[{\bf 12}],0}=\frac{c^{P}_{[{\bf 1}],0}c^{P}_{[{\bf 2}],0}}{c^{P}_{[\cdot],0}}
\, ,
& c^{P}_{[{\bf 12}],2}=
\frac{c^{P}_{[{\bf 1}],2}c^{P}_{[{\bf 2}],2}}{c^{P}_{[\cdot],2}}
\, .
\end{array}
}
\ee
In the case of BCFTs, which corresponds to setting $q=1$, the identity for the lowest component of the PSU was considered in \cite{Zhou:2020ptb, Gliozzi:2025xmr}. Our result matches with those ones.
Similarly for the TSU we have
\be
\begin{array}{ll}
D^T_{[\s]} \hat g^{(p,q)}_{\hat \Delta, s}=\hat c^{T}_{[\s],0} \  \hat g^{(p,q+2)}_{\hat \Delta, s} +\hat c^{T}_{[\s],-2} \ \hat g^{(p,q+2)}_{\hat \Delta, s-2} 
\, ,
\qquad\qquad &
[\s]=[\cdot],[{\bf 1}],[{\bf 2}],[{\bf 1 2}]\, ,
\\
D^T_{[1\bar 2]} \hat g^{(p,q)}_{\hat \Delta, s}=-s \ \hat g^{(p,q+2)}_{\hat \Delta, s-1} \, . & 
\end{array}
\ee
The coefficients $c^T_{{[\sigma],i,j}}$ are equal to $c^P_{{[\sigma],i,j}}$ after mapping $-\hat \Delta \leftrightarrow s$ and $p+2\leftrightarrow q$.\footnote{Up to a change of sign when $[\sigma]=[1\bar{2}],[\bf{1}],[\bf{2}]$. This is due to the normalization of two-point function.} This is not a coincidence.
Indeed we notice that the defect blocks of \eqref{defect_blocks}  are factorized into two identical functions, one dependent on parallel data, the other on transverse data (the data here consists of both cross ratios and  quantum labels of the operators).
The blocks are thus invariant under a map that switches the roles of parallel with transverse, which we describe in more generality in appendix \ref{App:T-Pmap}. 
The action of same map switches the roles of $D^{P}_{[\sigma]} \leftrightarrow D^{T}_{[\sigma]}$ as in \eqref{mapD_T_P}.
This fact clarifies why the defect channel blocks satisfy two sets of relations built using the  differential operator $D^{P}_{[\sigma]}$ and $D^{T}_{[\sigma]}$.

\paragraph{Explicit examples: bulk blocks }
We can also give examples of these fomulae for the bulk conformal blocks but they are more cumbersome so we will avoid showing all of them in the main text. 
For the components $[\cdot]$ and $[1{\bar 2}]$  we obtain
the following relations
{
\medmuskip=0mu
\thinmuskip=0mu
\thickmuskip=0mu
\be \nonumber
\begin{array}{l}
\phantom{-\partial_\up}g^{(p,q)}_{\Delta , \ell}=  g^{(p+2,q)}_{\Delta , \ell}+ c^{P}_{[\cdot],0,-2} g^{(p+2,q)}_{\Delta , \ell-2}+ c^{P}_{[\cdot],2,0} g^{(p+2,q)}_{\Delta+2 , \ell}+ c^{P}_{[\cdot],2,-2} g^{(p+2,q)}_{\Delta+2 , \ell-2} \, ,
\\
-\partial_\up g^{(p,q)}_{\Delta , \ell}= c^{P}_{[1 \bar{2}],0,0} \Sigma_{[1 \bar{2}]}g^{(p+2,q)}_{\Delta , \ell}+ c^{P}_{[1 \bar{2}],0,-2} \Sigma_{[1 \bar{2}]}g^{(p+2,q)}_{\Delta , \ell-2}+ c^{P}_{[1 \bar{2}],2,0} \Sigma_{[1 \bar{2}]}g^{(p+2,q)}_{\Delta+2 , \ell}+ c^{P}_{[1 \bar{2}],2,-2} \Sigma_{[1 \bar{2}]}g^{(p+2,q)}_{\Delta+2 , \ell-2} \, ,
\end{array}
\ee
}
with coefficients defined as
{
\medmuskip=0mu
\thinmuskip=0mu
\thickmuskip=0mu
\be
\begin{array}{rl}
c^{P}_{[\cdot],0,-2}&=\frac{\ell  (q+\ell -3)}{(p+q+2 \ell -4) (p+q+2 \ell -2)}\, , 
\\
c^{P}_{[\cdot],2,0}&=-\frac{(\Delta -1) (-\Delta +q-2) \left(\Delta +\Delta _1-\Delta _2+\ell \right) \left(\Delta -\Delta _1+\Delta _2+\ell \right)}{4 (\Delta +\ell -1) (\Delta +\ell +1) (-2 \Delta +p+q-2) (-2 \Delta +p+q)} c^{P}_{2,0}\, , 
\\
c^{P}_{[\cdot],2,-2}&=-\frac{(\Delta -1) (-\Delta +q-2) \left(-\Delta +\Delta _1-\Delta _2+p+q+\ell -2\right) \left(-\Delta -\Delta _1+\Delta _2+p+q+\ell -2\right)}{4 (-2 \Delta +p+q-2) (-2 \Delta +p+q) (-\Delta +p+q+\ell -3) (-\Delta +p+q+\ell -1)} c^{P}_{0,-2} \, ,
\\
c^{P}_{[1 \bar{2}],0,0}&=\frac{\left(\Delta -\Delta _1-\Delta _2-\ell \right)}{4}  
\,
,
\\
c^{P}_{[1 \bar{2}],0,-2}&=
\frac{\left(\Delta -\Delta _1-\Delta _2+p+q+\ell -2\right)}{4} 
c^{P}_{[\cdot],0,-2}
\, ,
\\
c^{P}_{[1 \bar{2}],2,0}&=
\frac{\left(-\Delta -\Delta _1-\Delta _2+p+q-\ell \right)}{4} c^{P}_{[\cdot],2,0}\,,
\\
c^{P}_{[1 \bar{2}],2,-2}&=
\frac{\left(-\Delta -\Delta _1-\Delta _2+2 p+2 q+\ell -2\right)}{4} 
c^{P}_{[\cdot],2,-2} \, .
\end{array}
\ee
}
As for the defect channel, when $q=1$ the identity for the lowest component coincides with the one computed in  \cite{Zhou:2020ptb}. 
Similar formulae work for the TSU case, e.g.
{
\medmuskip=0mu
\thinmuskip=0mu
\thickmuskip=0mu
\be
\nonumber
\begin{array}{l}
\phantom{-\partial_{\ut}}g^{(p,q)}_{\Delta , \ell}=  g^{(p,q+2)}_{\Delta , \ell}+ c^{T}_{[\cdot],0,-2} g^{(p,q+2)}_{\Delta , \ell-2}+ c^{T}_{[\cdot],2,0} g^{(p,q+2)}_{\Delta+2 , \ell}+ c^{T}_{[\cdot],2,-2} g^{(p,q+2)}_{\Delta+2 , \ell-2} \, ,
\\
-\partial_{\ut}g^{(p,q)}_{\Delta , \ell}=  c^{T}_{[1 \bar{2}],0,0} \Sigma_{[1 \bar{2}]}g^{(p,q+2)}_{\Delta , \ell}+ c^{T}_{[1 \bar{2}],0,-2} \Sigma_{[1 \bar{2}]}g^{(p,q+2)}_{\Delta , \ell-2}+ c^{T}_{[1 \bar{2}],2,0} \Sigma_{[1 \bar{2}]}g^{(p,q+2)}_{\Delta+2 , \ell}+ c^{T}_{[1 \bar{2}],2,-2} \Sigma_{[1 \bar{2}]}g^{(p,q+2)}_{\Delta+2 , \ell-2} \, .
\end{array}
\ee
}
Interestingly we find that, for all components $[\sigma]$, the coefficients for PSU and TSU are related as follows,
\be
\begin{array}{ll}
c^{T}_{[\sigma],0,0}=c^{P}_{[\sigma],0,0}
\, ,
\qquad
&c^{T}_{[\sigma],0,-2}=-\frac{q+\ell -3}{p+\ell -1} c^{P}_{[\sigma],0,-2}
\, ,
\\
 c^{T}_{[\sigma],2,0}=-\frac{-\Delta +q-2}{p-\Delta }c^{P}_{[\sigma],2,0} 
\, ,
\qquad
&
c^{T}_{[\sigma],2,-2}=\frac{c^{T}_{[\sigma],2,0}}{c^{P}_{[\sigma],2,0}}\frac{c^{T}_{[\sigma],0,-2}}{c^{P}_{[\sigma],0,-2}}c^{P}_{[\sigma],2,-2} \, .
\end{array}
\ee
We will not present here the rest of the formulae because they are lengthy but we include them in an ancillary Mathematica file. 

Let us mention that the bulk channel conformal blocks are not factorized in terms of a transverse function that depends on $q,\ut$ times a parallel function that depends on $p,\up$. Generically they are  very complicated functions which are not known in closed form. Because of this reason at first sight it looks very non-trivial that there are two sets of differential operators  $D^P_{[\s]}$, $D^T_{[\s]}$  which act by shifting respectively $p$ and $q$. This is even more puzzling since $D^P_{[\s]}$, $D^T_{[\s]}$  are mapped into each other by $(\ut,q)\leftrightarrow(\up,p+2)$ according to \eqref{mapD_T_P} while no such relation was known for the blocks. Motivated by this puzzle, we showed in appendix \ref{App:T-Pmap} that the bulk blocks are themselves  invariant (up to possible overall normalizations) under the same map \eqref{mapD_T_P}. In the same appendix we further show that this map is a piece of a much broader set of relations that the kinematical functions of DCFTs must satisfy under the swap of parallel and transverse data. See appendix \ref{App:T-Pmap} for more details.

\subsection{Bulk-defect-defect three-point function}
The bulk-defect-defect three-point function in the uplifted theory is fixed as
\bea
\label{BDD_PSU}
\langle O_{\Delta_1}(y_1)\hat{O}_{\hat{\Delta}_2}(\ypar_2)\hat{O}_{\hat{\Delta}_3}(\ypar_3)\rangle=
\frac{f(W)}{|y_1^\perp|^{\Delta_1}||y_1-y^{||}_2|^{\hat{\Delta}_2-\hat{\Delta}_3}|y_1-y^{||}_3|^{\hat{\Delta}_3-\hat{\Delta}_2}|y^{||}_2-y^{||}_3|^{\hat{\Delta}_2+\hat{\Delta}_3}} \, ,
\eea
up to a function $f(W)$ of a single  cross ratio given by
\be
W \equiv \frac{|y_1^\perp|^2 |y^{||}_{23}|^2}{|y_1-y^{||}_2|^2|y_1-y^{||}_3|^2} \, .
\ee
As usual, the form of $y_i,\ypar_i,\yperp_i$ differs for  PSU and TSU, according to the definitions of  section \ref{sec:PSU_TSU}. 
 Using \eqref{bDS_def}, we extract all primary components of this correlator by acting with differential operators $\bD_{[\sigma]}$
 and setting all fermionic variables to zero. Each component is then written as a function $f^{k}_{[\sigma]}(\z)$ of a single cross ratio as in \eqref{BDD}, which takes the form
\be
\label{def:flocal}
f^{k}_{[\s]}(\z) \equiv 
\hat D^{k}_{[\s]} f(\z) \, , \qquad k=P,T \, ,
\ee
where $\hat  D^{k}_{[\s]}$ are differential operators in the variable $\z$. As for the bulk two-point function, also here the cross ratio and the kinematic factor of \eqref{BDD} are different for PSU and TSU,
(in particular we should replace $\xpar \to \Xpar$ for PSU and $\xperp \to \Xperp$ for TSU)
but we will keep this implicit to shorten the notation.

Let us now define the operators $\hat  D^{k}_{[\s]}$.
As usual the lowest components are defined by $\hat D^{T}_{[\cdot]}=1$ and $\hat D^{P}_{[\cdot]}$=1. 
For the TSU, the defect operators do not live in superspace, so there is only one additional differential operator, 
\begin{align}
& {\hat D}^{T}_{[\bf{1}]}=\Delta _1 \left(p-\Delta _1\right)-2 \z \left((p+2 \z-2) \partial _\z+2 (\z-1) \z \partial _\z^2\right)+\hat{\Delta }_{23}^2 \z \, ,
\label{D_bdd_TSU}
\end{align}
which corresponds to the highest primary component of the bulk operator.

In the PSU case, instead, there are $13$ more components defined by non-trivial differential operators, e.g.
{
\medmuskip=0mu
\thinmuskip=0mu
\thickmuskip=0mu
\begin{align}
& {\hat D}^{P}_{[\bf{1}]}=\Delta _1 \left(\Delta _1-q+2\right)+2 \z \left( \left(q-2 \left(\Delta _1+\z\right)\right)\partial _\z-2 (\z-1) \z \partial _\z^2\right)+\hat{\Delta }_{23}^2 \z \, , \nonumber\\
&  {\hat D}^{P}_{[\bf{2}]}=-\left(\hat{\Delta }_{23} \left(\hat{\Delta }_2+\hat{\Delta }_3+\left(\hat{\Delta }_{23}+2\right) \z\right)\right)-4 \z \left(\left(\hat{\Delta }_3+\left(\hat{\Delta }_{23}+2\right) \z-1\right) \partial _\z+(\z-1) \z \partial _\z^2\right) \, , \quad \nonumber\\
&  {\hat D}^{P}_{[\bf{3}]}={D}^{P}_{[\bf{2}]}|_{\hat{\Delta }_2\leftrightarrow\hat{\Delta }_3}\, , \label{eqn:D_bdd} \\
&   {\hat D}^{P}_{[2\bar{3}]}=-\hat{\Delta }_2-\hat{\Delta }_3+2 \z \partial _\z \, , \nonumber\\
&  {\hat D}^{P}_{[1\bar{2}]}=\sqrt{\z}(-\hat{\Delta }_{23}-2 \z \partial _\z) \, .
\nonumber
\end{align}}
We do not provide explicit expressions for the other operators here, as their forms are considerably lengthy. Instead, we include them in the ancillary Mathematica file submitted with this paper. Note that the highest component, ${\hat D}^{P}_{[\mathbf{123}]}$, is a sixth-order differential operator.

\subsection{Relations for 3pt  function  conformal blocks} \label{subsection:3pt function relations}
Now we derive the relations between three-point blocks. We only have defect channel expansion in this case. So the expectation is similar to the discussion of \ref{subsection:blockrelation} in the defect channel, 
\be
\hat D^k_{[\sigma]} \;
\raisebox{
1.5em
}{
$\xymatrix@=10pt{
&&& 
\ar@{-}@<-0.1mm>@[newyellow][d] 
\ar@{-}@<0.1mm>@[newyellow][d] 
\ar@{-}@[newyellow][d] \\
\ar@{-}[rrr]^{\hat{\Delta}} \ar@{}[rrr]
& & & *+[o][F]{} 
\ar@{-}@<-0.1mm>@[newyellow][d] 
\ar@{-}@<0.1mm>@[newyellow][d] 
\ar@{-}@[newyellow][d] \\
&&&
*{\bullet}
\ar@{-}@<-0.1mm>@[newyellow][d] 
\ar@{-}@<0.1mm>@[newyellow][d] 
\ar@{-}@[newyellow][d] \\
&&& 
*{\bullet}
\ar@{-}@<-0.1mm>@[newyellow][d] 
\ar@{-}@<0.1mm>@[newyellow][d] 
\ar@{-}@[newyellow][d]\\
&&& \quad \quad \\
&&& *{\vphantom{\bullet}} \save "4,4"*+!U(-8){\scriptstyle (p)} \restore
}$
}
= \;  \sum_{i \in  \hat S^k_{[\sigma]}} \hat r^{k}_{[\sigma],i} \Sigma_{[\sigma]}\;\!
\raisebox{
1.5em
}{
$\xymatrix@=10pt{
&&& 
\ar@{-}@<-0.1mm>@[newyellow][d] 
\ar@{-}@<0.1mm>@[newyellow][d] 
\ar@{-}@[newyellow][d]\\
 \ar@{-}[rrr]^{\hat{\Delta} + i} & & & *+[o][F]{} 
\ar@{-}@<-0.1mm>@[newyellow][d] 
\ar@{-}@<0.1mm>@[newyellow][d] 
\ar@{-}@[newyellow][d]
 \\
&&& 
*{\bullet}
\ar@{-}@<-0.1mm>@[newyellow][d] 
\ar@{-}@<0.1mm>@[newyellow][d] 
\ar@{-}@[newyellow][d] \\
&&& 
*{\bullet}
\ar@{-}@<-0.1mm>@[newyellow][d] 
\ar@{-}@<0.1mm>@[newyellow][d] 
\ar@{-}@[newyellow][d]
\\
&&& \quad \quad \\
&&& *{\vphantom{\bullet}} \save "4,4"*+!U(-8){\; \scriptstyle (p)^k} \restore
}$
} \, ,
\vspace{-2.5em}
\ee
where $(p)^P=p+2$ while $(p)^T=p$.

Let us now spell out what happens for the TSU. 
In this case the defect is not defined in superspace and the defect channel expansion consists of single defect exchange. In other words the set $\hat {S}^T_{[\s]}$ contains a single element $\hat {S}^T_{[\s]}=\{0\}$ and the correspondent coefficient is 
\be
 \hat{r}^T_{[\cdot],0}=1 \, ,
 \qquad
  \qquad
  \hat{r}^T_{[{\bf{1}]},0}=\left(\hat \Delta -\Delta _1\right) \left(\hat \Delta +\Delta _1-p\right) \, .
\ee
Moreover we notice that for the TSU the label $p$ is not shifted and also $\Sigma_{[\s]}$ acts trivially since the block does not depend on $\Delta_1$. 
As a result, the relation for the component $[\cdot]$ is trivial, namely just equates the conformal block to itself.
On the other hand the relation for $[\mathbf{1}]$ turns out to be substantially just the Casimir equation for the conformal block. 

For the PSU the relations are more non trivial.
The set $\hat {S}^P_{[\s]}$ contains either one of two elements according to $\hat {S}^P_{[\s]}=\{0,2\}$ for all $[\sigma]$ except for  $[\sigma]=[1{\bar 2}],[1{\bar 3}],[221\bar 3],[331\bar 2]$ where $\hat {S}^P_{[\s]}=\{0\}$.
Let us exemplify the form of some coefficients, {
\medmuskip=0mu
\thinmuskip=0mu
\thickmuskip=0mu
\begin{equation}
\begin{array}{ll}
     \hat{r}^P_{[\cdot],0}=1,
   &
   \hat{r}^P_{[\cdot],2}= \frac{\hat{\Delta }_{23}^2-\hat\Delta ^2}{(p-2 \Delta ) (p-2 (\hat \Delta +1))}\, ,\\
     \hat{r}^P_{[{\bf{1}]},0}=\left(\hat \Delta -\Delta _1\right) \left(\hat \Delta -\Delta _1+q-2\right)\, , 
  \qquad \qquad 
   &
   \hat{r}^P_{[{\bf{1}]},2}=-\frac{(\hat \Delta ^2-\hat{\Delta }_{23}^2 ) \left(-\hat \Delta -\Delta _1+p\right) \left(-\hat \Delta -\Delta _1+p+q-2\right)}{(p-2 \hat \Delta ) (p-2 (\hat \Delta +1))}\, ,\\
    \hat{r}^P_{[1\bar{2}],0}= -\hat \Delta -\hat{\Delta }_{23}\, .
    &
    \end{array}
\end{equation}}
The remaining coefficients are defined in the ancillary Mathematica file included with the submission.
\section{Examples}
\label{sec:examples}

\subsection{Trivial defect}
Let us start by considering the trivial defect and show that one can find its conformal block decomposition using constraints of SUSY, akin to what was done in \cite{Trevisani:2024djr} for the four-point function of scalar operators in generalized free theory.

We start by considering the standard  two-point function $\langle \Ocal_1(x_1) \Ocal_1(x_2) \rangle = |x_{12}|^{-2\Delta_1}$ in a CFT$_d$ without any defect. Now we think of it as if a trivial $p$-dimensional defect was inserted. This observable is now written in terms of a function of two cross ratios according to \eqref{2ptD} which takes the form,
\be
\label{f_trivial_D}
f(\up,\ut)=[2(\up-\ut)]^{-\Delta_1} \, .
\ee
As a consistency check, we first apply the differential operators $D^k_{[\sigma]}$ to \eqref{f_trivial_D} and match the result of for the components of a two-point function computed in (3.11)-(3.13) of \cite{Trevisani:2024djr}. 
We can also check that $D^k_{[{\bf 1}]}$ and $D^k_{[{\bf 2}]}$ annihilate this two-point function. This has to be the case since it is computing the two-point function of different primary operators (one lowest and one highest component).
It turns out that we can use this simple information to compute the defect conformal block decomposition of the trivial defect (one could in principle apply the same logic to the bulk decomposition but this is too simple since it is solved in terms of the single identity block). In particular the following relations hold 
\be
\label{vanishing_component}
0=D^k_{[{\sigma}]} f(\up,\ut) = \sum_{\hD, s} a_{\hD, s} D^k_{[{\sigma}]} \hat g^{(p,q)}_{\hD, s}
=
\sum_{\hD, s}  a_{\hD, s} 
\sum_{(i,j) \in \hat S^k_{[\sigma]}} \hat c^{k}_{[\s],i,j} \hat g^{(p,q)^k}_{\hD+i , s+j} =  \sum_{\hD', s'} b^k_{\hD', s'} \hat g^{(p,q)^k}_{\hD' , s'}  ,
\ee
where in the third equality we used \eqref{def_rel} while the last equality we grouped together all blocks with the same labels (now called $\hat{\Delta}', s'$). The coefficients $b^k_{\hD', s'}$ are just linear combinations of $\hat c^{k}_{[\s],i,j}$ and $a_{\hD, s} $. 
Since blocks with different labels are linearly independent, we conclude that  all the coefficients $ b^k_{\hD', s'}$ vanish. 
The spectrum of the exchanged defect operators is defined by the dimensions  $\hD=\Delta_1+2m+s$ and spin $s$, which descends from the Taylor expansion of $\Ocal_1$ around the defect $(\partial^2_{\perp})^m  \partial^{i_1}_\perp \dots \partial^{i_s}_\perp \Ocal_1(\xpar)$. Using this information we can recast $b^k_{\hD', s'}=0$ as recurrence relations for $a_{\hD, s} $. In particular we present the PSU and TSU relation for $[{\bf{1}}]$ (the ones of $[{\bf{2}}]$ are equal), 
\be
\begin{array}{l}
0=(c^{P}_{[{\bf{1}}],0}|_{\hD\to \hD+2}) a_{\hD , s}  + c^{P}_{[{\bf{1}}],2} a_{\hD+2 , s}\bigg|_{\hD=\Delta_1+2m+s} 
\, , \\
0=c^{T}_{[{\bf{1}}],0} a_{\hD , s}+(c^{T}_{[{\bf{1}}],-2}|_{s\to s+2}) a_{\hD, s+2}  \bigg|_{\hD=\Delta_1+2m+s} \, ,
\end{array}
\ee
where we notice that the PSU one gives a recurrence relation in $\hD$, while the TSU a relation in $s$. 
These equations can be rewritten more explicitly as
\be
\begin{array}{l}
a_{\Delta_1+2m+s,s}=
\frac{(m+1) (2 m+q+2 s) \left(2 \Delta _1+4 m-p+2 s\right) \left(2 \Delta _1+4 m-p+2 s+2\right)}{2 \left(\Delta _1+2 m+s\right) \left(\Delta _1+2 m+s+1\right) \left(2 \Delta _1+2 m-p-q+2\right) \left(2 \Delta _1+2 m-p+2 s\right)}
a_{\Delta_1+2m+2+s, s} \, ,
\vspace{0.2cm}
\\
a_{\Delta_1+2m+s,s}=
\frac{(s+1) (s+2) (2 m+q+2 s) \left(2 \Delta _1+2 m-p-q\right)}{2 m (q+2 s) (q+2 s+2) \left(2 \Delta _1+2 m-p+2 s\right)}
a_{\Delta_1+2m+s, s+2} \, .
\end{array}
\ee
We can further solve them as
\be
\begin{array}{l}
a_{\Delta_1+2m+s,s}=
\frac{ \Gamma \left(\frac{q}{2}+s\right) \Gamma \left(2 m+s+\Delta _1\right) \Gamma \left(m+\Delta _1-\frac{p}{2}-\frac{q}{2}+1\right) \Gamma \left(m+s+\Delta _1-\frac{p}{2}\right)}{m! \Gamma \left(s+\Delta _1\right) \Gamma \left(m+\frac{q}{2}+s\right) \Gamma \left(-\frac{p}{2}+\Delta _1-\frac{q}{2}+1\right) \Gamma \left(2 m+s+\Delta _1-\frac{p}{2}\right)}
\times
\frac{2^s \left(c_1+c_2 (-1)^s\right) \left(\Delta _1\right)_s}{ s!} \, ,
\end{array}
\ee
where the first term is obtained by solving the recursion in $\hD$ keeping as a seed $a_{\Delta_1+s,s}$, while the second terms solves the recursion in $s$ and it is determined up to two seeds which can be fixed as $c_1=1, c_2=0$ by computing $a_{\Delta_1,0}=c_1+c_2=1$, $a_{\Delta_1+1,1}=2 (c_1-c_2) \Delta _1=2 \Delta _1$. 

This exercise shows that the existence of the two uplifts can be used to compute the defect block decomposition of the two-point function in the presence of a trivial defect. We find that the full trajectory of operator is fixed by SUSY (by knowing the first two OPE coefficients which are trivial to compute).

Finally we can also consider  the case of a bulk-defect-defect three-point function. In this case, since the defect is trivial, the result is just  a normal three-point function. By writing it in the parametrization \eqref{BDD} we find that 
for the trivial defect 
\be
f(\z)=  \z^\frac{\Delta_1}{2} \, .
\ee
Now we can act with the differential operators \eqref{eqn:D_bdd} on $f(\z)$ to compute all components. Since the defect is trivial, the result must match the components of a standard uplifted three-point function, which were computed in equations (3.14)-(3.15)  \cite{Trevisani:2024djr}.
We verified that this is indeed the case, which is a consistency check for the differential operators \eqref{BDD}. In a general theory, there is no vanishing component for this observable, so we do not find any relation of the type \eqref{vanishing_component} to use to extract recurrence relations for OPE coefficients. One could however consider special cases, e.g.  
$\Delta_1=\hat \Delta_2+ \hat  \Delta_3$ or $\hat \Delta_2= \Delta_1+ \hat  \Delta_3$ (these conditions are satisfied for a trivial defect in a free theory) where some components of the three-point function vanish, as shown in  \cite{Trevisani:2024djr}.  
One could then repeat the same logic as before and obtain recurrence relations for the OPE coefficients. We leave this exercise to the reader.

\subsection{Free theory defect}
Let us consider the simplest case of a defect obtained by integrating a local operator over an extended manifold. 
We focus on the theory of  a $d$-dimensional free scalar $\phi$ with the insertion of a defect obtained by integrating $\phi$ over  $p$-dimensions as in
\cite{Billo:2016cpy},
\begin{equation}
\label{S_free_defect}
S=\int
    \,d^dx ~\frac12 ( \partial{\phi})^2+ h_0\int
    \,d^p\xpar \,{\phi}(\xpar) \, .
\end{equation}
We choose $\Delta_{{\phi}}=\frac{d-2}{2}=p$ (e.g. $p=1$ and $d=4$) so that the defect is conformal for any value of  $h_0$.
In this theory the one-point function of bulk field $\phi(x)$ is given by,
\begin{equation} \label{1ptfree}
\begin{split}
    \langle \phi(x) \rangle_p 
    =\frac{\langle \phi(x) e^{h_0\int\,d\xpar\,\phi(\xpar) } \rangle_0}{\langle e^{h_0\int\,d\xpar\,\phi(\xpar)} \rangle_0}
    =h_0 \int d^p\xpar_1 \langle \phi(x)\phi(\xpar_1) \rangle_0
      =\frac{a_{\phi}}{|\xperp|^{\Delta_{\phi}}} \, ,
    \end{split}
\end{equation}
where, $a_{\phi}= h_0 \mathcal{N}_d\frac{\Gamma(\frac{p}{2})\Gamma(\Delta_{\phi}-\frac{p}{2})}{2\Gamma(\Delta_{\phi})}\Omega_{p-1}$ and $\mathcal{N}_d$ is the same as in \eqref{def:GPhiPhi}.
We denote the free theory correlators in the absence of the defect with the subscript $0$ and, according to \eqref{S_free_defect}, the free propagator is normalized as  $\langle \phi(x) \phi(0) \rangle_0=\mathcal{N}_d |x|^{2-d}$.
The two-point function takes the following form,
\begin{equation}
    \langle \phi(x)\phi(y) \rangle=\frac{1}{|x-y|^{2\Delta_{\phi}}}+\frac{a_{\phi}^2}{|\xperp|^{\D_{\phi}}|\yperp|^{\D_{\phi}}} \, .
\end{equation}
Let us now show how to uplift this setup to realize both the PSU and the TSU.  
Following the conventions of \eqref{PSU_coords} and \eqref{TSU_coords}, the uplifts can be written in terms of the following two actions
\begin{align}
    S_{\text{PSU}} & = \int
    \, d^{d+2}X  \, [d\theta][d\bar{\theta}] ~\frac12\left( \partial\Phi\right)^2  + h_0\int
    \,d^{p+2}\Xpar\,[d\theta][d\bar{\theta}]\Phi(\Xpar,\theta,\bar{\theta}) \, ,
    \label{PSU_free}
    \\
     S_{\text{TSU}} & = \int
     \,d^{d+2}X [d\theta][d\bar{\theta}]\, ~\frac12\left( \partial\Phi\right)^2  + h_0\int
     \,d^p\xpar \,\varphi(\xpar) \, ,
      \label{TSU_free}
\end{align}
where the superfield $\Phi$ has the following expansion in components,
\begin{equation}
\label{Phi_components}
\Phi(X,\theta,\bar{\theta})=\varphi(X)+\theta \bar{\psi}(X)+\bar{\theta}\psi(X)+\theta \bar{\theta} \omega(X) \, .
\end{equation}
The free propagators  in superspace are written in \eqref{def:GPhiPhi}, while the non-vanishing propagators of the components \eqref{Phi_components} take the following form,
\begin{equation}
    \langle\varphi(X)\varphi(0) \rangle_0=\frac{\mathcal{N}_d}{|X|^{d-2}} \, ,
    \qquad \langle\varphi(X)\omega(0) \rangle_0=\frac{\mathcal{N}_d(d-2)}{|X|^{d}}=\langle\psi(X)\bar{\psi}(0) \rangle_0 \, .
\end{equation}
Let us now describe a couple of simple examples. For the PSU, the one-point function is given by,
\begin{equation}
\begin{split}
    \langle \varphi(X)\rangle_{\text{PSU}} 
    =\frac{h_0}{2\pi} \int d^{p+2}\Xpar_1 \langle \varphi(X)\omega(\Xpar_1)\rangle_0 
    =\frac{h_0}{2\pi} \mathcal{N}_d(d-2)\int \frac{d^{p+2}\Xpar_1 }{\big|X-\Xpar_1\big|^d}
    =\frac{a_{\phi}}{|\xperp|^{\Delta_{\phi}}} \, ,
\end{split}
\end{equation}
which agrees with \eqref{1ptfree}, as expected from dimensional reduction. Moreover one can see that the one-point functions of the other fields  $\psi,\psib,\omega$ all vanish (since their two-point functions with $\omega$ vanish), therefore $\langle \Phi(y)\rangle_{\text{PSU}}=\langle \varphi(X)\rangle_{\text{PSU}}$ as expected from \eqref{eq:1pt_PSU}.

For the TSU, the one-point function  $\langle \varphi(x) \rangle_{\text{TSU}}$ trivially agrees with \eqref{1ptfree} since the two computations are defined in terms of the same integral.
In particular we stress that the propagator of $\varphi$ and the one of $\phi$ are exactly the same.
It is also interesting to notice that in the TSU, the highest component $\omega$  has a non-trivial one-point function and it is easy to  check that $\langle \Phi(y)\rangle_{\text{TSU}}$ is indeed consistent with \eqref{eq:1pt_TSU}.

This thus exemplifies dimensional reduction at the level of one-point function. The reduction for the two-point function can be easily shown since this equals a connected piece (which is the same of a trivial defect that we considered above) plus the square of the one-point function which we just considered.

This setup simply shows how to explicitly realize the PSU and TSU for defects defined by integrating local operators. Given an original defect defined by integrating $\phi$ in $p$ dimensions, we consider the uplifted superfield $\Phi$ and we define
the PSU by integrating its highest component in $p+2$ dimensions,  while the TSU by integrating its lowest component in $p$ dimensions.
Another example of this prescription will be presented below.

Finally, let us comment on an amusing feature of the PSU. In this uplift one can compute the vacuum expectation value of the defect itself 
\be
\left\langle e^{\int d^{d+2}\Xpar \omega(\Xpar) }\right\rangle_0 =1 \, ,
\ee
which is equal to one, since $\langle\omega(X)\omega(0)\rangle_0=0$. This is in contrast with what happens with the action  \eqref{S_free_defect} or the TSU one, where the same observable gives rise to a divergent result. 
Notice that this is a property of supersymmetry that holds non-perturbatively also in cases where the integrated field is not free. 
The PSU setup thus provides a convenient framework for defect computations, where the defect divergencies are automatically regulated.
\subsection{
 Line defect in Wilson-Fisher
}
In this section we give an example of PSU and TSU for a weakly interacting theory. This will serve also as a concrete application of the Feynman diagram reduction discussed in section \ref{sec:reduction}.

We consider the line defect in the perturbative Wilson-Fisher CFT, which is defined by the following action in $d=4-\epsilon$ spacetime dimensions \cite{Allais:2014fqa, Cuomo:2021kfm},
\begin{equation}
\label{WF_line}
   S = \int\,d^dx \left[\frac12\left( \partial\phi(x)\right)^2 + \frac{\lambda_0}{4!}\phi^4(x)\right] + h_0\int\,d x^1\,\phi(x^1) \, .
\end{equation}
When $\epsilon=0$ the coupling $\lambda_0$ is marginally irrelevant and in the IR the defect becomes equivalent to the one described in \eqref{S_free_defect}. For small $\epsilon$, we
choose $\l$ such that its beta function vanishes, so that the bulk theory in the IR describes the Wilson-Fisher fixed point which is accessible in perturbation theory.
The field $\phi$ receives a negative anomalous dimension and the defect coupling $h_0$ runs. By setting the relative beta function to zero, this model defines a non-trival conformal defect line in the Wilson-Fisher fixed point, which was analyzed in detail in \cite{Cuomo:2021kfm}.
Let us show how to uplift this setup.

First we define the PSU which takes the same form as  \eqref{PSU_free} with $p=1$ with the addition of a quartic interaction in superspace,
\begin{equation}
     S_{\text{PSU}} = \int d^{d+2}X [d\theta][d\bar{\theta}]\left[\frac12\left( \partial\Phi(y)\right)^2 + \frac{\lambda_0}{4!}\Phi^4(y)\right] + h_0\int d^3\Xpar\,[d\theta][d\bar{\theta}]\Phi(\ypar) \,  ,
\end{equation}
where, following the usual conventions of  \eqref{PSU_coords}, we define $y=(X,\theta,\thetab)$, $\ypar=(\Xpar,\theta,\thetab)$. As before, the field $\Phi$ can be expanded in components according to \eqref{Phi_components} and, 
 if we integrate out the fermionic defect coordinates, the defect reduces to a three-dimensional integral of the highest component $\omega$.
\begin{figure}[htbp]
    \centering
 \begin{subfigure}[b]{0.4\textwidth}
       \centering
        \begin{tikzpicture}
            \draw[-,ultra thick,newyellow] (3,0) -- (5,0);
            \filldraw [newyellow] (4,0) circle (3pt) node[below] { $\Phi(\ypar_1)$};
            \filldraw (4,2) circle (3pt) node[above,yshift=0.05cm] { $\Phi(y)$};   
            \draw (4,0) -- (4,2);
        \end{tikzpicture}
        \caption{
      Order   $O(\lambda_0 h_0)$
        }
        \label{fig:leading}
     \end{subfigure}
       \begin{subfigure}[b]{0.45\textwidth}
 \centering
        \begin{tikzpicture}
          \draw (2,0) -- (4,1);
            \draw (4,0) -- (4,1);
            \draw (6,0) -- (4,1);
            \draw (4,2) -- (4,1);
            \draw[-,ultra thick,newyellow] (1,0) -- (7,0);
            \filldraw [newyellow] (2,0) circle (3pt) node[below] {
            $\Phi(\ypar_1)$};
            \filldraw [newyellow] (4,0) circle (3pt) node[below] {
            $\Phi(\ypar_2)$};
            \filldraw [newyellow] (6,0) circle (3pt) node[below] { $\Phi(\ypar_3)$};
            \filldraw [black, thick](4,1) circle (3pt) node[right,yshift=0.08cm] { \,$y_I$}; 
            \filldraw (4,2) circle (3pt) node[above,yshift=0.05cm] { $\Phi(y)$}; 
        \end{tikzpicture}
      \caption{
       Order $O(\lambda_0 h^3_0)$
       }
        \label{fig:subleading}
   \end{subfigure}
   \hfill
   \caption{Diagrams contributing to $\langle \Phi(y) \rangle$ up to order $O(\lambda_0)$.}
    \label{fig:1order}
\end{figure}
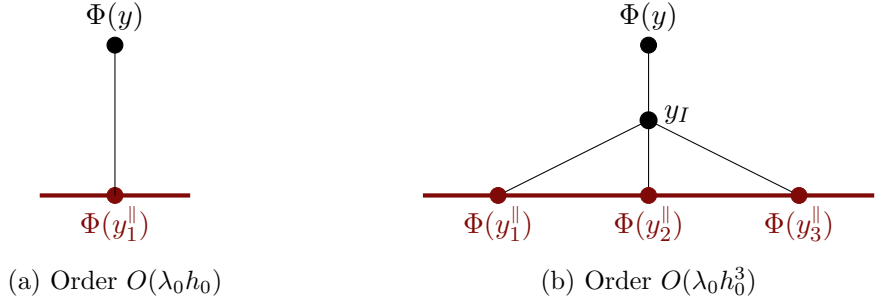
For the sake of clarity, let us showcase some simple perturbative computations and explain how they reduce to the ones of \cite{Cuomo:2021kfm}. 
We shall focus on the one-point function of $\Phi$ field to first order in bulk coupling (higher orders can also be considered, as shown in \cite{Cuomo:2021kfm}), which is given by the Feynman diagrams shown in figure \ref{fig:1order}. The leading order diagram in  figure \ref{fig:leading} is simply given by
\begin{align}
\langle\Phi(y)\rangle|_{h_0}
&=-h_0  \, \mathcal{N}_d\int d^3\Xpar_1 [d\theta_1][d\bar{\theta_1}] \frac{1}{\left( (\xperp)^2 +
(\Xpar-\Xpar_1)^2
-2(\theta-\theta_1)(\bar{\theta}-\bar{\theta}_1)\right)^{\frac{d-2}{2}}}
\nonumber
\\
& =-h_0\,\mathcal{N}_d\int d^3 \Xpar_1 [d\theta_1][d\bar{\theta_1}] \int du\, u^{\frac{d-2}{2}-1} e^{-u\left((\xperp)^2 +(X^1-X^1_1)^2+(X^\alpha-X^\alpha_1)^2-2(\theta-\theta_1)(\bar{\theta}-\bar{\theta}_1)\right)}\nonumber
\\
&=-h_0\,\mathcal{N}_d\int d X^1 \int du\, u^{\frac{d-2}{2}-1}e^{-u\left((\xperp)^2 +(X^1)^2\right)}
\nonumber\\
&=-h_0 \frac{\mathcal{N}_d U(1,\frac{d-2}{2})}{|\xperp|^{d-3}},
\label{PhiPSU_WF_line_lead}
\end{align}
where $U(p,k) \equiv  \Gamma(p/2)   \Gamma \left(k -p/2\right)/ \Gamma \left(k\right)$. 
In the second line, we used Schwinger parametrization and we split the parallel three-vectors as $\Xpar=(X^1,X^\alpha)$ with $\alpha=d+1,d+2$. The third line is obtained  by shifting $\Xpar_1,\theta_1,\thetab_1$ and performing the integrals over $\mathbb{R}^{2|2}$, as explained in \eqref{eq:2|2integral}. The final result precisely matches  the computation in $d$ dimensions in \cite{Cuomo:2021kfm}. \\
Similarly the subleading diagram reads
\begin{equation}
    \begin{split}
\langle\Phi(y)\rangle|_{\lambda_0 h^3_0}&
=\frac{\lambda_0 h^3_0}{3!} \int d^{d+2} X_B [d\theta_B][d\bar{\theta}_B]\prod_{i=1}^3d^3\Xpar_i [d\theta_i][d\bar{\theta}_i]G_{\Phi\Phi}(y-y_B)\prod_{i=1}^3G_{\Phi\Phi}(y_B-\ypar_i)
\\
&=\frac{\lambda_0 h_0^3}{3!}\mathcal{N}_d^4 U(1,\frac{d-2}{2})^4 C(d-1,\frac{3d-9}{2},\frac{d-3}{2}) \frac{1}{|\xperp|^{3d-11}} \, , 
\\
  C(d,a,b)&\equiv\frac{\pi ^{d/2} \Gamma \left(\frac{d}{2}-a\right) \Gamma \left(\frac{d}{2}-b\right) \Gamma \left(a+b-\frac{d}{2}\right)}{\Gamma (a) \Gamma (b) \Gamma (-a-b+d)} \, .
\end{split}
\label{PhiPSU_WF_line_sublead}
\end{equation}
The derivation follows exactly the same steps as before, i.e. we write the diagram using Schwinger parameters and then perform all the integrals over $\mathbb{R}^{2|2}$ directions. This also matches \cite{Cuomo:2021kfm}.
Notice that, as expected from \eqref{eq:1pt_PSU}, in both formulae \eqref{PhiPSU_WF_line_lead} and \eqref{PhiPSU_WF_line_sublead}, the dependence on $\theta$,$\bar{\theta}$ has completely disappeared so that only $\varphi$ acquires a vev but not the higher components.

These computations show how the coupling $h_0$ renormalizes and one can canonically use them to extract the beta function for $h_0$. 
The result matches the lower dimensional counterpart of \cite{Cuomo:2021kfm}, since all the diagrams match. In particular at this order one finds $
 \beta_h= -h\frac{\epsilon}{2}+  \frac{\lambda h^3}{6(4\pi)^2} $ and by setting $\lambda$ to the critical value for the Wilson-Fisher fixed point $\lambda^*= \frac{(4\pi)^2}{3}\epsilon$ (the beta function for $\lambda$ also matches the dimensionally reduced one, as  reviewed in \cite{Kaviraj:2020pwv}), one finds the defect fixed point at $(h^*)^2=9$. As it is known, the value of $h^*$ is not of order $O(\epsilon)$, which means that at a given order of $\lambda$ we should  consider all possible orders in $h$ as we did in the computations above. 
One can further compute anomalous dimensions of operators and, since the dimensional reduction works diagram by diagram according to section \ref{sec:reduction}, the results must match the ones of the original action \eqref{WF_line}.

Let us now turn to the TSU. 
This defect setup is obtained by integrating the lowest component of $\Phi$ on a line. In $d+2=6-\epsilon$ dimensions the resulting model takes the form 
\begin{equation}
   S_{\text{TSU}} = \int
   \,d^{d+2}X [d\theta][d\bar{\theta}] ~~\left[\frac12\left( \partial\Phi(x,\theta,\thetab)\right)^2 + \frac{\lambda_0}{4!}\Phi^4(x,\theta,\thetab)\right] + h_0\int
   \,dx^1\,\varphi(x^1) \, .
\end{equation}
In this case, when considering correlators of the lowest component $\varphi$, we find that all propagators in the Feynman diagrams are just the ones of $\langle\varphi(x)\varphi(0)\rangle_0$, which coincide with the lower dimensional ones.
For example let us spell out the leading order diagram,
\begin{equation}
\begin{split}
\langle\Phi(y)\rangle|_{h_0}&=-h_0\int dx^1_1  \frac{1}{\left((x^1-x^1_1)^2+|\xperp|^2+|X^\alpha|^2-2\theta\bar{\theta}\right)^{\frac{d-2}{2}}} \, ,
\end{split}
\label{PhiTSU_WF_line_lead}
\end{equation}
where we followed the conventions of \eqref{TSU_coords} and we split $\Xperp=(\xperp,X^\alpha)$ with $\alpha=d+1,d+2$.
First we notice that the dependence on $\theta$,$\bar{\theta}$ does not disappear (it cannot be shifted away as in the PSU examples above) and therefore the $\omega$ field acquires a vev as expected from \eqref{eq:1pt_TSU}.
Moreover here we see that by setting the $\mathbb{R}^{2|2}$ coordinates $X^{\alpha},\theta,\thetab$ to zero, we trivially recover the $d$-dimensional Feynman diagram. Similarly any other computation will simply dimensionally reduce.
In particular, also in this case, we can define beta functions for $\lambda$ and $h$ and by setting them to zero we can define a conformal defect, where we can compute any observable which must automatically dimensionally reduce to the ones computed by the action \eqref{WF_line}.
\subsection{Conformal boundaries in minimal models} \label{Subsection:bcft 2D}
In this section, we consider an infinite set of non-perturbative examples of BCFTs in $d=2$ and  show the exact form of some uplifted four-dimensional correlators. 
In particular we focus on minimal models in the presence of conformal boundaries, where the bulk two-point functions  is known in a closed form. 
We uplift these to four dimensions and extract different components which, by construction, are decomposed in terms of the conformal blocks of the uplifted CFT. This thus provides an infinite sequence of examples of  higher-dimensional DCFTs where correlators can be computed exactly.

 Let us start by considering standard two-dimensional  $(m,m+1)$ minimal models \cite{DiFrancesco:1997nk}. The primary operators of these models can be labelled by two integers, $(r,s)$ with $1\leq r\leq m-1$ and $1\leq s\leq m$. Conventionally $(1,2)$ operator is labelled as $\sigma$ (spin operator) and $(1,3)$ operator is labelled as $\epsilon$ (energy operator). Their dimensions\footnote{It is also conventional to write the holomorphic dimension as $h_{r,s}=\frac{((m+1)r-m s)^2-1}{4m(m+1)}$. Then the scaling dimension,  $\Delta_{r,s}=h_{r,s}+\bar{h}_{r,s}$.} are given by,
\begin{equation}
\label{Delta_sigma_eps}
    \Delta_\sigma=\frac{1}{2}-\frac{3}{2(m+1)} \, ,
    \qquad\qquad \Delta_\epsilon=2-\frac{4}{m+1}\, .
\end{equation}
We further study minimal models in the presence of a boundary which respects the conformal boundary conditions, originally classified by Cardy in  \cite{Cardy:1984bb, CARDY1989581,Cardy:2004hm}. In particular we will focus on the identity boundary condition (where the boundary spectrum is fixed in terms of the Virasoro identity multiplet) and study the two-point function of $\sigma$. 
By conformal symmetry the latter is fixed as
\begin{equation}
   \langle\sigma(x_1) \sigma(x_2)\rangle=\frac{1}{|\xperp_{1}|^{\Delta_\sigma}|\xperp_{2}|^{\Delta_\sigma}} G_{\sigma\sigma}(\up) \, ,
\end{equation}
where $G_{\sigma\sigma}$ is a function of the single cross ratio $\up$, since $\ut=1$ in boundary setups.
In this specific example the function $G_{\sigma\sigma}$ is known in closed form as follows \cite{Zamolodchikov:2001ah,Liendo:2012hy}
\begin{align}
G_{\sigma\sigma}(\up)&=P_{\Delta_\sigma}(\up) \; {}_2F_1(a,b,c,\tfrac{2}{1+\up}) \, ,
\\
P_{\Delta_\sigma}(\up)&\equiv 2 (\up -1)^{\frac{\Delta _{\sigma }}{3}+\frac{1}{3}} (1+\up )^{-\frac{\Delta _{\sigma }}{3}-\frac{1}{3}} \sin \left(\tfrac{\pi}{6}   \left(4 \Delta _{\sigma }+1\right)\right) \, ,
\end{align}
where $a\equiv \frac{1}{3} (1-2 \Delta_\sigma ),b\equiv \frac{2 (\Delta_\sigma +1)}{3},c\equiv \frac{1}{3} (2-4 \Delta_\sigma )$. In the boundary channel, only scalars global primaries of dimension $\hD=0,2,4,\dots $ are exchanged, while in the bulk channel, global primaries with dimensions $\Delta= 4n$ and $ \Delta=\Delta_{\epsilon} + 4n $ appear. This reflects the fact that only the identity Virasoro block contributes in the boundary channel, whereas in the bulk, both the Virasoro multiplet of the identity and of $\epsilon$ are exchanged.

Let us consider first consider the PSU. Since  $G_{\sigma\sigma}$ is explicitly known, it is straightforward to compute all the uplifted correlators by  applying the differential operators $D^P_{[\sigma]}$ defined in section \ref{sec:BB_2pt}. 
Let us show a couple of examples.
The $[1\bar{2}]$ component  takes the following form,
{
\medmuskip=0mu
\thinmuskip=0mu
\thickmuskip=0mu
\be
G_{[1\bar{2}]}(\up)= \frac{2 \left(\Delta _{\sigma }+1\right)\, P_{\Delta_\sigma}(\up) }{3 (\up -1) (\up +1)^2}
\left[(\up+1) {}_2F_1(a,b,c,\tfrac{2}{1+\up})-(\up-1){}_2F_1(a+1,b+1,c+1,\tfrac{2}{1+\up})\right] \, .
\ee}
When we decompose this correlator in terms of uplifted boundary channel blocks, we see that the exchanged global primaries have dimensions $\hD=2n+1 $, with $n$ being non-negative integers.
We find that the square of the OPE coefficients in the boundary channel can be negative signaling the non-unitarity of PSU theory.
E.g. for the exchange of the operator of dimension $\hD=3$ we get $b_{3}^2=8 \Delta _{\sigma } \left(\Delta _{\sigma }+1\right) \sin \left(\frac{\pi}{6}   \left(4 \Delta _{\sigma }+1\right)\right)/(4 \Delta _{\sigma }-5)<0$ for all values of $\Delta _{\sigma }$ allowed by \eqref{Delta_sigma_eps}. In the bulk channel, there are two towers of global primaries with dimensions $ \Delta = 2n $ and $ \Delta = \Delta_\epsilon + 2n $, as can be anticipated from the structure of the original theory.
\\
The $[\bf{1}]$ component takes the form,
{
\medmuskip=0mu
\thinmuskip=0mu
\thickmuskip=0mu
\begin{align} 
    G_{[{\bf 1}]}(\up)&= \frac{  (\Delta _{\sigma }+1) \, P_{\Delta_\sigma}(\up) }{9 (\up-1) (\up +1)^3 (4 \Delta _{\sigma }-5)} \bigg[
    (\up +1)^2 \left(4 \Delta _{\sigma }-5\right) \left((3 \up +5) (3 \up -1) \Delta _{\sigma }+4\right)  {}_2F_1(a,b,c,\tfrac{2}{1+\up})\nonumber\\
    & 
    \qquad\qquad\qquad
    -4 \left(\up ^2-1\right) \left(4 \Delta _{\sigma }-5\right) \left((3 \up +2) \Delta _{\sigma }+5\right){}_2F_1(a+1,b+1,c+1,\tfrac{2}{1+\up})
    \phantom{\bigg]}
    \nonumber\\
    & 
    \qquad\qquad
    +8 (\up -1)^2 \left(\Delta _{\sigma }-2\right) \left(2 \Delta _{\sigma }+5\right){}_2F_1(a+2,b+2,c+2,\tfrac{2}{1+\up})\bigg] \, .
\end{align} 
}
By decomposing it in terms of uplifted boundary blocks we find that it contains identity operator and then global even primaries with $\hD=2n$, with $n$  positive integers. We again notice that the square of the boundary channel OPE coefficient can be negative. The first instance appears when $\hD=4$ and takes the following form,
\begin{equation}
  b^2_4=  \frac{4 \left(\Delta _{\sigma }-2\right){}^2 \left(\Delta _{\sigma }-1\right) \Delta _{\sigma } \left(\Delta _{\sigma }+1\right){}^2 \sin \left(\frac{ \pi}{6}  \left(4 \Delta _{\sigma }+1\right)\right)}{\left(4 \Delta _{\sigma }-11\right) \left(4 \Delta _{\sigma }-5\right)} \, ,
\end{equation}
which is negative when $\Delta_{\sigma}$ respects the condition \eqref{Delta_sigma_eps}. The bulk exchanges are the same as we saw in the uplift of $[1\bar{2}]$ component.

We notice that the defect spectrum of the PSU contains only operators with integer dimensions, which indeed arise as the uplift of the Virasoro multiplet of the identity. The defect operators for the component $[1\bar{2}]$ are odd dimensional, while the ones of the component $[{\bf 1}]$ are even dimensional. This is related to the fact that the former are fermionic, while the latter are bosonic. It is also interesting to notice that the bulk decomposition has an interesting feature when $\Delta_\epsilon=1$. In this case the bulk channel block for the exchange of  the uplifted counterpart of $\epsilon$ is singular. However the singularity is exactly cancelled by the contribution of the highest primary component of its multiplet. This phenomenon was already noted in \cite{Trevisani:2024djr} in four-point functions without boundaries and it is related to the existence of null states that do not decouple. At a more formal level it can be also explained in terms of exotic conformal multiplets known as 
``extended'' or ``staggered'',  see \cite{Brust:2016gjy, Herzog:2024zxm, Kytola:2009ax}. 

Let us now comment on the TSU.
Unlike the PSU case, where a BCFT is mapped to another higher-dimensional BCFT, here the uplift is to a codimension‑3 defect setup. As a result, reconstructing the full two-point function is not possible. This is because the two-point function of a scalar operator on a codimension-3 defect depends on two independent cross ratios, whereas in a BCFT it depends on only one. Nevertheless, the $d=2$ BCFT correlator fully reconstructs the higher codimension defect correlator in the kinematic limit of
$\ut=1$. In this special configuration, both the bulk and defect channel contributions in the uplifted theory come exclusively from scalar operators.
At the level of blocks, the relations of section \ref{subsection:blockrelation} still hold by setting $\ut=1$ and all spins to zero.
In this limit, the bulk block is known in closed form for every $q$ in terms of a ${}_3F_2$ hypergeometric function, as  shown in equation \eqref{eq:bulk_block_bdy} in appendix \ref{appendix:3F2}.
Thanks to the TSU uplift, a single bulk channel block for $q=1$ ---given by a ${}_2F_1$ as shown in \eqref{eq:bulk_block_q_v1}--- can be expressed as a sum of two ${}_3F_2$ hypergeometric functions. 
This is fairly interesting, as  identities relating different classes of transcendental functions, such as ${}_2F_1$ and ${}_3F_2$, are quite rare. In the mathematics literature, these are known as finite-term identities for hypergeometric functions. We provide further discussion and derivations in appendix \ref{appendix:3F2}. Notably, the TSU construction offers an independent route to deriving these identities.

Finally, let us consider the uplift of the bulk-defect-defect  $3-$point function. It can be derived using the method of images for a boundary primary operator with dimension $h_{r,s}$ and bulk primary operator with dimension $2h_{1,2}$ \cite{Antunes:2024hrt},
\begin{equation}
\begin{aligned}
    \langle\phi_{1,2}(x_1)\psi_{r,s}(x_2) \psi_{r,s}(x_3)\rangle&=\frac{1}{|x_1|^{2h_{1,2}}|x_{23}|^{2h_{r,s}}}f(\z)\\
    &=\frac{1}{|x_1|^{2h_{1,2}}|x_{23}|^{2h_{r,s}}}(b_1 V_{1,1}(\z)+b_2 V_{1,3}(\z)) \, ,
\end{aligned}
\end{equation}
where the boundary conditions are labelled by $(a_1,a_2)_m$. For example, for $m=3$ Ising model $\mathbb{Z}_2$ preserving boundary condition is labelled as $(1,2)_3$.\footnote{See \cite{Antinucci:2023uzq} for a detailed discussion and derivation of this correlator.} The Virasoro blocks take the following form,
\begin{align}
V_{(1,1)}(\z) &= 
{}_2F_{1}\!\left(
\frac{m(r-s)+r+1}{2m+2},\,
\frac{m(s-r)-r+1}{2m+2};\,
\frac{m+3}{2m+2};\,
\z
\right), \nonumber\\[6pt]
V_{(1,3)}(\z) &= 
\z^{\,h_{1,3}/2}\,
{}_2F_{1}\!\left(
\frac{m(r-s)+m+r}{2m+2},\,
\frac{m(s-r)+m-r}{2m+2};\,
\frac{3m+1}{2m+2};\,
\z
\right).
\end{align}
and 
\begin{equation}
\begin{aligned}
& b_1=2(-1)^{a_1}\cos\big(\frac{\pi a_2 m}{m+1}\big)\sqrt{-\frac{\sin\big(\frac{\pi m}{m+1}\big)}{\sin\big(\frac{2\pi m}{m+1}\big)}}\\
& b_2=-b_1\frac{
2^{\tfrac{2}{m+1}-1}
\, \Gamma\!\left(\tfrac{1}{2} + \tfrac{1}{m+1}\right)
\, \Gamma\!\left(\tfrac{rm - sm + m + r}{2m+2}\right)
\, \Gamma\!\left(\tfrac{-rm + sm + m - r}{2m+2}\right)
}{
\Gamma\!\left(\tfrac{3}{2} - \tfrac{1}{m+1}\right)
\, \Gamma\!\left(\tfrac{mr + r - ms + 1}{2m+2}\right)
\, \Gamma\!\left(\tfrac{-(m+1)r + ms + 1}{2(m+1)}\right)
}.
\end{aligned}
\end{equation}
Decomposing this correlator in the boundary channel, we obtain global primary operators with dimensions $\hD = 2n$ and $\hD = h_{1,3} + 2n$, as expected from the structure of the Virasoro exchanges. We next consider both the PSU and TSU uplifts of this correlator, and we have verified that they are consistent with the relations derived in \ref{subsection:3pt function relations}. As an illustration, we focus on the PSU uplift obtained by acting with $\hat{D}^p_{[1\bar{2}]}$. This operation renders the bulk operator and one boundary operator fermionic. Consequently, the bulk operator cannot exchange the identity block in the boundary channel, since the two-point function between a fermionic and a bosonic operator vanishes. Also this implies the one point function of a fermionic operator is zero in the bulk as expected. We indeed find that only fermionic exchanges are allowed, with global primaries of dimensions $\hD = 3 + 2n$ and $\hD = h_{1,3} + 2n + 3$.

\section{Discussion}
\label{sec:discussion}
In this work, we initiated the study of the PS uplift and reduction in the presence of extended probes. 
Our analysis gives support to the conjecture that every $d$-dimensional QFT can be embedded into a corresponding $d+2$-dimensional PS QFT. In particular, we demonstrated that this uplift remains consistent even in the presence of  defects of any dimension $p$.

A striking result of our study is the discovery of two distinct yet consistent uplifts for such $p$-dimensional defects: one with dimension $p+2$ and codimension $q\equiv d-p$, and the other with dimension $p$ and codimension $q+2$. We refer to these, respectively, as the parallel space uplift (PSU) and the transverse space uplift (TSU). The uplifted $d+2$-dimensional theory thus contains two extended operators, which reduce to the same original $p$-dimensional defect upon dimensional reduction.
Specifically, we showed that the correlators of local operators in both the PSU and TSU frameworks—when evaluated with insertions restricted to $\mathbb{R}^d$—coincide with those in the original $d$-dimensional theory.

It is instructive to examine the theories that live on the PSU and TSU defects. The theory on the PSU defect is a CFT$_{p+2}$ with global symmetry $\so(q)$. This can itself be viewed as the PS uplift of the original CFT$_p$ with $\so(q)$ global symmetry, which lives on the dimensionally reduced defect.
The situation is markedly different for the TSU. In this case, the defect theory is a CFT$_p$ with global symmetry  $OSp(q+2|2)$. 
This lives in the same $p$-dimensional spacetime as the original defect, but has an enlarged global symmetry.
Nonetheless we argue that these two theories are in fact equivalent once a certain global symmetry reduction is applied. Specifically, we define this reduction by decomposing representations of $OSp(q+2|2)$ into those of $\so(q) \times OSp(2|2)$ and keeping only the singlets under $OSp(2|2)$.
This type of global symmetry reduction has appeared in various contexts in the literature, but the TSU construction offers infinitely many new examples. It would be highly interesting to determine the precise conditions under which a model with $O(q)$ global symmetry can be embedded into one with $OSp(q+2n|2n)$ symmetry. Our analysis suggests that this might always be possible.

We present two independent proofs of dimensional reduction for both the PSU and TSU constructions. The first is non-perturbative: it demonstrates that an infinite number of operators decouples in the lower-dimensional theory as a consequence of PS supersymmetry. This decoupling underpins, for instance, an essentially one-to-one correspondence between higher-dimensional superfields and their lower-dimensional counterparts.
The second proof is perturbative and holds to all loop orders. It shows that each Feynman integral in the higher-dimensional theory reduces to a single Feynman diagram in the lower-dimensional model. 

We also provide strong evidence that both PSU and TSU uplifts can be constructed from the axiomatic CFT perspective. Specifically, we consider an arbitrary two-point function of bulk operators in a CFT$_d$ with a $p$-dimensional defect. Such a correlator is characterized by a function $f(u,v)$ of the conformal cross ratios $u$ and $v$, and it satisfies the bootstrap axioms—i.e., it admits consistent conformal block decompositions in both the bulk and defect channels.
We show that this correlator can be uplifted to a corresponding set of two-point functions in the PSU and TSU frameworks. These uplifted correlators are obtained by acting with specific differential operators $D_{[\sigma]}$ on the original function $f(u,v)$.
Crucially, we prove that the resulting uplifted two-point functions automatically satisfy the conformal block decomposition in both the bulk and defect channels of the higher-dimensional theory. This establishes that, at least for this class of observables, the uplift procedure is always consistent.

As a byproduct of our analysis, we derive a set of relations for conformal blocks in both the bulk and defect channels involving shifts in the labels $p$ and $q$. Specifically, we show that the action of $D_{[\sigma]}$ on a block labeled by $(p, q)$ yields a linear combination of at most four blocks: either with labels $(p+2, q)$ in the PSU case, or $(p, q+2)$ in the TSU case. The coefficients in these linear combinations are computed explicitly in closed form.
Remarkably, the same $D_{[\sigma]}$ acts on both the bulk and defect blocks, despite the fact that these blocks obey distinct Casimir equations and differ significantly in structure—while defect channel blocks are equal to a product of two $_2F_1$ functions, generic bulk channel blocks are considerably more intricate and are not known in closed form.

Another byproduct of our analysis is the identification of a general kinematic property that must be satisfied by all defect setups. The key idea is that, by appropriately exchanging the roles of the transverse and parallel directions, the overall kinematic structure remains essentially invariant.
More precisely, this symmetry can be understood as interchanging the roles of $\so(p+1,1)$ and $\so(q)$ within the preserved symmetry group $\so(p+1,1) \times \so(q)$ of a $p$-dimensional defect in a $d$-dimensional CFT. The explicit form of this map is presented in Appendix~\ref{App:T-Pmap}.
One striking consequence of this symmetry is that the bulk conformal blocks exhibit a nontrivial invariance under the combined transformation of swapping the cross ratios \eqref{def_CR} and exchanging $p+2 \leftrightarrow q$. This generalizes intriguing relations previously observed for $q = 1, 2$ in \cite{Giombi:2020xah} and \cite{Lauria:2017wav}, extending them to arbitrary $q$.
This symmetry also plays a significant role in the uplift constructions, as it effectively exchanges the differential operators associated with the PSU and TSU cases.

We also applied the uplift procedure to bulk--defect--defect three-point functions, which are determined by a single function \( f(w) \) of one cross ratio \(\w\).
We demonstrated that \( f(\w) \) can be uplifted---both in the PSU and TSU frameworks---into a set of functions of \(\w\), obtained by acting with new differential operators on the original function \( f(\w) \). Moreover, if \( f(\w) \) admits a block decomposition in the lower-dimensional theory, the corresponding higher-dimensional functions automatically inherit a block decomposition as well.
This provides yet another consistency check for the uplift procedure and, as before, leads to a novel set of differential relations between the three-point conformal blocks.

Finally, we present several examples to illustrate the PSU and TSU constructions. 
First, we consider the uplift of a trivial defect. We show that some of the uplifted correlators vanish, which imposes constraints on the defect channel decomposition. This leads to recurrence relations for the OPE coefficients that can be explicitly solved. Thus, the uplifts serve as a tool to constrain the dynamical data of the CFT.

Next, we examine a free bulk theory with a defect defined by integrating the free field over \(\mathbb{R}^p\), ensuring that the defect is conformal by construction. A canonical example is the scalar Wilson line in \(d=4\), which we discuss now for simplicity. In this setting, we explicitly define both the PSU and TSU defects. The free field can be uplifted to a superfield \(\Phi = \varphi + \dots + \theta \bar{\theta} \, \omega\), with lowest component \(\varphi\) of scaling dimension \(\Delta = 1\) and highest component \(\omega\) of dimension \(\Delta = 3\). Within the uplifted theory, one can define a conformal defect by integrating \(\varphi\) along a line (corresponding to the TSU) or by integrating \(\omega\) over a three-dimensional surface (corresponding to the PSU).

We then proceed to perturbation theory around this setup. In the original \(d=4\) theory, we introduce a quartic bulk coupling while preserving the same defect. It is known that in \(4 - \epsilon\) dimensions this system exhibits a tractable RG flow: the bulk theory flows to the Wilson–Fisher fixed point, while the defect flows to the so-called magnetic line defect. We illustrate how to express this setup within the PSU and TSU frameworks.

Finally, we consider a fully non-perturbative example given by boundary conditions in minimal models. In this case, the two-point functions in the presence of a boundary are known in closed form. We demonstrate how to uplift these correlators to \(4d\) using the differential operators introduced earlier, and we comment on some notable features of these models.

Together, these examples provide  explicit realizations of the PSU and TSU constructions, complementing the general framework and structural results developed in the rest of the paper. Collectively, this work offers strong nontrivial evidence that the PS uplift is consistent even in the presence of extended probes.

There are many possible outlooks.
It would be very interesting to study the uplift of more observables, e.g. higher point functions and correlators 
of local operators with spin both in  presence and absence of a defect. This would lead to new checks of the uplift and to new sets of relations for the relative conformal blocks. 
Another direction worth investigating is the PS uplift in curved backgrounds such as AdS and dS. Such uplifts may yield novel kinematical relations, as studied in the AdS case for standard Witten diagrams \cite{Zhou:2020ptb}.

It is also interesting to better understand the structure of the uplifted minimal models, as well as the set of their allowed conformal defects. We think that these models could be defined independently from a four-dimensional description, which would be worth investigating. 

A separate question is whether the uplift may offer new handles for the bootstrap. For example, one could bootstrap the four-point function of scalar operators in the super multiplet of a conserved current or a stress tensor in the uplifted theory. This would contain some information of the spinning bootstrap in the original theory ---which is typically quite constraining but very complicated \cite{Dymarsky:2017xzb, Dymarsky:2017yzx, Reehorst:2019pzi, Chang_2025}--- at the price of a much simpler scalar setup. 

Finally, it is natural to ask whether the PSU and TSU frameworks can be realized in physical models. We are currently investigating cases where these uplifts emerge as IR fixed points of random field models with extended operators. Using the RG framework of \cite{Kaviraj:2021qii}, we have indeed found examples where the PSU appears in the IR, which we plan to report on in forthcoming work.


\section*{Acknowledgements}

We would like to thank Apratim Kaviraj, Shota Komatsu, Edoardo Lauria, Marco Meineri, Ritam Sinha, Xinan Zhou for valuable discussions. 
The research of ET was partially supported by the European Research Council (ERC) under the European Union’s Horizon 2020 research and innovation programme (grant agreement number 949077). KG is supported by  Royal Society under grant RF \textbackslash ERE\textbackslash    231142.

\appendix
\section{Reduction of Feynman 
diagrams}\label{app:RedFeyn}
In this appendix, we provide the detailed steps involved in the reduction of Feynman diagrams, as outlined in section \ref{sec:reduction} of the main text. Let us focus on the PSU case and begin with the general expression for a generic Feynman diagram,
\begin{equation} \label{eqn:Feynman_app}
   G(y_1,\dots y_n)= \int [dy] \prod_{I<J}(G^{BB}_{\Phi\Phi}(y_{IJ}))^{q_{IJ}}\prod_{K,i}(G^{BD}_{\Phi\Phi}(y_{Ki}))^{q_{Ki}}\prod_{j<k}(G^{DD}_{\Phi\Phi}(y_{jk}))^{q_{jk}},
\end{equation}
where the notations are defined below \eqref{eqn:Feynman}. 
Using the Schwinger identity we can write it as 
{
\medmuskip=0mu
\thinmuskip=0mu
\thickmuskip=0mu
\begin{equation}
\begin{split}
   & G(y_1,\dots,y_n)= \mathcal{N}_d^{N} A \int \prod_{I<J} \prod_{K,i} \prod_{j<k}  du_{IJ} du_{Ki} du_{jk} (u_{IJ} )^{q_{IJ}\frac{(d-2)}{2}-1}(u_{Ki} )^{q_{Ki}\frac{(d-2)}{2}-1}(u_{jk} )^{q_{jk}\frac{(d-2)}{2}-1}
   \\
    &  \qquad\qquad \int [dy] \exp[-\sum_{I,J}(M_{IJ} X_{I}\cdot X_{J}+2M_{IJ}\bar{\theta}_{I} \theta_{J})]
    \exp[-\!\!\!\!\!\sum_{a,b\in \mathcal{I}_{BD}}\!\!\!\!\!(M_{ab} X_{a}\cdot X_{b}+2M_{ab}\bar{\theta}_{a} \theta_{b})]
    \\
    &  \qquad \qquad \exp[-\sum_{j,k}(M_{jk} X_{j}\cdot X_{k}+2M_{jk}\bar{\theta}_{j} \theta_{k})] \, ,
\end{split}
\end{equation}
}
where the matrix elements are given as follows
\be
     M_{ab}=
     \left\{
\begin{array}{ll}
\sum_c u_{ac},\, & \text{if }a=b\, ,\\
-u_{ab},\, &\text{otherwise,}
\end{array}
\right.
\ee
$a,b$ can either be in the bulk or on the defect and  $\mathcal{I}_{BD}$ is the set of bulk and defect points $\{K,i\}$ that label bulk-defect propagators. 
For the diagonal elements $M_{aa}$, the sum runs over all vertices $c$ such that there is a propagator from $c$ to $a$. $N$ is the total number of propagators appearing in the Feynman diagram, and $A$ is the product of Gamma functions arising from the Schwinger parametrization.
 At this stage now we set $x^{\alpha}=\theta=\bar{\theta}=0$ for external points and separate these components from all the internal propagators,
{
\medmuskip=0mu
\thinmuskip=0mu
\thickmuskip=0mu
\begin{equation}
\begin{split}
  G(x_1,\dots,x_n)= &  \mathcal{N}_d^{N} A \int \prod_{I<J} \prod_{K,i} \prod_{j<k}  du_{IJ} du_{Ki} du_{jk} (u_{IJ} )^{q_{IJ}\frac{(d-2)}{2}-1}(u_{Ki} )^{q_{Ki}\frac{(d-2)}{2}-1}(u_{jk} )^{q_{jk}\frac{(d-2)}{2}-1}  \\
   &\int [dx] \, \exp[-\sum_{I,J}(M_{IJ} 
   x_{I}\cdot x_{J})
   ]
    \,
    \exp[-
    \!\!\!\!\!
    \sum_{a,b\in \mathcal{I}_{BD}}
    \!\!\!\!\!
    (M_{ab} x_{a}\cdot x_{b})]
    \,
    \exp[-\sum_{j,k}(M_{jk} x_{j}\cdot x_{k})]
    \\
    &
    \int \prod_{c\in\mathcal{I}} dX^{\alpha}_{c} [d\bar{\theta}_c] [d\theta_c] \exp[-\sum_{a,b\in \mathcal{I}} (\tilde{M}_{ab}  X_a^{\alpha} X_{b\alpha}+2 \tilde{M}_{ab}  \bar{\theta}_a \theta_b)].
   \end{split}
\end{equation}}
$\tilde{M}_{ab}$ is the matrix we get when all the labels correspond to internal lines.
Here $\mathcal{I}$ belongs to all the internal bulk and defect points. 
Now we can perform the gaussian integrals over these coordinates and using \eqref{eq:2|2integral}  we can see that the bosnic and fermionic fields nicely cancel each other's contribution to produce a factor of 1.
So finally, we arrive at the following expression as we also showed in the main text,
\begin{equation}
    G(x_1,\cdots,x_n)=\int [dx] \prod_{I<J}(G^{BB}_{\phi\phi}(x_{IJ}))^{q_{IJ}}\prod_{K,i}(G^{BD}_{\phi\phi}(x_{Ki}))^{q_{Ki}}\prod_{j<k}(G^{DD}_{\phi\phi}(x_{jk}))^{q_{jk}}.
\end{equation}

This is precisely the Feynman diagram for $p$-dimensional defect. Since in this case both the defect and bulk theories are uplifted to supersymmetric theories, the argument for the reduction is the same as the usual statement: integrating out fermions eliminates two bosonic directions.

\section{
\texorpdfstring{$OSp(m|n)$}{OSp(m|n)}
and 
\texorpdfstring{$O(m-n)$}{O(m-n)}
Wilson-Fisher fixed points}
\label{app:global_sym_uplift}
TSU implies that a CFT$_p$  with global symmetry $OSp(q+2|2)$ should reduce to a CFT$_p$ with global symmetry $\so(q)$ (or  $O(q)$ as explained in footnote \ref{footnote:Z2}). In this appendix, we perform  one-loop computations of  observables in 
the Wilson-Fisher fixed point with $OSp(m|n)$ global symmetry and show that they match with those of the $O(m-n)$ theory.

Let us start by writing the action
\be
\int d^d x \left[ \frac{1}{2}\partial^\mu \Phi^I(x) \partial_\mu \Phi_I(x) + m^2  (\Phi^I(x) \Phi_I(x)) +  \frac{\lambda_B}{4!} (\Phi^I(x) \Phi_I(x))^2 \right] \, ,
\ee
where $\Phi^I(x)$ is a field that lives in $\mathbb{R}^{d}$ and transforms in the vector representation of $OSp(m|n)$, namely it can be written as 
  \be
  \label{Phi_to_phipsi}
\Phi^I(x)=\{\phi_{1}(x),\dots,\phi_{m}(x) ,\psi_1(x),\psib_1(x),\dots, \psi_{{n/2}}(x),\psib_{n/2}(x)\} \, ,
  \ee 
where $\phi_{i}$ (with $i=1,\dots,m$) are usual commuting scalars, while $\psi_j$ and $\psib_j$ (with $j=1,\dots,n/2$)  are Grassmann scalars (symplectic fermions). 
To lower the indices $I$ we use the orthosymplectic metric $g_{IJ}$, so that e.g. $\Phi^I \Phi_I=\phi_i^2-2 \psi_j\psib_j $ (repeated indices are summed over their respective ranges).
The action is invariant under $OSp(m|n)$ global symmetry. We can study it in $\epsilon$-expansion around $4-\epsilon$ dimensions where the quartic term is weakly relevant and triggers a perturbative RG flows. In order to reach a CFT we need to tune the mass parameter (which in usual dim-reg can then be suppressed from the start). 

We can take two different approaches to study this model.
First let us work directly in terms of the variables $\Phi^I$ and define propagators and vertices for them.
In this case it is straightforward to see that everything would just match usual perturbation theory of $O(m-n)$. 
Indeed the propagator $G_{\Phi^I \Phi^J}= G_\phi g^{I J}$, where $G_\phi $ is a free scalar propagator. It is easy to see that all loops diagrams will look exactly like the ones of a usual $O(N)$ vector model, but where now instead of using the Kronecker delta $\delta_{ij}$ (which is the invariant tensor of  $O(N)$), we should appropriately use $g^{IJ}$ (the invariant tensor of $OSp(m|n)$). The result is that all computations match with the ones of $O(N)$ by replacing the value of the trace of $\delta_{ij}$ by the supertrace of $g^{IJ}$. It turns out that the supertrace of $g^{IJ}$ is equal to $m-n$ and therefore this model computes exactly the same observables as the $O(m-n)$ vector model.

While this approach is very general and clearly explains why the two theories are essentially the same, it is also instructive to see what happens when we work in components. Writing the action in components, it takes the following form:
\be
\int d^d x \left[ \frac{1}{2}\partial^\mu \phi_i \partial_\mu \phi_i - \partial^\mu \psi_j \partial_\mu \psib_j
+ m^2  (\phi_i^2-2 \psi_j \psib_j ) + \frac{\lambda_B}{4!} (\phi_i^2-2 \psi_j \psib_j )^2 \right] \, .
\ee 
Let us compute the one-loop beta function for $\lambda_B$  using the OPE method \cite{Cardy_1996}. Let us consider a generic correlator $\langle\cdots\rangle$and compute its correction to second order in bare coupling, 
\begin{equation}
\begin{aligned}
\Big\langle\cdots \Big\rangle=&\Big\langle\cdots \Big\rangle_0-\frac{\lambda_B}{4!} \int d^d x \; \Big\langle \big(\phi_i^2 - 2 \psi_j \bar{\psi}_j \big)^2(x) \, \cdots \Big\rangle_0 \\
& + \frac{1}{2} \left(\frac{\lambda_B}{4!}\right)^2 \int d^d x \, d^d y \; \Big\langle \big(\phi_{i_1}^2 - 2 \psi_{j_1} \bar{\psi}_{j_1} \big)^2(x) \, \big(\phi_{i_2}^2 - 2 \psi_{j_2} \bar{\psi}_{j_2} \big)^2(y) \, \cdots \Big\rangle_0 + O(\lambda_B^3) \, ,
\end{aligned}
\end{equation}
where $\langle \dots \rangle_0$ is the free theory correlator.
To compute how $\lambda_B$ renormalizes, we work in dimensional regularization by setting $d = 4 -\epsilon$ and we focus on extracting the divergent piece as $\epsilon \to 0$. This divergence arises when the separation $x$ approaches $y$ in the bottom line. Therefore, we analyze the OPE limit of the two perturbing operators. Expanding the brackets produces several terms, and below we compute the OPE for each of them. 
Let us start by focusing on the OPEs that generate $(\phi_i^2)^2$. 
First, there is the contribution from the following OPE
\begin{align}
(\phi_{i_1}^2\phi_{i_2}^2)(x)\times (\phi_{i_3}^2 \phi_{i_4}^2)(0)\sim&
 +{ 8} \langle \phi_{i_1}(x)\phi_{i_3}(0)\rangle_0 \langle \phi_{i_1}(x)\phi_{i_3}(0)\rangle_0 
(\phi_{i_2}^2 \phi_{i_4}^2)(0) \nonumber \\
&+ 32 \langle \phi_{i_1}(x)\phi_{i_3}(0)\rangle_0 \langle \phi_{i_2}(x)\phi_{i_3}(0)\rangle_0 
(\phi_{i_2}\phi_{i_1}\phi_{i_4}^2)(0)\nonumber\\
&+ 32 \langle \phi_{i_1}(x)\phi_{i_3}(0)\rangle_0 \langle \phi_{i_2}(x)\phi_{i_4}(0)\rangle_0 
(\phi_{i_1}\phi_{i_2}\phi_{i_3}\phi_{i_4})(0)\nonumber
\\
=&({ 8 m} +64) G_{\phi}(x)^2 (\phi_i^2)^2(0) \, . \label{OPE_Om_WF}
\end{align}
This computation exactly matches the one of the usual $O(m)$ vector model,
where the free $O(m)$ propagator is $G_{\phi_{i_1} \phi_{i_2}}=\delta_{{i_1},{i_2}} G_{\phi}$ and $G_{\phi}$ is the  two-point function of a single free boson. The factor of $m$ appears because of $\delta_{i_1,  i_3}^2$.
In the $OSp(m|n)$ model there is another OPE that generates the operator $(\phi_i^2)^2$ which we need to take into account,
\begin{align}
(-4\phi_{i_1}^2\psi_{j_1} \psib_{j_1})(x)\times (-4 \phi_{i_2}^2 \psi_{j_2} \psib_{j_2})(0)&\sim
 16 \langle \psi_{j_1}(x)\psib_{j_2}(0)\rangle_0 \langle \psib_{j_1}(x) \psi_{j_2}(0)\rangle_0 
(\phi_{i_1}^2 \phi_{i_2}^2)(0) \nonumber\\
&=( -8n ) G_{\phi}(x)^2 (\phi_i^2)^2(0) \, , \label{OPE_Sp_WF}
\end{align}
where the propagator for symplectic fermions reads $G_{\psi_{j_1} \psib_{j_2}}=-G_{\psib_{j_2} \psi_{j_1}}=-\delta_{j_1 j_2} G_\phi $ with  $\delta_{j_1, j_2}^2=\frac{n}{2}$. 
Only the OPEs in \eqref{OPE_Om_WF} and \eqref{OPE_Sp_WF} generate the operator $(\phi_i^2)^2$ and, by summing their contribution, they give $({ 8 m}-8n +64) G_{\phi}(x)^2 (\phi_i^2)^2(0)$, which can then be used to compute the $\beta$ function. Since in the OPE the factor of $m$ is shifted by $-n$ we are bound to get the same result of the $O(m-n)$ vector model. Let us see this in more detail.
We first write the relation between bare and renormalized coupling as $
\lambda_B=\mu^\epsilon Z_\lambda \lambda$, 
where $Z_\lambda=1+\frac{\delta\lambda}{\epsilon}$ to this order.
 By integrating the free theory propagator in $x$ we find, in dimensional regularization, that this gives a singular piece $ \frac{1}{2}\frac{\lambda^2}{(4!)^2}\frac{\Omega_{d-1}}{\e}$, where $\Omega_{d-1}$ is the area of $d$-dimensional surface. This contribution can be cancelled by renormalizing the coupling with the following counter term $ \delta \lambda=\lambda^2\frac{4(m-n+8)\Omega_{d-1}}{ 4! }$.
 Then we have 
 \begin{equation}
    \beta_\lambda=-\e \lambda +\l^2\frac{4(m-n+8)\Omega_{d-1}}{ 4!} =0
\quad     
\implies
\quad
\lambda^*=\frac{6 \epsilon}{(m-n+8)\Omega_{d-1}} \, .
 \end{equation}
As expected, the resulting beta function matches the one of the $O(m-n)$ vector model. To be precise, in this computation we focused only on the term $(\phi_i^2)^2$, while the full quartic interaction contains also $(4 \psi_{j_1} \bar{\psi}_{j_1}\psi_{j_2} \bar{\psi}_{j_2})$ and $(-4\phi_{i}^2\psi_{j}\psi_j)$. 
 Of course our computation was sufficient, since the relative coefficients of the two extra terms are fixed by symmetry. However, for the sake of clarity, let us show that this is indeed the case. 
First, let us analyze the OPEs which  generate the quartic perturbation $(4 \psi_{j_1} \bar{\psi}_{j_1}\psi_{j_2} \bar{\psi}_{j_2})$,
\begin{equation}
\begin{split}
   & (-4 \phi^2_{i_1} \psi_{j_1}\bar{\psi}_{j_1})(x)\times   (-4 \phi^2_{i_2} \psi_{j_2}\bar{\psi}_{j_2})(0) 
    \sim 8m G_{\phi}(x)^2 (4 \psi_{j_1} \bar{\psi}_{j_1}\psi_{j_2} \bar{\psi}_{j_2})(0)\,,\\
    &(4 \psi_{j_1} \bar{\psi}_{j_1}\psi_{j_2} \bar{\psi}_{j_2})(x)\times  (4 \psi_{k_1} \bar{\psi}_{k_1}\psi_{k_2} \bar{\psi}_{k_2})(0) 
  \sim (-8n+64) G_{\phi}(x)^2 (4 \psi_{j_1} \bar{\psi}_{j_1}\psi_{j_2} \bar{\psi}_{j_2})(0) \, .
    \end{split}
\end{equation}
By summing the two contribution we get the factor $(8m-8n+64) G_{\phi}(x)^2$ as before. 
Now, let us  consider the  three OPEs which generate $(-4\phi_{i}^2\psi_{j}\psi_j)$,
\begin{equation}
\begin{split}
    &(-4 \phi_{i_1}^2\psi_{j1}\bar{\psi}_{j_1})(x)\times  (-4 \phi_{i_2}^2\psi_{j_2}\bar{\psi}_{j_2})(0)
     \sim  32 G_{\phi}^2(x) (-4 \phi_{i_1}^2 \psi_{j_1}\bar{\psi}_{j_1})(0)\, ,\\
   & (-4 \phi_{i_1}^2\psi_{j_1}\bar{\psi}_{j_1})(x)\times  (4 \psi_{k_1}\bar{\psi}_{k_1} \psi_{k_2}\bar{\psi}_{k_2})(0)
     \sim(8-4n)G^2_{\phi}(x)(-4\phi_{i_1}^2\psi_{k_1}\bar{\psi}_{k_1})(0)\, ,\\
    &(-4 \phi_{i_1}^2\psi_{j1}\bar{\psi}_{j_1})(x)\times  (\phi_{k_1}^2\phi_{k_2}^2)(0)
 \sim(4m+8) (-4\phi_{i_1}^2\psi_{j_1}\bar{\psi}_{j_1})(0) \, .
    \end{split}
\end{equation}
Note that the OPEs of different operators need to be multiplied by a factor of 2 since we can also switch their places. Summing all contributions up, as expected, we get again the same factor as above. 
 
Next, let us compute the anomalous dimension of some operators. 
We shall start by considering the singlet  
\be
\label{def:Phi2p}
(\Phi^{I}\Phi_{I})^p=  
\sum_{q=0}^p \binom{p}{q}(\phi_i^2)^{p-q} (-2\psi_j\bar{\psi}_j)^{q}
=(\phi_i^2)^p - 2p(\phi_i^2)^{p-1}\psi_j\bar{\psi}_j +\dots 
\, ,
\ee
 which corresponds to $ (\phi_i ^2)^p $ in the $O(m-n)$ model.
 We study a correlator that contains $(\Phi^{I}\Phi_{I})^p$ at first order in the coupling,
\begin{equation}
\!
\langle(\Phi^{I}\Phi_{I})^p(0) \dots \rangle=\langle(\Phi^{I}\Phi_{I})^p(0) \dots \rangle_0  -\frac{\lambda}{4!}  \int d^dx\langle (\Phi^{I_1}\Phi_{I_1})^2(x) \     (\Phi^{I}\Phi_{I})^p(0) \dots \rangle_0 +O(\lambda^2)\, .
\end{equation}
To find the anomalous dimension of $(\Phi^{I}\Phi_{I})^p(0)$ we thus compute the OPE of $(\Phi^{I_1}\Phi_{I_1})^2(x) \times (\Phi^{I}\Phi_{I})^p(0)\sim \# (\Phi^{I}\Phi_{I})^p(0) $. By rewriting the operators using \eqref{def:Phi2p}, the OPE contains various pieces. 
Let us start by the one that only contains bosonic fields,
\begin{equation}
\begin{split}
(\phi_{i_1}^2)^2(x)  \times (\phi_{i}^2)^p(0) 
&=4p(m+2+6(p-1))G^2_{\phi}(x) (\phi_i^2)^p(0).
\end{split}
\label{Om_phi2pOPE}
\end{equation}
This is exactly the same computation of the $O(m)$ vector model, so we expect that the rest of the terms will conspire to shift $m$ to $m-n$. Let us see this. For simplicity we will focus only on the OPEs that generate $(\phi_i^2)^p$ since the other terms of $(\Phi^{I}\Phi_{I})^p$ will be generated with the same coefficient for symmetry reasons, as we explained above.
The only additional 
OPE that generates $(\phi_i^2)^p$ is 
\begin{equation}
\begin{split}
& (-4 \phi_{i_1}^2\psi_{j_1} \bar{\psi}_{j_1})(x)  \times (-2p(\phi_i^2)^{p-1}\psi_j\bar{\psi}_j)(0) 
  =-4pn G^2_{\phi}(x) (\phi_i^2)^p(0).
 \end{split}
\end{equation}
By adding this to \eqref{Om_phi2pOPE}, we notice that the dependence in $m$ is shifted to $m-n$ as expected. 
To compute the anomalous dimension $\gamma_p$ of $(\Phi^{I}\Phi_{I})^p$ we integrate in $x$, which gives a pole in $1/\epsilon$ that must be cancelled by the wavefunction renormalization, $  Z_p=1-\frac{1}{\epsilon}4p(m-n+2+6(p-1))\Omega_{d-1} \frac{ \lambda}{4! }$. The anomalous dimension is then given by,
\begin{equation}
\begin{split}
    \gamma_p&=4p(m-n+2+6(p-1))\Omega_{d-1} \frac{\lambda^*}{4!}=\frac{p (m-n+6 (p-1)+2)}{m-n+8} \, ,
    \end{split}
\end{equation}
which agrees with the standard computation in the $O(m-n)$ model, see e.g. \cite{KleinertPhi4,Henriksson:2022rnm,Rychkov:2015naa}. 

 So far, we have discussed singlet operators, but the same story holds for all representations.
  For instance, let us consider the  bilinear 
  operator $\Phi_{\{I_1}\Phi_{I_2\}}$, where the brackets implement graded simmetrization and tracelessness of the indices.\footnote{Notice that the operator $\Phi_{\{I_1}\Phi_{I_2\}}$ has many more components than its $O(m-n)$ counterpart, however one can branch  $OSp(m|n)\to O(m-n)\times OSp(n|n)$ and take singlet of  $OSp(n|n)$. This prescription generates the correct reduced representation, which in our case is given by $\phi_{i_1}\phi_{i_2}-\frac{\delta_{i_1, i_2}}{m-n}\Phi^{I}\Phi_{I} $ with indices $i_k=1,\dots,m-n$ (notice that $\Phi^{I}\Phi_{I}$ renormalizes as $\phi_i^2$ as we showed above).  } 
Instead of studying the full graded symmetric and traceless representation, we can focus on  the components $I_1=i_1$, $I_2=i_2$ with $i_1\neq i_2$, so that, to compute its anomalous dimension, we only consider the  OPE
$ (\Phi^{I}\Phi_{I})^2\times \phi_{i_1}\phi_{i_2} \sim \# \phi_{i_1}\phi_{i_2} $.
It is easy to see that the only term in $(\Phi^{I}\Phi_{I})^2$ which generates the correct right hand side is $(\phi_{i}^2)^2$, so the computation reduces to
\begin{equation}
(\phi_{i}^2)^2(x) \times (\phi_{i_1}\phi_{i_2})(0)  \sim 8 G^2_{\phi}(x)\,(\phi_{i_1}\phi_{i_2})(0)  \, ,
\end{equation}
which indeed exactly matches the correspondent computation in the $O(m-n)$ model (in this case there is no explicit dependence on $m$ in the OPE, so the shift $m\to m-n$ acts trivially).
By repeating the same logic as above, the anomalous dimension of this operator is  $\frac{2}{m-n+8}$, in agreement with \cite{KleinertPhi4,Henriksson:2022rnm, Dey:2016mcs}.

\section{Parallel 
\texorpdfstring{$\leftrightarrow$}{<->} Transverse map of DCFTs}
\label{App:T-Pmap}
In this appendix we want to point out a map which we call ``parallel $\leftrightarrow$ transverse map'' or $P\leftrightarrow T$,  that works at kinematic level in all DCFT contexts (it does not rely on PS SUSY). We decided to investigate such map motivated by the relation \eqref{mapD_T_P} which maps parallel and transverse differential operators $D^P_{[\sigma]}$ and $D^T_{[\sigma]}$. 
However, the map
$P\leftrightarrow T$ has a considerably broader scope, constraining the kinematics of observables of all defect conformal field theories.

The idea is very simple. A DCFT has symmetry $\so(p+1,1)\times \so(q)$, where $p$ ($q$) is the (co-)dimension of the defect. We can call $\so(p+1,1)$ the parallel symmetry and $\so(q)$ the transverse one. 
We can think of studying a parallel symmetry $\so(p+2)$ instead of  $\so(p+1,1)$ if we allow the compact group $\so(p+2)$ to have representations labelled by Young tableaux with the first row analytically continued to  real values $\hl_0$. The prescription is then that $\hD=-\hl_0$ as shown in \cite{Lauria:2018klo}. If this point of view is taken, then the symmetry group becomes $\so(p+2)\times \so(q)$ and it seems quite obvious that there should be a map of all kinematic observables when we exchange the roles of $\so(p+2)$ and $\so(q)$. How do we realize this in practice? First  we should map $q \leftrightarrow p+2$. Then we should also map everything that was transverse to parallel and vice versa. 
This is particularly simple in embedding space using the conventions of \cite{Lauria:2018klo}. Indeed in embedding space the defect setup can be realized by breaking the embedding metric into a parallel metric $g^P$ of $\mathbb{R}^{p+1,1}$ and a transverse metric $g^T$ of $\mathbb{R}^{q}$. Now another part of the kinematic map is to just substitute $g^P \leftrightarrow g^T$. Finally we should also take into account the operators living on the defect itself. They also transform in representations of $\so(p+1,1)\times \so(q)$ where $\hD,\hl$ are the $\so(p+1,1)$ labels and $s$ are the $\so(q)$ ones. Here $s=(s_1,\dots,s_{[q/2]})$ and $(-\hD,\hl)=(-\hD,l_1,\dots,l_{[p/2]})$   
where we want to consider $-\hD=\hl_0$ in order to have $\so(p+2)$ representations as mentioned before. 
In particular we can consider a defect operator $\hOcal$ inserted at a point $\hat P\in \mathbb{R}^{p+1,1}$ in embedding space, contracted with polarization vectors $\hat Z^{(k)}\in \mathbb{R}^{p+1,1}$ (with $k=1,\dots,[p/2]$) for the $\so(p)$ spin, and $W^{(j)}\in \mathbb{R}^{q}$ for the $\so(q)$ spin (with $j=1,\dots,[q/2]$). 
Now the prescription is that we should exchange the sets $(\hat P, \hat Z^{(k)}) \leftrightarrow (W^{(j)})$ together  with $(-\hD,\hl) \leftrightarrow s$. These prescriptions altogether take the form
{
\be
\label{P-T_map}
\boxedB{
\;
 \begin{array}{c}
\phantom{\Big|} p+2    \phantom{\Big|} \\ 
 \phantom{\Big|}   g^P \phantom{\Big|}\\  
  \phantom{\Big|}  -\hD,\hl_1,\dots,\hl_{[p/2]} \phantom{\Big|} \\
\phantom{\Big|} \hat P,\hat Z^{(1)},\dots,\hat Z^{([p/2])}
\phantom{\Big|}
\end{array}
{\xleftrightarrow{\; P\leftrightarrow T \;}}
\begin{array}{c}
\phantom{\Big|} q  \phantom{\Big|}   \\
 \phantom{\Big|}  g^T \phantom{\Big|}\\   
\phantom{\Big|}   s_1,\dots,s_{[q/2]} \phantom{\Big|}\\
 \phantom{\Big|}    W^{(1)},\dots,   W^{([q/2])} \phantom{\Big|}
\end{array}
\;
}
\ee
}
which defines the $P\leftrightarrow T$ map, that exchanges the roles of the parallel and transverse spaces.
To be precise we should consider observables
where the number of defect operators parallel quantum numbers matches the orthogonal ones e.g. a defect operator with dimensions $\hD$ and zero parallel spin $\hl=0$ should be considered with generic value of $s_1$. Accordingly any given defect operator will be contracted to parallel polarization vectors $\hat Z^{(1)}, \dots \hat Z^{(n)}$ and to orthogonal polarization vectors $W^{(1)}, \dots W^{(n+1)}$ for some given value of $n$.

Let us give examples of the $P\leftrightarrow T$ map. 
One of the simplest examples is the  two-point function  of a scalar bulk operator  $\Ocal$ of dimension $\Delta$ and $\hOcal$ with a scalar defect operator of dimension $\hD$ and transverse spin $s$. This two-point function is fixed in terms of a single tensor structure,
 \be
 \langle \Ocal(P) \hOcal(\hat P, W) \rangle = b_{\Ocal \hOcal} \frac{1}{(P\cdot g^T \cdot P )^{\frac{\Delta}{2}}} \left(\frac{P\cdot g^T \cdot W }{P\cdot g^T \cdot P}\right)^{s} \left(\frac{P\cdot g^P \cdot \hat P }{P\cdot g^P \cdot P}\right)^{-\hD} \, ,
 \ee
 which maps to itself under \eqref{P-T_map}. To be precise it maps to itself up to an overall phase due to the fact that $P\cdot g^T \cdot P$ maps to $P\cdot g^P \cdot P=-P\cdot g^T \cdot P$, but this is just a choice of normalization of the structure which could be removed by rescaling the coefficient $b_{\Ocal \hOcal}$. Here we also see that is was crucial to have a generic value for $s$ to perform the map $s \leftrightarrow -\hD$. Another very simple example is the two-point function of the same defect operator $\hOcal$ defined above. In this case
 $\langle \hOcal(\hat P_1, W_1) \hOcal(\hat P_2, W_2) \rangle= (\hat P_1 \cdot g^P \cdot \hat P_2 )^{-\hD} (W_1 \cdot g^T \cdot W_2 )^{s}$ which again trivially maps to itself under \eqref{P-T_map}. 
 
Let us now turn to the more interesting case of a two-point function of bulk operators where the kinematical information is contained in the conformal blocks.
First we notice that the cross ratios in \eqref{def_CR} can be written in embedding space as
\be
\up=\frac{P_1 \cdot g^P \cdot P_2}{(P_1 \cdot g^P \cdot P_1)^{\frac{1}{2}}(P_2 \cdot g^P \cdot P_2)^{\frac{1}{2}} }
\, ,
\qquad 
\ut=\frac{P_1 \cdot g^T \cdot P_2}{(P_1 \cdot g^T \cdot P_1)^{\frac{1}{2}}(P_2\cdot  g^T \cdot P_2)^{\frac{1}{2}} } \, ,
\ee
where the points $P_i$ are the embedding space insertions of the operators $\Ocal_i$ in the two-point function. The minus sign in the first definition arises by  using $P_i \cdot g^P \cdot P_i=- P_i \cdot g^T \cdot P_i$.
Therefore under the $P \leftrightarrow T$ map these cross ratios map as $\up \leftrightarrow \ut$.
Let us now apply this to conformal blocks. The simplest case is the one of defect blocks which are defined in \eqref{defect_blocks}.
The blocks are manifestly invariant under the $P \leftrightarrow T$ map which here simply acts by $ s,q,\ut  \leftrightarrow -\hD,p+2,\up $. 
This factorization was already noted in \cite{Billo:2016cpy} and was used in 
 section 4.2 of \cite{Lauria:2018klo} to determine more generic defect blocks for two-point functions of spinning operators. 

A very interesting example of the map appears in the case of bulk channel conformal blocks, 
 which is far less trivial since these blocks are not known in a closed form. 
 To justify the \( P \leftrightarrow T \) map, we need to consider the Casimir differential equation that defines the bulk channel blocks, given by 
$\mathcal{C} \, g_{\Delta \ell}^{(p,q)} = 0$,
where the differential operator $\Ccal$ is defined as  
\be
\begin{array}{ll}
\Ccal\equiv &
2 \fchiu  \left(c_{\Delta ,l}-\Delta _1^2+\Delta _1 \left(-2 \Delta _2 \ut  \up +p+q\right)+\Delta _2 \left(-\Delta _2+p+q\right)\right)
\\
&+4 \left(-\left(\Delta _1+\Delta _2\right)  \left(\up ^2-1\right) \ut  -\ut  \up ^2+\up- (-\up+\ut )p \right) \partial _{\up }  \fchiu 
\\
&+4 \left(-\left(\Delta _1+\Delta _2\right) \left(\ut ^2-1\right) \up -\up  \ut ^2 +\ut+(-\up+\ut)(q-2) \right) \partial _{\ut }  \fchiu 
\\
&-8 \left(\ut ^2-1\right) \left(\up ^2-1\right) \partial _{\up }  \partial _{\ut }  \fchiu 
+4 \left(\ut ^2-1\right) (-\ut  \up +1) \partial _{\ut }^2  \fchiu 
+4 \left(\up ^2-1\right) (-\ut  \up +1) \partial _{\up }^2  \fchiu \, .
\end{array}
\ee
The Caisimir eigenvalue $c_{\Delta ,\ell}=\Delta  (\Delta -d)+\ell (d+\ell-2)$ is included in $\Ccal$ for convenience.
 It is easy to see that the differential equation  
 is invariant under $P\leftrightarrow T$, namely  \be
\Ccal|_{(\ut,q)\leftrightarrow(\up,p+2)} = \Ccal \, ,
\ee
which in particular means that there exists a normalization of the blocks such that 
$g^{(p,q)}_{\Delta,\ell}(\up,\ut)=g^{(q-2,p+2)}_{\Delta,\ell}(\ut,\up) $.
Since the Casimir is a homogeneous equation for the blocks, they might have an  overall normalization dependent on $p$ and $q$ which  transforms non-trivially
under this map.
 Indeed in our normalization the bulk conformal blocks satisfy the following property 
\be
\label{block_rel_PT}
g^{(p,q)}_{\Delta,\ell}(\up,\ut)=(-1)^{\frac{\Delta_1+\Delta_2+\ell}{2}}g^{(q-2,p+2)}_{\Delta,\ell}(\ut,\up) \, .
\ee
Nevertheless we can conclude that the blocks are invariant under  $(\ut,q)\leftrightarrow(\up,p+2)$ up to an overall constant which only depends on the normalization.

Since the map switches $(p,q)\leftrightarrow (q-2,p+2)$, an interesting case appears for $q=p+2$, when these labels are fixed by $P \leftrightarrow T$ and thus the map gives a constraint on the form of bulk conformal blocks.
Other interesting cases appear for boundaries, namely $q=1$, which are mapped to configurations with $p=-1$,  as noticed in \cite{Giombi:2020xah}. 
The case of $q=2$ maps to $p=0$, where the defect corresponds to two-points. This was used in \cite{Lauria:2017wav} to find a relation between the bulk  channel block and a standard four-point  block (in the absence of a defect). The case $p=-2$  to our knowledge never appeared in the literature but from the $P \leftrightarrow T$ should be encoded into a $q=0$ defect setup which is typically understood as absence of the defect.

Let us now turn to the  differential operators  $D^P_{[\s]}$ and $D^T_{[\s]}$ which motivated in first place this analysis. 
We notice that the map \eqref{mapD_T_P} that relates $D^P_{[\s]} $ to $D^T_{[\s]}$ is 
$(\ut,q)\leftrightarrow(\up,p+2)$, which exactly corresponds to the $P\leftrightarrow T$ map. 
The existence the PSU differential operator is then implied by the existence of the TSU ones and vice versa,  as consequence of this map. 
In particular if $D^P_{[\s]}$  acts on the two-point function defining a new two-point function in a new theory where $p$ is shifted by two units, it makes sense that under the $P\leftrightarrow T$ map there should exist a new differential operator that shifts $q$ by two units. Indeed we can argue that the $P\leftrightarrow T$ map is mapping at kinematic level  PSU $\leftrightarrow$ TSU directly from \eqref{P-T_map} applied to the uplifted theory. Indeed the differential operators only arise by expanding the Grassmann  dependence of the insertion points. 
In the PSU we have $\hat P,\hat Z^{(k)}$ which contain Grassmann variables while $W^{(j)}$ is bosonic and vice versa for the TSU. 
Now the map \eqref{P-T_map} effectively exchanges the place where the Grassmann variables appear from the parallel space to the transverse space and vice versa. This exactly exchanges the role of PSU and TSU.

Finally let us comment that one could also study more general observables, e.g. the bulk-defect-defect three-point function under this map. However to do so one should consider defect operators $\hOcal_2$, $\hOcal_3$ with transverse spin $s_2$ and $s_3$, since they map to the dimension $-\hD_2$ and $-\hD_3$. This could be done but it is not the in the scope of the present paper where we focused on the simplest example $s_2,s_3=0$. For this reason we cannot provide here a simple map of the whole set of PSU and TSU differential operators in \eqref{eqn:D_bdd} and \eqref{D_bdd_TSU}. 
\section{
More identities for conformal blocks and special functions 
} \label{appendix:3F2}
The goal of this appendix is twofold. First, we want to discuss extra relations for conformal blocks in the two cases of bulk-bulk two-point functions for BCFT and bulk-bulk-defect three-point functions. 
Secondly, in both such setups, we  want to discuss how the relations for conformal blocks found from the uplift might be interesting even from a mathematical standpoint, to obtain possibly new identities for special functions.
\subsection{BCFT and information loss
}
In the main text we explained that any bulk-bulk two-point function $f(u,v)$ can  always be uplifted.
It is possible to generate the uplifted correlators using the differential operators $D^{T/P}_{[\s]}$ of \eqref{def:fks} and the resulting functions can be expanded in block thanks to the relations of  section \ref{subsection:blockrelation}.  
In this appendix, we shall give more details about the case of BCFTs,  discussed in section \ref{Subsection:bcft 2D}, where there is only one independent cross ratio, since $v=1$. 
Because of 
 $v=1$, the uplift is more subtle 
 as we will explain in the following. 
 Nevertheless, we will show that the block relations of section \ref{subsection:blockrelation} can still be used
 and give rise to interesting identities for hypergeometric functions.

Let us think of a BCFT bulk-bulk two point function as a function $f(u,v)$ in the collinear limit $v=1$. To uplift it, we use again the differential operators of \eqref{def:fks}. 
 The uplift of the lowest component $[\cdot]$  is trivial to achieve (for both PSU and TSU) since $D^{T/P}_{[\cdot]}=1$; moreover the fact that it can be decomposed in blocks descends automatically from the relations of section  \ref{subsection:blockrelation} by setting the spin  to zero and $\ut=1$. 
The case of higher components is much less straightforward, since the differential operators  typically contain derivatives in direction $\ut$, which are not defined in the BCFT (the BCFT is defined strictly at $\ut=1$).
To exemplify this phenomenon, let us consider the action of $D^{T/P}_{[\bf{1}]}$ on a function $f(\up ,\ut)$ when we set $\ut=1$,
\begin{equation}
\begin{aligned}
D^{P}_{[\mathbf{1}]} f(\up,\ut)|_{v=1} =\; & \up \big(2 \Delta_1 - q + 3\big) 
\partial_u f(\up,1) 
+ (1 - q) 
\partial_v f(\up,1) 
\\&
+ (\up^2 - 1) 
\partial_u^2 f(\up,1) 
+ \Delta_1 (\Delta_1 - q + 2) f(\up,1)\,,
\\[10pt]
D^{T}_{[\mathbf{1}]} f(\up,\ut)|_{v=1} =\; & \big(-2 \Delta_1 + p - 1\big) 
\partial_v f(\up,1) 
+ (p + 1) u 
\partial_u f(\up,1) 
\\
& + (\up^2 - 1) 
\partial_u^2 f(\up,1) 
+ \Delta_1 (p - \Delta_1) f(\up,1)\,.
\end{aligned}
\end{equation}
For the PSU, the coefficient in front of $ \partial_v f(\up ,1)$ vanishes for $q=1$, thus   
we can again reconstruct the uplifted BCFT correlator and similarly we can prove its conformal block decomposition using the relations of section \ref{subsection:blockrelation} setting $q=1$ and spin to zero. For the TSU it is not the case. 
 The derivative in $v$ does not cancel and there is no way to reconstruct the uplifted correlator. 
 This is actually related to the fact which was dubbed as ``loss of information'' in \cite{Trevisani:2024djr}. Since the
$q=3$ two-point function has 
more cross ratios in $q=3$, it cannot be fully reconstructed from the BCFT  correlator. 
Only the lowest component can be constructed and uniquely in the limit of  $\ut=1$. 
This is already non trivial, since it can be used to compute all the uplifted scalar exchanges exactly, however all other exchanged sectors remain undetermined.
It would be interesting to see how much of the information about other spinning sectors can be reconstructed by considering the crossing of two-point function. Since introducing a single exchange in one channel requires infinite exchanges in other channel (as follows  from the lightcone limit), other sectors should also be constrained.

Although the correlators of higher components cannot be fully reconstructed in the TSU, we can still employ the formulae of section \ref{subsection:blockrelation} to derive  relations between BCFT blocks and $q=3$ blocks evaluated at $v=1$. 
We shall exemplify this for the bulk channel, where these relations take a particularly interesting form.
The bulk channel block for a $q$ codimensional defect is known in a closed form in the collinear limit $\ut=1$,
\be
\label{eq:bulk_block_q_v1}
g^{(d-1,q)}_{\Delta,0}(u,1)=2^{\Delta } (1+\up)^{\Delta /2}{}_3F_2\bigg[\frac{\Delta }{2}-\frac{q}{2}+1,\frac{\Delta }{2}+\frac{\Delta _{12}}{2},\frac{\Delta }{2}-\frac{\Delta _{12}}{2};\frac{\Delta }{2}+\frac{1}{2},-\frac{d}{2}+\Delta +1;-1-\up\bigg] \, .
\ee
This reduces to the bulk channel block of BCFT when $q$ is set to $1$ and it simplifies to a ${}_2F_1$,
\be 
\label{eq:bulk_block_bdy}
g^{(d-1,1)}_{\Delta,0}(u,1)=2^{\Delta } (1+\up)^{\Delta /2}{}_2F_1\bigg[\frac{1}{2} \left(\Delta +\Delta _{12}\right),\frac{1}{2} \left(\Delta -\Delta _{12}\right);-\frac{d}{2}+\Delta +1;-1-\up\bigg]\, .
\ee
Let us consider the relation \eqref{bulk_rel} for the component $[{\bf 1}]$ in TSU. As explained above, this contains a derivative in $v$ of the collinear blocks, which in principle we cannot compute. Nonetheless, we realize that the component $[1 \bar{2}]$ can be used to express any derivative in $v$ of the collinear blocks for $q=1$ in terms of a sum of of collinear blocks for $q=3$, namely
\be
-\partial_v g^{(d-1,1)}_{\Delta,0}(u,1)=c^T_{{[1\bar{2}]},0,0} g^{(d-1,3)}_{\Delta,0}(u,1)+c^T_{{[1\bar{2}]},2,0} g^{(d-1,3)}_{\Delta+2,0}(u,1) \, .
\ee
This equation can be substituted in all differential operators to generate identities between the collinear blocks.
Explicitly, for   $D^T_{[{\bf 1}]}$ we obtain the following identity,
{
\medmuskip=0mu
\thinmuskip=0mu
\thickmuskip=0mu
\begin{align}
   & \!\!\!  (p + 1) u \partial_{\up}g^{(d-1,1)}_{\Delta,0} + (u^2 - 1)  \partial^2_\up g^{(d-1,1)}_{\Delta,0}+ \Delta_1 (p - \Delta_1) g^{(d-1,1)}_{\Delta,0} =\\
&
\qquad 
=c^T_{{[\bf{1}]},0,0}\Sigma_{[\bf{1}]} g^{(d-1,3)}_{\Delta,0}+c^T_{{[\bf{1}]},2,0}\Sigma_{[\bf{1}]} g^{(d-1,3)}_{\Delta+2,0}
+\big(p - 1-2 \Delta_1 \big)\big(c^T_{{[1\bar{2}]},0,0} g^{(d-1,3)}_{\Delta,0}+c^T_{{[1\bar{2}]},2,0} g^{(d-1,3)}_{\Delta+2,0}\big) \, ,
\nonumber
\end{align}
}
where we suppressed the dependence on the collinear cross ratios $(u,v=1)$.
Similarly, the other derivatives also yield relations of this type between boundary and codimension-3 collinear blocks, which in turn imply relations between ${}_3F_2$ and ${}_2F_1$.

Now, let us generalize this type of relations  for any value of  $q$, in the collinear limit. The relation between $q$ and $q+2$ dimensional bulk blocks when cross ratio $v=1$ is given by,
{
\begin{align} 
&\!\!\!\!\!\!\!-\frac{\left(\Delta +\Delta _1-\Delta _2\right) \left(\Delta -\Delta _1+\Delta _2\right) z  (d-\Delta -q) \, {}_3F_2(a_1+1,a_2+1,a_3+1,b_1+1,b_2+1;z)}{ (\Delta +1) (d-2 \Delta -2) (d-2 \Delta )}\nonumber\\
& 
\qquad\qquad 
+\, {}_3F_2(a_1,a_2,a_3;b_1,b_2;z) -\,{}_3F_2(a_1+1,a_2,a_3;b_1+1,b_2;z)=0 \, ,
\end{align}}
where we have set $\up=1-2z$ to compare it with the known result and 
\begin{align}
& a_1=\frac{\Delta}{2}-\frac{q}{2},\, \quad
a_2=\frac{\Delta+\Delta_{12}}{2}, \,\quad
a_3=\frac{\Delta-\Delta_{12}}{2}, \,\quad
b_1=\Delta-\frac{d}{2},\,\quad
b_2=\frac{\Delta+1}{2}.
\end{align}
This relation is precisely the known three term recursion between ${}_3F_2$ hypergeometric function as documented in \href{https://functions.wolfram.com/HypergeometricFunctions/Hypergeometric3F2/17/02/01/0010/}{Wolfram website}. When we set $q=-1$ and $d=-6$, we can reproduce Lavoie's equation \cite{milgram2006hypergeometric}.
For $q = 1$, the third hypergeometric function reduces to a ${}_2F_1$, reproducing the relation discussed in the main text. Considering PSU uplift we find,
\begin{align}
   &-\frac{ \left(\Delta +\Delta _1-\Delta _2\right) \left(\Delta -\Delta _1+\Delta _2\right) z (-\Delta +q-2) {}_3F_2(2+a_1,1+a_2,1+a_3;b_1+2,b_2+1;z)}{(\Delta +1) (d-2 \Delta -2) (d-2 \Delta )} \nonumber \\
   &-{}_3F_2(1+a_1,a_2,a_3;b_1,b_2;z)+{}_3F_2(1+a_1,a_2,a_3;b_1+1,b_2;z)=0 \, .
\end{align}
By acting with differential operators, one can generate special cases of other known finite-term identities among these hypergeometric functions.

\subsection{Bulk–bulk–defect correlators and PS uplift: further relations}
Finally, let us consider the bulk--bulk--defect correlator. This setup allows us to derive relations for  Appell functions and ${}_4F_3$ hypergeometric functions, which are not as well studied in the literature as the ${}_3F_2$ or ${}_2F_1$. Consequently, the relations we obtain here could represent genuinely new results. For a defect of codimension $q>1$, the bulk--bulk--defect three-point function depends on three cross ratios, which reduce to two in the BCFT case since the angle between the transverse coordinates trivializes. The three cross ratios are given by \cite{Lauria:2020emq,Buric:2020zea}
\begin{equation}
    \xi_1=\frac{(\xpar_{12})^2+(\xperp_{1})^2+(\xperp_{2})^2}{2|\xperp_1||\xperp_2|}\,,\qquad 
    \xi_2=\frac{\big((\xpar_{13})^2+(\xperp_1)^2\big)|\xperp_2|}{\big((\xpar_{23})^2+(\xperp_2)^2\big)|\xperp_1|}\,, \qquad 
    \xi_3=\frac{\xperp_1\cdot \xperp_2}{|\xperp_1||\xperp_2|}\,.
\end{equation}
It is also useful to introduce an alternative set of cross ratios 
\begin{equation}
    \vt_2=\xi_2^2 \vt_1\,,\qquad \vt_1=\frac{1}{\xi _2^2-2\xi _1 \xi _2+1}\, .
\end{equation}
The correlator can then be expressed in terms of these cross ratios as
\begin{equation}
\label{eq:BBD}
    \langle O_{\Delta_1}(x_1)O_{\Delta_2}(x_2)\hat{O}_{\hat{\Delta}_3}(\xpar_3)\rangle
    =\frac{\big((\xpar_{23})^2+(\xperp_2)^2\big)^{-\hat{\Delta}_3}}{|\xperp_1|^{\Delta_1}\,|\xperp_2|^{\Delta_2-\hat{\Delta}_3}}\, 
    F(v_1,v_2,\xi_3)\, .
\end{equation}
In the following we focus on the block expansion of this correlator, when both $\Ocal_{\Delta_1}$ and $\Ocal_{\Delta_2}$ are taken close to the defect,
\be
F(\vt_1,\vt_2,\xi_3)=\sum_{\hOcal,\hOcal'}
\lambda_{\hOcal,\hOcal'\!,\hat{\Delta}_3}
\,
b_{\Delta_1, \hOcal}
\,
b_{\Delta_2, \hOcal'}
\,
F^{(p,q)}_{\hat{\Delta},\hat{\Delta}'\!,s}(\vt_1,\vt_2,\xi_3) \, ,
\ee
where the exchanged operators $\hOcal, \hOcal'$ have respectively dimensions $\hat{\Delta}$ and $\hat{\Delta}'$ and possess the same transverse spin $s$. The function $F^{(p,q)}_{\hat{\Delta},\hat{\Delta}'\!, s}$ is the defect channel bulk-bulk-defect conformal block which we will derive below. 

Before proceeding, let us note how this setup reduces to the familiar cases discussed in the main text. When the third operator is the identity, we recover the bulk--bulk two-point function with cross ratios $\up=\xi_1$ and $\ut=\xi_3$. On the other hand, if the bulk operator at $x_2$ is pushed to the boundary, we obtain the bulk--defect--defect three-point function with cross ratio $\z=\xi_1^{-1}\xi_2^{-1}$. Thus, the relations derived here consistently reduce to those in the main text in the appropriate limits. 

We now turn to the derivation of conformal blocks in the defect channel using the uplift construction. Even though the block is already known in terms of Appell functions, we will rederive it using the uplift framework. In particular, one can view TSU as an alternative means of deriving the Casimir equation for $F^{(p,q)}_{\hat{\Delta},\hat{\Delta}'\!, s}$. 
While this approach does not necessarily offer an advantage over the Casimir method, it illustrates a useful application and lends further support to the uplift conjecture. 
First, by exploiting the $SO(q-1)$ invariance of the problem, we can immediately see that the dependence on $p$ and $q$ factorizes. Moreover, when the third operator on the defect is the identity, we recover the bulk two-point function, which implies that the $q$-dependent part is identical to the defect-channel block of the bulk two-point function. Accordingly, we can write the defect-channel block  as 
\be
F^{(p,q)}_{\hat{\Delta},\hat{\Delta}'\!,s}(\vt_1,\vt_2,\xi_3)=
\vt_1^{\frac{\hat\Delta}{2}}\vt_2^{\frac{\hat{\Delta}'-\hat\De_3}{2}}  F^p_{\hat{\Delta},\hat{\Delta}'}(\vt_1,\vt_2)\,C^{\frac{q-2}{2}}_s(\xi_3) \, ,
\ee
where $C_s^q(\xi_3)$ is the Gegenbauer polynomial and we extracted a power of $\vt_1,\vt_2$ for convenience.

Now the idea is to uplift \eqref{eq:BBD} to superspace (which as usual amounts to replacing $x \to y$) and use the differential operators $\bD_{[\sigma]}$ of \eqref{bDS_def} to extract all primary components. Each component can then be written as a function of the three cross ratios,
\begin{equation}
\label{def:flocal1}
F^{k}_{[\sigma]}(\vt_1,\vt_2,\xi_3) \equiv 
 \mathfrak{D}^{k}_{[\s]} F(\vt_1,\vt_2,\xi_3)\,, \qquad k=P,T\, ,
\end{equation}
which can be computed by acting on  the original function $F(\vt_1,\vt_2,\xi_3)$ with a differential operator $\mathfrak{D}^{k}_{[\s]}$ that acts on the three cross ratios. 
Using the same logic as  in the main text, we find relations between defect blocks which take the generic form
\begin{equation}
\label{eq:DF=sumF_BBD}
\mathfrak{D}^{k}_{[\sigma]}F^{(p,q)}_{\hat\Delta,\hat\De'\!,s}(\vt_1,\vt_2,\xi_3)
=\sum_{
i,j,k
}\mathfrak{r}^k_{[\sigma],i,\,j,\,k}\,\Sigma_{[\sigma]}F^{(p,q)^k}_{\hat\Delta+i,\hat\De'+j,s+k}(\vt_1,\vt_2,\xi_3)\, ,
\end{equation}
where the sum over $i,j,k$ always contains finite number of terms, as exemplified below.

Now let us show how, using the TSU,  we obtain an homogeneous  differential equation for $F^p_{\hat{\Delta},\hat{\Delta}'}(\vt_1,\vt_2)$. 
 In the  TSU only $q$ is shifted, so the label $p$ stays the same in \eqref{eq:DF=sumF_BBD}. In principle the operators $\Sigma_{[\sigma]}$ should shift the labels of the blocks, but since the bulk-defect OPE depends on those labels only through the overall scaling in $x$, it is easy to argue that the blocks do not depend on $\Delta_1, \Delta_2$, thus these shifts act trivially  (notice that $[\sigma]$ for the TSU cannot contain the label $3$ because it lives on the defect). 
With this in mind, we consider equation \eqref{eq:DF=sumF_BBD} for the component $[\s]=[{\bf 1}]$ and $k=T$, which can be recasted as the following  differential equation    
\be
\nabla_1 F^{p}_{\hat{\De},\hat{\De}'}(\vt_1,\vt_2) = 0 \, ,
\ee
for $F^{p}_{\hat{\De},\hat{\De}'}$, where $\nabla_1$ takes the form\footnote{The actual differential operator is
$\mathfrak{D}^{T}_{[{\bf 1}]}=4\vt_1^2(1-\vt_1)\partial^2_{\vt_1}-4\vt_1 \vt_2^2\partial^2_{\vt_2}+2(2+p(\vt_1-1)-4\vt_1)\vt_1\partial_{\vt_1}+2(p-4)\vt_1\vt_2\partial_{\vt_2}-8\vt_1^2\vt_2\partial_{\vt_1}\partial_{\vt_2}+\xi_3(p-1-2\Delta_1)\partial_{\xi_3}-(\xi_3^2-1)\partial^2_{\xi_3}+(p-\Delta_1)\Delta_1.$
The $\xi_3$-dependent part acts on Gegenbauer polynomials, whose properties fix the form of the coefficients as follows,
\be
\mathfrak{r}^T_{[{\bf 1}],0,0,0}=-\tfrac{(q-2) \left(-\hat{\Delta }+\Delta _1+s\right) \left(\hat{\Delta }+\Delta _1-p+s\right)}{q+2 s-2} \,,
\qquad 
\mathfrak{r}^T_{[{\bf 1}],0,0,-2}=\tfrac{(q-2) \left(\hat{\Delta }-\Delta _1+q+s-2\right) \left(-\hat{\Delta }-\Delta _1+p+q+s-2\right)}{q+2 s-2} \, .
\ee
 After factoring this piece, we obtain an equation for 
$F^{p}_{\hat{\Delta},\hat{\Delta}'}(\vt_1,\vt_2)$.}
{
\medmuskip=0mu
\thinmuskip=0mu
\thickmuskip=0mu
\begin{equation}
\nabla_1 \equiv \vt_1 (1-\vt_1)\partial_{\vt_1}^2 
- \vt_2^2 \partial^2_{\vt_2}  
- 2 \vt_1 \vt_2  \partial_{\vt_1} \partial_{\vt_2} 
+ (c_1 - (a+b+1) \vt_1) \partial_{\vt_1}
- (a+b+1) \vt_2 \partial_{\vt_2} 
- a b  \, ,
\end{equation}
}
where the coefficients  $a,b,c_2$ are defined as follows 
\begin{equation}
    a\equiv \frac{\hat{\Delta}+\hat{\De}'-\hat{\De}_3}{2}\,,\qquad 
    b\equiv\frac{\hat{\Delta}+\hat{\De}'-\hat{\De}_3+2-p}{2}\,,\qquad 
     c_1\equiv\hat{\Delta}-\frac{p}{2}+1
    \,.
\end{equation}
Similarly, considering  the component $[{\bf 2}]$ of TSU in \eqref{eq:DF=sumF_BBD},  yields the equation
\begin{equation}
\nabla_2 F^{p}_{\hat{\De},\hat{\De}'}(\vt_1,\vt_2)=0
     \, ,
\end{equation}
where $\nabla_2  \equiv\nabla_1|_{c_1\rightarrow c_2,\,v_1\leftrightarrow v_2}$ and $ c_2\equiv \hat{\Delta}'-\frac{p}{2}+1$.
Functions of two variables which solve these differential equations, were extensively studied by Paul Appell and are known as Appell functions $F_4$ \cite{DLMF},
\begin{equation}
F^{p}_{\hat{\De},\hat{\De}'}(\vt_1,\vt_2)=F_4(a,b,c_1,c_2,v_1,v_2).
\end{equation}

Let us now turn to the PSU, which can be used to extract non trivial identities for the Appell function. 
First, by considering the lowest component $[\cdot]$ in \eqref{eq:DF=sumF_BBD}  we find 
\begin{equation}
\label{id_appell_known}
\begin{aligned}
&-F_4(a; b-1; c_1-1, c_2-1; \vt_1, \vt_2) + F_4(a; b; c_1, c_2; \vt_1, \vt_2) \\
&\quad + \frac{a}{(c_1-1)c_1 (c_2-1)c_2} \Bigg[ 
(c_2-1)c_2\, \vt_1 (b-c_1)\, F_4(a+1; b; c_1+1, c_2-1; \vt_1, \vt_2) \\
&\qquad - (a+1) b\, \vt_1 \vt_2 (b-c_1-c_2+1)\, F_4(a+2; b+1; c_1+1, c_2+1; \vt_1, \vt_2) \\
&\qquad + (c_1-1)c_1\, \vt_2 (b-c_2)\, F_4(a+1; b; c_1-1, c_2+1; \vt_1, \vt_2) 
\Bigg] = 0 \, .
\end{aligned}
\end{equation}
Although we did not find \eqref{id_appell_known} explicitly stated in the literature, the relations obtained by Wang~\cite{Wang2012} can be used to derive it. 
Let us now consider the component $[1\bar{3}]$ for the PSU. By massaging this relation we reach  the following  identity
\begin{align}
\begin{split}
&F_4(a,b;1+c_1,c_2;v_1,v_2)  \\
&= \frac{c_1}{(a-1)c_2(c_1+c_2-2b)v_1} \Bigg[
   (b-c_2)c_2\,F_4(a-1,b-1;c_1,c_2;v_1,v_2) \\
&\quad+ c_2(a-b+c_2-1)\,F_4(a-1,b;c_1,c_2;v_1,v_2)  \\
&\quad+ (a-1)\Big(c_2(1+v_1-v_2)\,F_4(a,b;c_1,c_2;v_1,v_2) 
       + 2b v_2\,F_4(a,b+1;c_1,1+c_2;v_1,v_2)\Big)
   \Bigg].
   \label{id_appell_new}
   \end{split}
\end{align}
As discussed in \cite{Kimurathesis}, when either $c_1$ or $c_2$ is shifted by a positive integer, the usual contiguous relations do not follow. 
The identity \eqref{id_appell_new}, to our knowledge, might therefore be genuinely new. 

We can easily generate the differential operators appearing in \eqref{eq:DF=sumF_BBD} in both the PSU and TSU frameworks, which allow us to write down such finite-term relations. A few of them are given explicitly below:
\begin{equation}
\begin{aligned}
    &\mathfrak{D}^{P}_{[\bf{3}]}=-4 \left(\xi _2^2-2 \xi _1 \xi _2+1\right)\partial^2_{\xi_2}+8 \left(\hat \Delta _3+1\right) \left(\xi _1-\xi _2\right)\partial_{\xi_2}-4 \hat\Delta _3 \left(\hat\Delta _3+1\right)\,,\\
    &\mathfrak{D}^{P}_{[1\bar{2}]}=\partial_{\xi_1}\,,\qquad 
      \mathfrak{D}^{P}_{[1\bar{3}]}=2\partial_{\xi_2}\,,\qquad 
      \mathfrak{D}^{P}_{[2\bar{3}]}=-2\xi_2\partial_{\xi_2}-2\,, \qquad 
      \mathfrak{D}^{T}_{[1\bar{2}]}=-\partial_{\xi_3}\,.
\end{aligned}
\end{equation}
We also computed the relative coefficient, which we shall exemplify below for the  component $[{\bf 3}]$ of the PSU,
{
\medmuskip=0mu
\thinmuskip=0mu
\thickmuskip=0mu
\begin{equation}
\begin{aligned}
\mathfrak{r}^P_{[{\bf 3}],0,0,0} &= \hat{\Delta}_3^2 - (\hat{\Delta}-\hat{\Delta}')^2,\\
\mathfrak{r}^P_{[{\bf 3}],2,0,0} &= 
 -\frac{(-\hat{\Delta}+\hat{\Delta}'+\hat{\Delta}_3)
        (-\hat{\Delta}+\hat{\Delta}'+\hat{\Delta}_3+2)
        (-\hat{\Delta}-\hat{\Delta}'-\hat{\Delta}_3+p)
        (\hat{\Delta}-\hat{\Delta}'-\hat{\Delta}_3+p-2)}
       {(p-2\hat{\Delta}')(p-2(\hat{\Delta}'+1))},\\
\mathfrak{r}^P_{[{\bf 3}],0,2,0} &= \mathfrak{r}^P_{[{\bf 3},2,0]} \Big|_{\hat{\Delta}\leftrightarrow \hat{\Delta}'},\\
\mathfrak{r}^P_{[{\bf 3}],2,2,0} &=
 -\frac{(-\hat{\Delta}+\hat{\Delta}'-\hat{\Delta}_3)(-\hat{\Delta}+\hat{\Delta}'+\hat{\Delta}_3)
        (\hat{\Delta}+\hat{\Delta}'-\hat{\Delta}_3)}
       {(-2\hat{\Delta}'+p-2)(p-2\hat{\Delta}')(-2\hat{\Delta}+p-2)(p-2\hat{\Delta})} \\
 &\quad\times 
 (-\hat{\Delta}-\hat{\Delta}'-\hat{\Delta}_3+p-2)
       (-\hat{\Delta}'-\hat{\Delta}-\hat{\Delta}_3+p)
       (-\hat{\Delta}'-\hat{\Delta}-\hat{\Delta}_3+2p-2).
\end{aligned}
\end{equation}
}
Note that Appell functions reduce to a ${}_4F_3$ hypergeometric function in special kinematics \cite{Lauria:2020emq}, therefore these relations also apply to such hypergeometric functions. The bulk channel block for this correlator is also known when $q=1$ \cite{Chen:2023oax}, so our method can likewise be used to derive relations for those blocks.
Moreover, the bulk two-point relations studied in the main text could be used to generate identities for Harish-Chandra functions (since the bulk-channel blocks can be written in terms of these functions).
Let us conclude this appendix by restating the importance of the uplift, also as a tool to derive identites for special functions, which might be worthwhile to investigate further also from a purely mathematical perspective.

\bibliographystyle{JHEP}
\bibliography{mybib}

    \end{document}